\documentclass[10pt,journal,compsoc]{IEEEtran}

\usepackage{booktabs} 
\usepackage[linesnumbered,ruled]{algorithm2e} 
\usepackage[numbers,sort&compress]{natbib}
\usepackage{amsmath}
\usepackage{amssymb}
\usepackage{caption}
\usepackage{subcaption}
\usepackage{xspace}
\usepackage{multirow}
\usepackage{color}
\usepackage{graphics}
\usepackage{epstopdf}
\usepackage{epsfig}
\usepackage{amsfonts}
\usepackage{graphicx}
\usepackage{enumitem}
\usepackage{wrapfig}
\usepackage{picinpar}
\usepackage{xcolor}
\usepackage{xpatch}
\usepackage{array}
\usepackage{url}
\usepackage{bbm}
\usepackage{wrapfig}
\usepackage{diagbox}
\usepackage{textcomp}
\usepackage{color}
\usepackage{picins}
\usepackage[normalem]{ulem}
\useunder{\uline}{\ul}{}

\def\BibTeX{{\rm B\kern-.05em{\sc i\kern-.025em b}\kern-.08em
    T\kern-.1667em\lower.7ex\hbox{E}\kern-.125emX}}

\begin{document}

\title{G\textsuperscript{3}R: Generating Rich and Fine-grained mmWave Radar Data from 2D Videos for Generalized Gesture Recognition}

\author{Kaikai~Deng,~\IEEEmembership{Student~Member, ~IEEE}, ~Dong~Zhao, ~\IEEEmembership{Member,~IEEE} ,~Wenxin~Zheng, ~Yue~Ling, ~Kangwen~Yin, and~Huadong~Ma,~\IEEEmembership{Fellow,~IEEE}
\IEEEcompsocitemizethanks{
\IEEEcompsocthanksitem K. Deng, D. Zhao, W. Zheng, Y. Ling, K. Yin, and H. Ma are with Beijing Key Laboratory of Intelligent Telecommunication Software and Multimedia, School of Computer Science, Beijing University of Posts and Telecommunications, Beijing, 100876, China. \protect
E-mail:$\{$dkk, dzhao, zhengwenxin, lingyue, 393473185, and mhd$\}$@bupt.edu.cn}

\thanks{(Corresponding authors: Huadong Ma and Dong Zhao.)}
}

\IEEEtitleabstractindextext{
\begin{abstract}
Millimeter wave radar is gaining traction recently as a promising modality for enabling pervasive and privacy-preserving gesture recognition. However, the lack of
rich and fine-grained radar datasets hinders progress in developing generalized deep learning models for gesture recognition across various user postures (e.g., standing, sitting), positions, and scenes. To remedy this, we resort to designing a software pipeline that exploits wealthy 2D videos to generate realistic radar data, but it needs to address the challenge of simulating diversified and fine-grained reflection properties of user gestures. To this end, we design \texttt{G\textsuperscript{3}R} with three key components: (i) a \textit{gesture reflection point generator} expands the arm's skeleton points to form human reflection points; (ii) a \textit{signal simulation model} simulates the multipath reflection and attenuation of radar signals to output the human intensity map; (iii) an \textit{encoder-decoder model} combines a \textit{sampling module} and a \textit{fitting module} to address the differences in number and distribution of points between generated and real-world radar data for generating realistic radar data. We implement and evaluate \texttt{G\textsuperscript{3}R} using 2D videos from public data sources and self-collected real-world radar data, demonstrating its superiority over other state-of-the-art approaches for gesture recognition.
\end{abstract}

}

\maketitle

\section{Introduction}\label{sec1}
In recent years, millimeter wave (mmWave) radar has gained growing interest among researchers as a reliable sensing modality for supporting pervasive and privacy-preserving human sensing \cite{deng2022geryon, deng2022global, liu2020real, liu2022mtranssee, ahuja2021vid2doppler, deng2023midas, zhang2022synthesized}, as it remains unaffected by adverse conditions (e.g., foggy weather, poor illumination) due to its strong permeability while not revealing texture information of users. Radar-based studies are also gradually developing from \textit{coarse-grained} (i.e., sensing tasks typically focus on users' whole-body movements) activity recognition \cite{ahuja2021vid2doppler} and object detection \cite{deng2022geryon} to \textit{fine-grained} (i.e., sensing tasks typically focus on fine-grained changes in user gestures) gesture recognition \cite{liu2022mtranssee, liu2021m}, enabling applications such as assisting users to effortlessly control common household appliances according to their gestures under poor illumination conditions \cite{palipana2021pantomime, shen2022ml}, providing users with a better interactive experience in AR/VR according to recognized gestures \cite{ling2020ultragesture, waghmare2023z}, and assisting workers to operate machines according to recognized gestures in an industrial plant \cite{xia2022time, deng2023midas}.

In practice, it is difficult to achieve a generalized gesture recognition system, as it necessitates the consideration of various practical factors, such as user postures (standing, sitting, etc.), user positions, and collection scenes of gesture data. 
For example, the gesture recognition accuracy significantly drops from 96.7\% to 47\%, when training on a large-scale radar dataset of users performing gestures while standing, and testing with radar data of users performing gestures while sitting; similarly, changes in user positions and scenes also result in a drop in recognition accuracy by 18.08\% and 5.26\%, respectively (see Section \ref{sec2}).
To address this problem, a direct method is to employ rich data across various user postures, positions, and scenes to train a generalized deep learning model. Unfortunately, compared to video and sound datasets, there are very few corresponding radar gesture datasets \cite{liu2020real, liu2022mtranssee}, which are further constrained by a single human body posture, data collected at fixed positions, and limited scenes. Meanwhile, collecting and labeling a large-scale radar dataset require substantial human efforts, slowing down the research and development \cite{chen2021sensecollect, zhang2022synthesized}. Fortunately, there are many public 2D video datasets \cite{kuehne2011hmdb, abu2016youtube, perera2018uav, soomro2012ucf101, escalante2016chalearn} and websites (e.g., YouTube\footnote{https://www.youtube.com/}, Bilibili\footnote{https://www.bilibili.com/}) that offer abundant human gesture data across various user postures, positions, and scenes. Furthermore, existing studies have utilized 2D videos to generate other sensor data, such as Inertial Measurement Unit (IMU) \cite{kwon2020imutube, kwon2021approaching} and sound data \cite{liang2019audio, santhalingam2023synthetic}, for training deep learning models, which motivates us to explore the possibility of using 2D videos to generate realistic radar data.

Exciting studies have explored the generation of realistic radar data from different data sources, such as motion capture (MoCap) data \cite{lin2017performance, seyfioglu2018diversified} and depth camera data \cite{erol2015kinect, li2019kinect}. However, either the generated radar data is coarse or it lacks some common gestures. The generative adversarial networks (GANs) \cite{erol2019gan, rahman2021physics} are widely used to augment existing radar datasets, but it could easily lead to data confusion when performing similar gestures. Recently, some studies \cite{deng2023midas, ahuja2021vid2doppler, zhang2022synthesized} explore using 2D videos to generate radar data. However, these methods primarily focus on generating coarse-grained radar data by just capturing users' whole-body movements, which makes it difficult to characterize fine-grained changes in their gestures. In contrast, we focus on utilizing wealthy 2D videos to generate fine-grained radar gesture data, yet we will face a unique challenge.

\emph{How to simulate diversified and fine-grained reflection properties of user gestures}? Existing works \cite{deng2023midas, ahuja2021vid2doppler, zhang2022synthesized} can generate 2D Range-Doppler signals or point clouds using 2D videos. However, either the signal is difficult to accurately characterize the 3D spatial representation of gestures
or the point cloud is coarse-grained.
Given the corresponding regions of each human body part, we can calculate the necessary components of radar data, i.e., the radar cross-section (RCS) and radial velocity of every vertex, in each region with respect to a radar sensor. However, there are three different influencing factors: user postures, user positions, and various scenes, all of which would lead to diversified differences in the gesture data produced by users. If each vertex is directly extracted to obtain the corresponding RCS and radial velocity without considering the multipath reflection and attenuation of radar signals during the transmitting and receiving process, this diversified and fine-grained difference will be ignored, resulting in lower fidelity of the generated radar data.

To address this challenge, we design a unique data generation system, \texttt{G\textsuperscript{3}R}, that allows converting wealthy 2D videos into fine-grained radar data. Specifically, a \textit{human parsing} model, a \textit{skeleton extraction} model, and a \textit{depth prediction} model extract human constituent parts, skeleton points, and depth information, respectively, to prepare for subsequent modules. Considering that the main reflection points of the human body focus on the arm, we design a \textit{gesture reflection point generator}, which expands the reflection points of the arm by using random interpolation based on the arm's skeleton points. However, interpolated reflection points may experience depth value shifts during arm movement. To this end, we map human constituent parts to the generated reflection points, correcting any overflow points and ensuring that the generated points completely belong to the arm.

Based on the obtained reflection points, we can calculate their corresponding RCS and radial velocity, but directly using them will result in lower quality of the generated radar data due to neglecting the multipath reflection and attenuation of radar signals.
To address this problem, a \textit{signal simulation model} is designed, which takes RCS and depth information as inputs to simulate the propagation characteristics of radar signals during the transmitting and receiving process, followed by outputting the human intensity map. Due to the differences in various user postures, positions, and scenes, the generated radar data differs from the real-world radar data in number and distribution of points. Therefore, an \textit{encoder-decoder model} combines a \textit{sampling} module
and a \textit{fitting} module to generate realistic radar data through graph convolution and matrix transformation. Besides, we utilize both a large amount of generated and a small amount of real-world radar data to train gesture recognition models to further enhance their generalization.

In summary, our contributions are as follows:  
\begin{itemize}
\item To our knowledge, this is the first work on a system called \texttt{G\textsuperscript{3}R} that utilizes wealthy 2D videos to generate rich and fine-grained radar data for developing a generalized gesture recognition model across various user postures, positions, and scenes. 
\item We propose a suit of novel and effective techniques: (i) a \textit{gesture reflection point generator} utilizes the extracted skeleton points to expand the reflection points of the arm; (ii) a \textit{signal simulation model} simulates the multipath reflection and attenuation of radar signals during the transmitting and receiving process, followed by outputting human intensity map; (iii) an \textit{encoder-decoder model} combines a \textit{sampling} module and a \textit{fitting} module to generate realistic radar data.
\item We implement and extensively evaluate \texttt{G\textsuperscript{3}R} for gesture recognition with 2D video data from five public datasets \cite{kuehne2011hmdb, abu2016youtube, perera2018uav, soomro2012ucf101, escalante2016chalearn}, YouTube, and Bilibili, as well as self-collected real-world radar data from 32 volunteers (a total of 23,040 samples collected over 3 months). 
The experimental results show that \texttt{G\textsuperscript{3}R} achieves 90.51\% accuracy, which surpasses three state-of-the-art approaches, \textit{Vid2Doppler} \cite{ahuja2021vid2doppler}, \textit{SynMotion} \cite{zhang2022synthesized}, and \textit{Midas} \cite{deng2023midas} by 67.41 percentage points (pp), 66.28 pp, and 51.36 pp, respectively, when training a model solely with all generated radar data; if we add a small amount of real-world radar data for training, \texttt{G\textsuperscript{3}R} achieves 97.32\% accuracy,
61.80 pp, 54.67 pp, and 34.45 pp higher than that of \textit{Vid2Doppler}, \textit{SynMotion}, and \textit{Midas}, respectively.
Moreover, when facing a new user, the model trained on all generated data only requires 6 samples per gesture to achieve the recognition accuracy of 96.76\%, while it only requires 6 samples per gesture to achieve 98.53\% accuracy by combining a small amount of real-world radar data for training.
Besides, we deeply evaluate the impact of various factors using self-collected real-world radar data from 5 volunteers (a total of 8400 samples). \texttt{G\textsuperscript{3}R} achieves 90.06\% and 96.99\% accuracy across various user postures, positions, and scenes when training with all generated radar data only and all generated radar data with a small amount of real-world radar data, respectively.
\end{itemize}
\section{Motivation and Challenges}\label{sec2}
\subsection{Necessity of Generating Rich Radar Data}\label{sec2.1}
\begin{figure*}[t]
  \centering
  \centerline{\includegraphics[width=140mm]{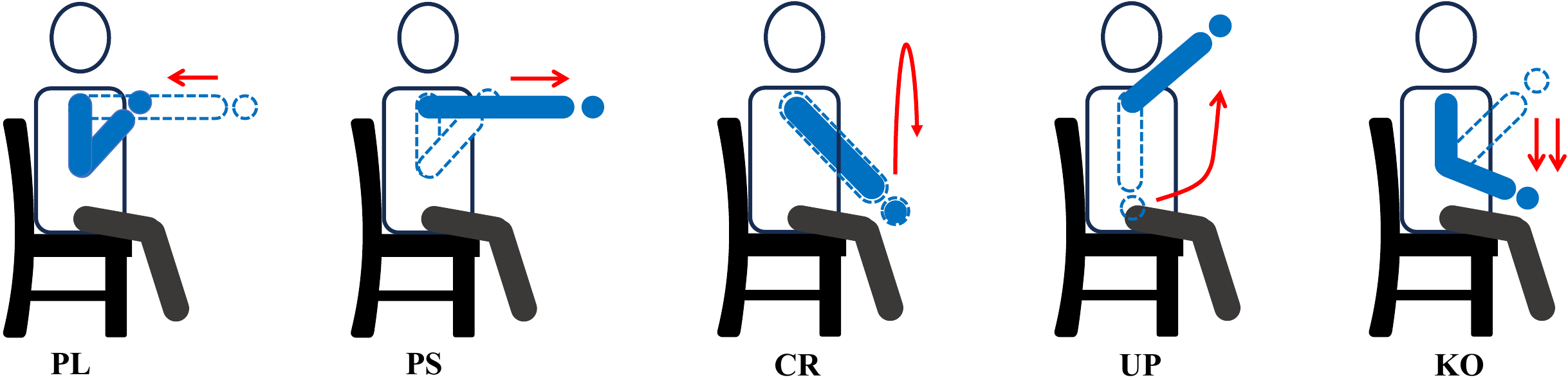}}
    \vspace{-3mm}
  \caption{Five different gestures. Red arrows represent the gesture's movement directions, while blue dotted and solid lines represent the starting and ending of different gestures, respectively.}\label{fig1}
  \vspace{-3mm}
\end{figure*}

Liu et al. \cite{liu2022mtranssee} have collected and openly shared a large-scale radar dataset of users performing gestures with a single posture (standing). To verify the necessity of generating rich radar data, we collect a radar dataset of users performing gestures while sitting, including 32 volunteers, 5 gestures (pull a hand (PL), push a hand (PS), draw a circle (CR), lift up a hand (UP), and knock a virtual table twice (KO), as shown in Fig. \ref{fig1}), 9 positions, and 2 scenes. Each gesture is performed 8 times, resulting in a total of 32 $\times$ 5 $\times$ 9 $\times$ 2 $\times$ 8 = 23040 samples. Using this dataset, we evaluate the influence of various user postures, positions, and scenes. Note that we utilize the control variable method to maintain two factors constant while verifying the impact of the remaining one factor.
\begin{figure*}[t]
  \centering
  \centerline{\includegraphics[width=160mm]{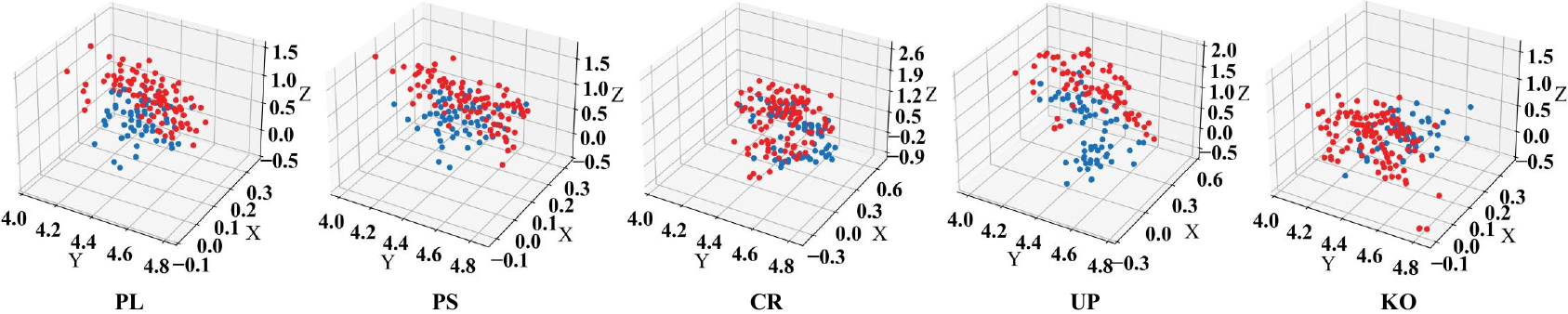}}
    \vspace{-3mm}
  \caption{Examples of real-world radar data for five gestures. Red and blue points represent the data distribution of a user performing the same gesture while standing and sitting, respectively.}\label{fig2}
  \vspace{-5mm}
\end{figure*}

\begin{figure}[t]
  \centering
  \centerline{\includegraphics[width=49mm]{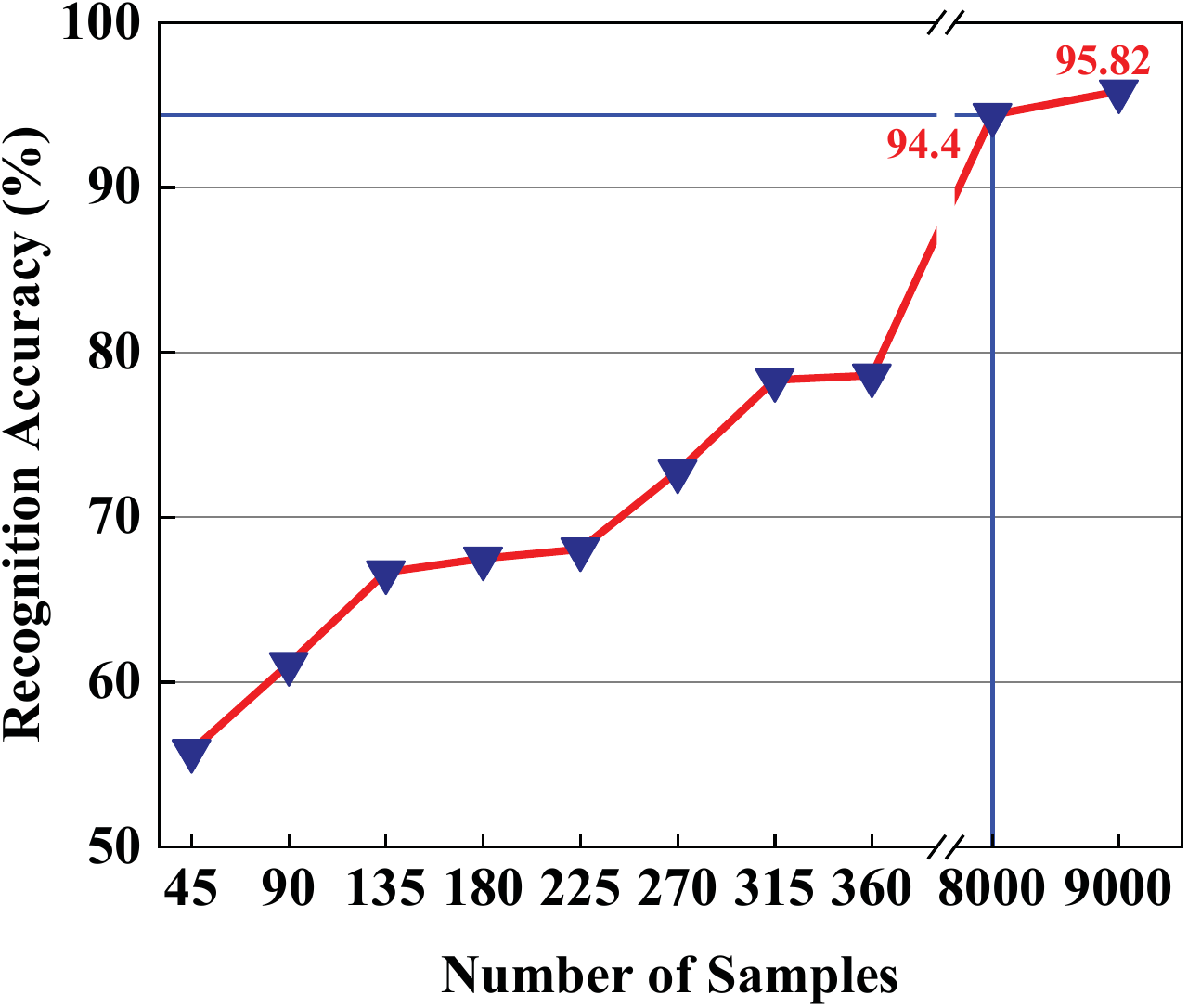}}
  \vspace{-2mm}
  \caption{Recognition accuracy as sitting posture samples increase.}\label{fig3}
   \vspace{-5mm}
\end{figure}

\textbf{Impact of User Postures}. Different user postures would cause the reflection of radar signals to change. Fig. \ref{fig2} shows the data distribution of a user performing the same gesture while standing and sitting, respectively. To verify the influence of different user postures on gesture recognition accuracy, we train a state-of-the-art radar-based gesture recognition model, \textit{mTransSee} (hereinafter referred to as the model) \cite{liu2022mtranssee}, using a large-scale radar dataset of users performing gestures while standing, and test it with radar data of users performing gestures while sitting. The gesture recognition accuracy significantly drops from 96.7\% to 47\%, which indicates significant disparities in the radar data produced by different user postures. Moreover, we augment the existing gesture recognition dataset with real-world data of users performing gestures while sitting to train the model. As depicted in Fig. \ref{fig3}, we observe that the recognition accuracy has an upward trend with the increase of data. When the number of samples increases to 8000, the recognition accuracy reaches 94.4\%. Therefore, it is imperative to continuously add a substantial amount of sitting posture samples for training to further improve the model's performance.
\begin{figure}[t]
  \centering
  \centerline{\includegraphics[width=49mm]{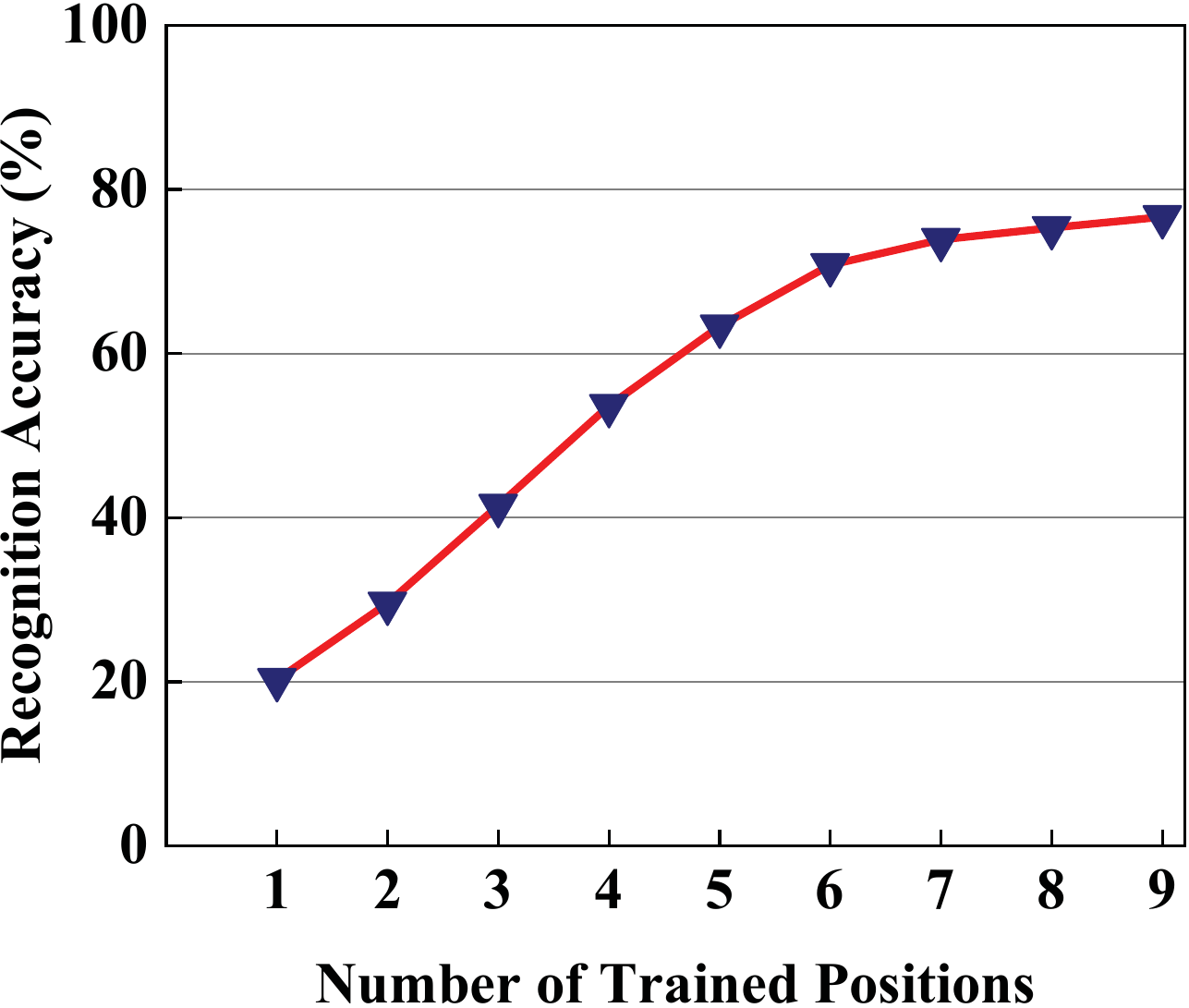}}
  \vspace{-2mm}
  \caption{Recognition accuracy as the number of positions increases.}\label{fig4}
   \vspace{-5mm}
\end{figure}

\textbf{Impact of User Positions}. The number of points and signal intensity are different when a user performs gestures in various positions, and this observation has also been validated by \textit{mTransSee} \cite{liu2022mtranssee}. Based on the collected dataset, we conduct corresponding experiments to demonstrate that the recognition accuracy is affected by user positions when a user performs gestures while sitting. We evaluate the recognition accuracy of user gestures at deviations of 50 cm from 9 positions. As shown in Fig. \ref{fig4}, we observe that: (i) the recognition accuracy is continuously improved with increased training data from different positions; (ii) even when training the model with data from 9 positions, the recognition accuracy only reaches a maximum of 76.66\%. Note that the model can achieve 94.74\% accuracy when training and testing on radar data from 9 positions. Therefore, the model requires more data from different positions for training to achieve higher accuracy that satisfies users' requirements.

\begin{figure}[t]
  \centering
  \centerline{\includegraphics[width=35mm]{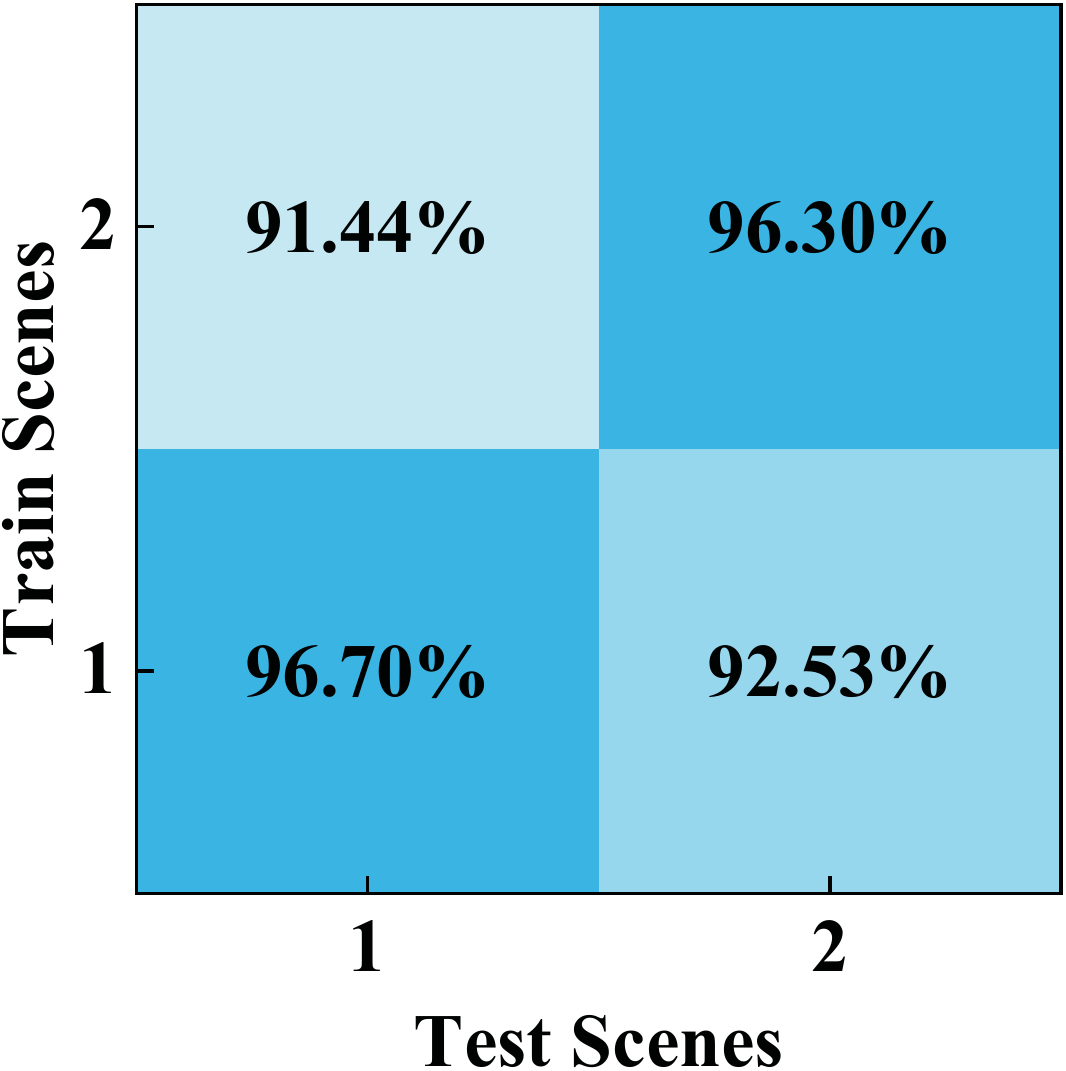}}
  \vspace{-1mm}
  \caption{Recognition accuracy with different scenes.}\label{fig5}
   \vspace{-4mm}
\end{figure}

\textbf{Impact of Different Scenes}. There are various reflectors in different scenes. When a user performs different gestures close to a reflector (i.e., within 60 cm), the reflector generates multipath reflection points that cannot be effectively processed using common noise filtering methods, such as CFAR \cite{rohling1983radar} and DBSCAN \cite{ester1996density}, making it difficult to differentiate between the signal reflection generated by user gestures and that of the reflector in various scenes. In this case, if the model does not learn sufficient knowledge, it will suffer a drop in recognition accuracy. To clearly show this, we perform a leave-one-scene-out cross validation. As shown in Fig. \ref{fig5}, we observe that: (i) when training using data from scene 1 and testing with scene 2, the recognition accuracy drops by 5.26 pp; (ii) when training using data from scene 2 and testing with scene 1, the recognition accuracy drops by 3.77 pp; the main reason is that the scene 2 is more complex, enabling the model to learn more about reflectors. Therefore, it is necessary to mix a large amount of data from different scenes for training to improve the model’s performance.

\textbf{Data Collection Cost}. To collect the above dataset, we spend 3 months. Similarly, Liu et al. \cite{liu2022mtranssee} collect a gesture dataset, enabling the model to accurately recognize gestures performed by users while standing, but it takes approximately 6 months. Apart from time cost, we must consider other complex factors that impact costs: (i) researchers need possess professional expertise and adeptly operate equipment while executing proficient data processing tasks. Note that some areas, particularly remote and underdeveloped areas, may suffer from a scarcity of professional talents, rendering data collection more difficult; (ii) data quality may be poor during collection, necessitating re-collection; (iii) some volunteers may quit after performing gestures for several hours due to boredom and fatigue, resulting in the interruption of data collection; (iv) unconscious gesture noise can be easily produced during data collection process; (v) labeling data is burdensome and prone to errors; (vi) there are many more user postures, user positions, and scenes in real life, which requires collecting more data to make deep learning models more generalized. Moreover, according to a survey \cite{chen2021sensecollect}, all researchers surveyed agree that collecting data is very difficult.

\textbf{Summary}. Based on the above analysis, we observe that the performance of gesture recognition is affected by many factors, requiring rich training data to continually enhance the model's performance, but collecting large-scale radar data is costly. 
Therefore, this motivates us to explore an efficient approach to address the lack of radar data and high collection cost.

\subsection{Opportunities}\label{sec2.2}
To obtain rich radar data, we tend to use other data sources to generate radar data. Inspired by this idea, we investigate existing data generation efforts and find that radar data can be generated from various data sources. However, different data sources have different pros and cons. \textit{First}, some works \cite{lin2017performance, seyfioglu2018diversified} utilize MoCap data to generate radar data, but such data is relatively sparse, resulting in very coarse radar data. 
\textit{Second}, some works \cite{erol2015kinect, li2019kinect} use depth camera data to generate radar data, but these datasets only have few gesture classes. 
In contrast, the ubiquity of mobile devices has produced vast amounts of 2D videos daily \cite{statista}, which are either collected as public datasets \cite{kuehne2011hmdb, abu2016youtube, perera2018uav, soomro2012ucf101, escalante2016chalearn} or accessible on some websites (e.g., YouTube, Bilibili), covering rich gesture content. Meanwhile, the computer vision technology enables the accurate extraction of 3D postures and skeletal points of humans from 2D videos, offering a foundation for exploring and generating large-scale radar data. Therefore, we set out to design a software pipeline that utilizes wealthy 2D videos to generate rich radar data, addressing the lack of radar data and reducing the cost of collecting real-world radar data.

\begin{figure}[t]
\centering
\subfloat[User postures]{\vspace{-1mm}
\begin{minipage}[b]{0.3\textwidth}
\centering
\centerline{\includegraphics[width=53mm]{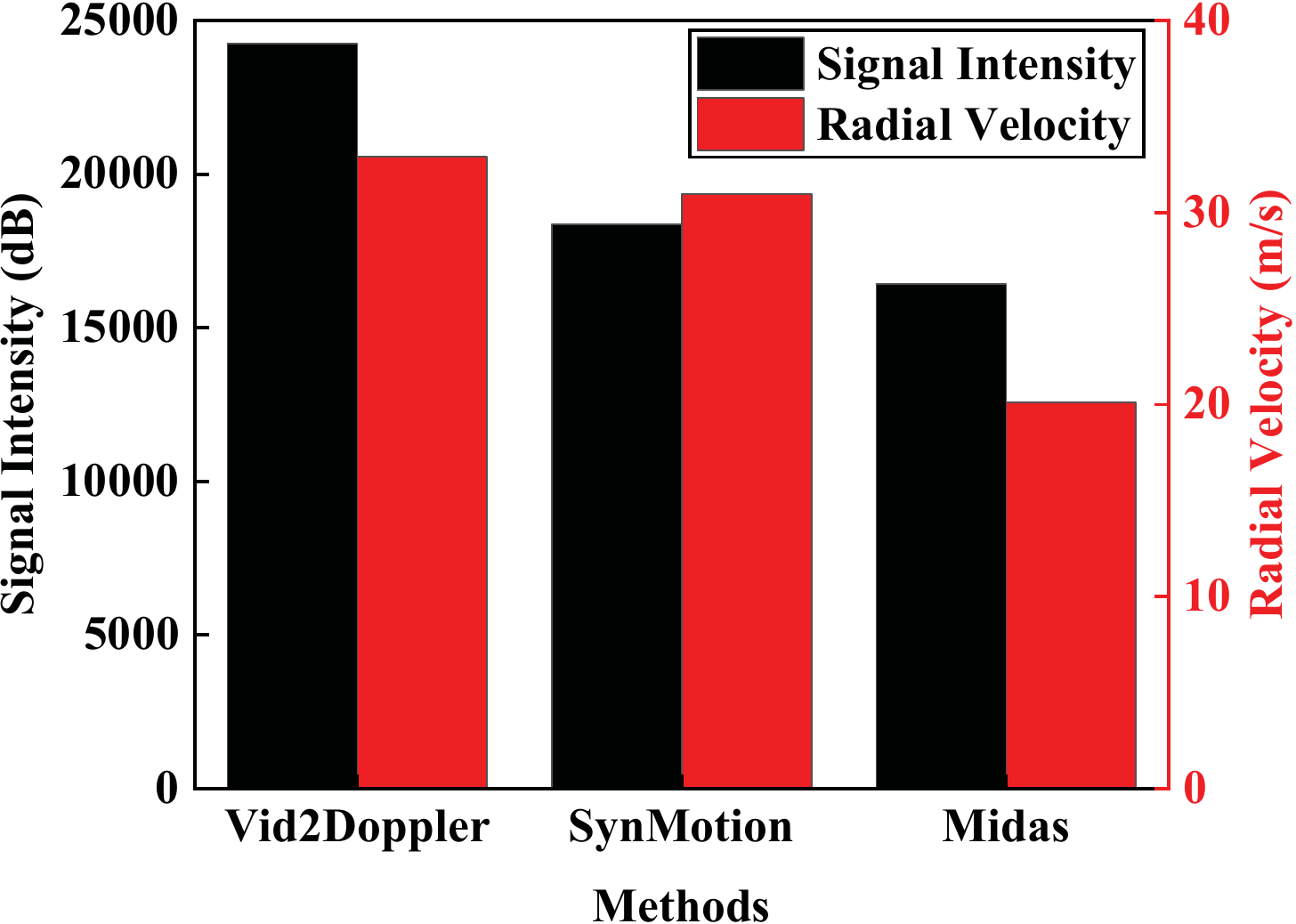}}
\end{minipage}
}
\vfill
\subfloat[User positions]{\vspace{-1mm}
\begin{minipage}[b]{0.3\textwidth}
\centering
\centerline{\includegraphics[width=53mm]{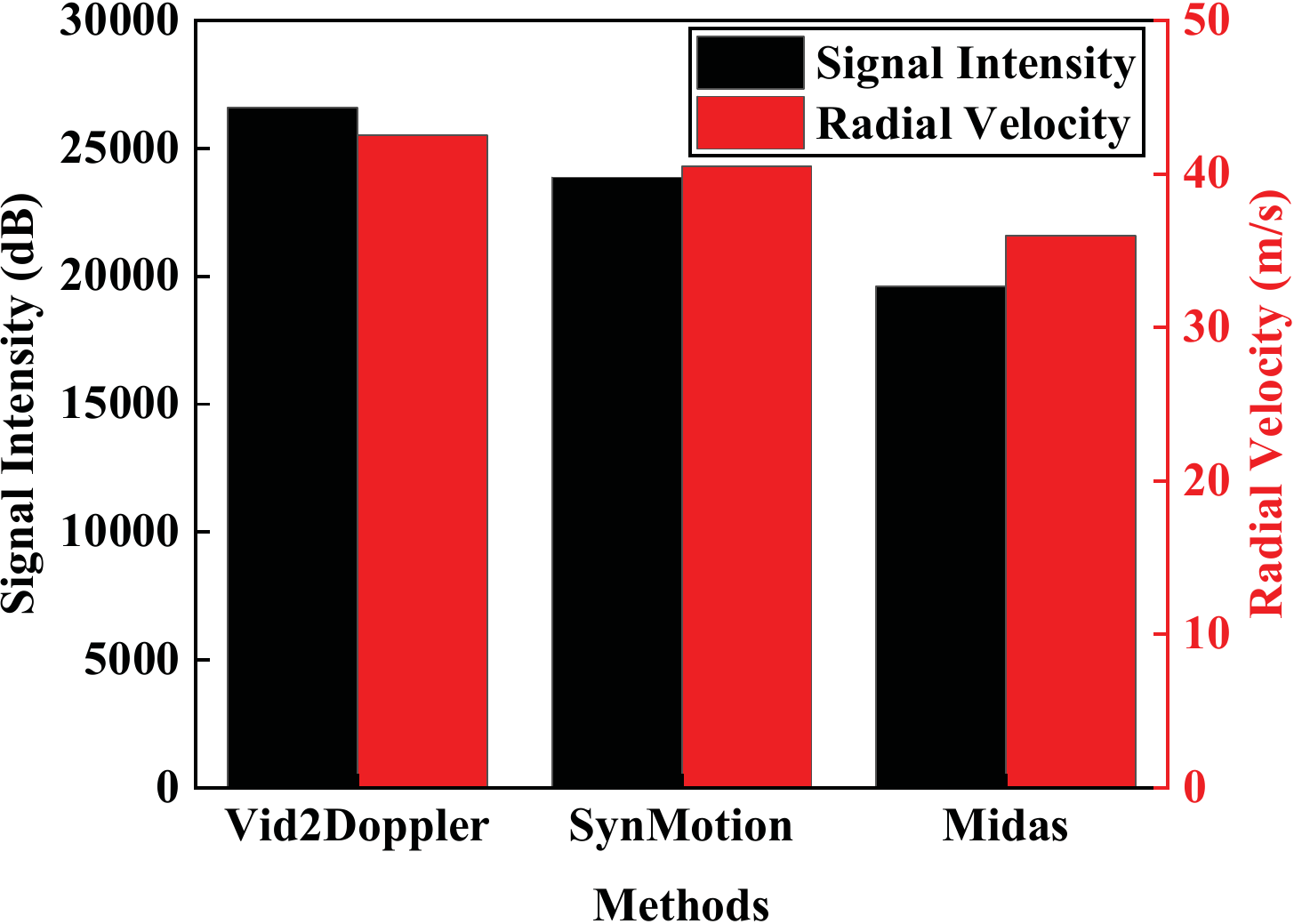}}
\end{minipage}
}
\vfill
\subfloat[Scenes]{\vspace{-1mm}
\begin{minipage}[b]{0.3\textwidth}
\centering
\centerline{\includegraphics[width=53mm]{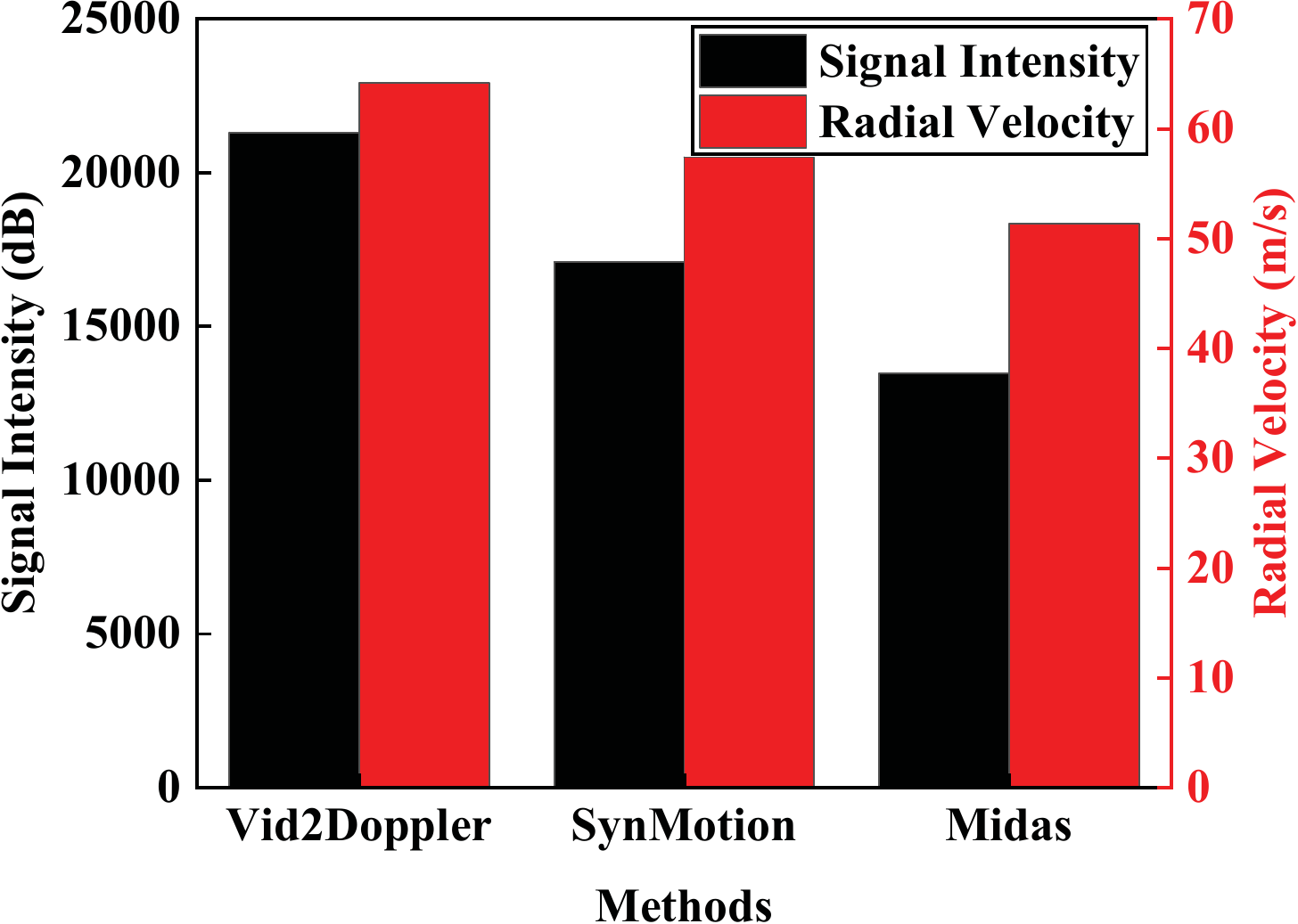}}
\end{minipage}
}
\vspace{-2mm}
\caption{Average cumulative errors of signal intensity and radial velocity for different methods with various factors.}\label{fig6}
\vspace{-5mm}
\end{figure}

\subsection{Design Challenges}\label{sec2.3}
The main challenge in designing the \texttt{G\textsuperscript{3}R} system lies in simulating diversified and fine-grained reflection properties of user gestures. Although recent works \cite{deng2023midas, zhang2022synthesized} consider the reflection properties of radar signals when generating radar data, they mainly simulate the coarse-grained motion reflections of humans, and the generated data is difficult to accurately characterize the diversified and fine-grained gesture features. 
We evaluate the quality of data generated by existing works under various factors, as shown in Fig. \ref{fig6}.
Note that each gesture sample typically comprises dozens or hundreds of reflection points, with each one being linked to a corresponding signal intensity value. The signal intensity range for each reflection point (each chirp) typically lies within the range of -150 to +150 dB. However, it is meaningless to compare the signal intensity for each reflection point, so we calculate the average cumulative errors in the signal intensity and radial velocity for all collected gesture samples.
Compared to the real values, there are significant average cumulative errors in the signal intensity and radial velocity obtained through three state-of-the-art methods \cite{deng2023midas, zhang2022synthesized, ahuja2021vid2doppler} under 2 user postures (standing and sitting), 9 user positions, and 2 scenes. For example, for the latest method \textit{Midas} \cite{deng2023midas}, the errors for signal intensity/radial velocity reach 16408 dB/20.12 m/s, 19608 dB/35.96 m/s, and 13472 dB/51.35 m/s, under three different factors, respectively. Note that according to our experience, we need these two errors to be less than 4000 dB and 15 m/s, respectively, to satisfy users' requirements (see Section \ref{sec5.2}). This shows that it is an important problem to simulate diversified and fine-grained reflection properties of user gestures. If the errors are not effectively addressed, it will hinder the model from learning the real features of each gesture, causing a drop in recognition performance.
\begin{figure*}[t]
\begin{minipage}[htb]{0.6\linewidth}
\centering
\includegraphics[width=10cm,height=4.2cm]{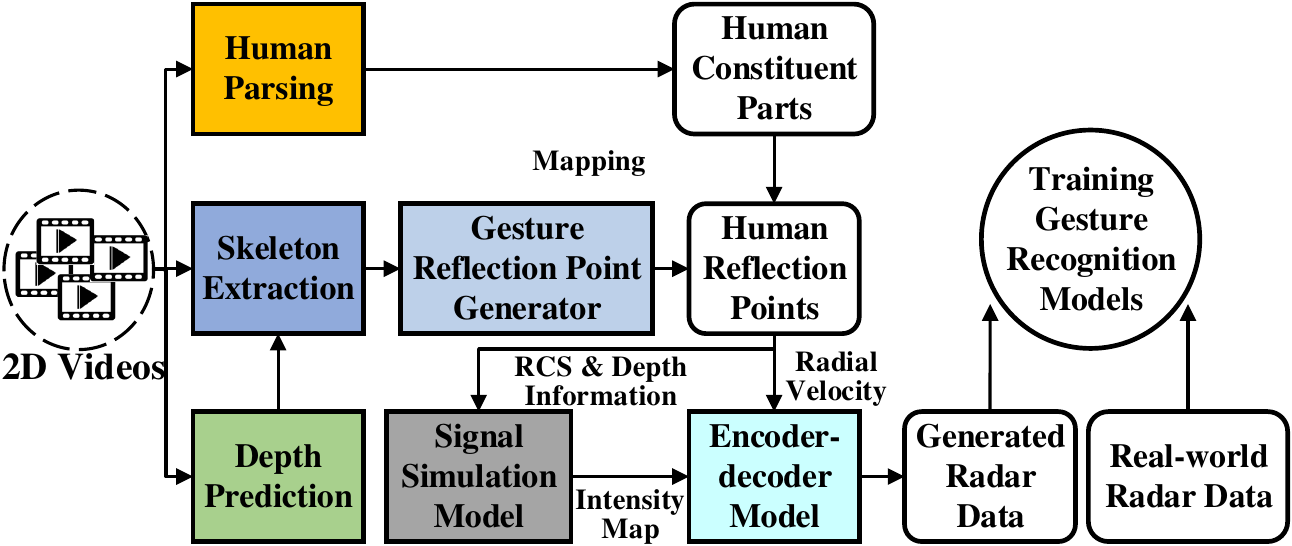}	
\vspace{-2mm}
\caption{System overview of \texttt{G\textsuperscript{3}R}.}\label{fig7}
\end{minipage}
\quad 
\begin{minipage}[htb]{0.35\linewidth}
\centering
\includegraphics[width=7cm,height=3.5cm]{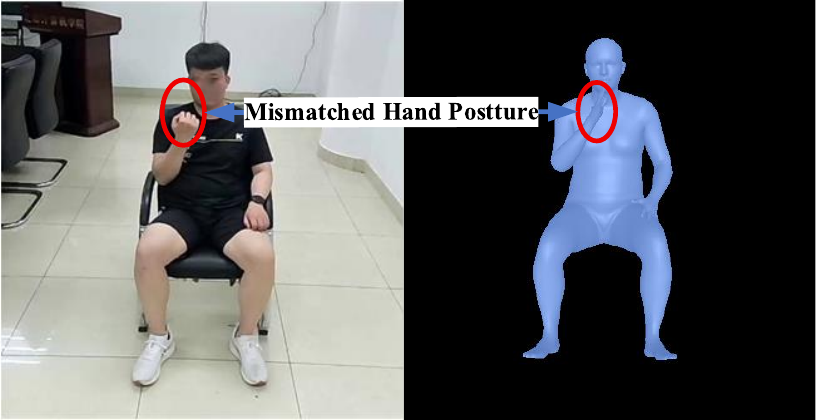}	
\vspace{0.5mm}
\caption{Illustration of hand mesh extraction.}\label{fig8}
\end{minipage}
\vspace{-4mm}
\end{figure*}
\section{System Overview}\label{sec3}
As shown in Fig. \ref{fig7}, \texttt{G\textsuperscript{3}R} consists of six modules. Specifically, there are differences in the reflections of various human body parts when a user performs gestures, with the primary reflection concentrated on the arm (including the hand) rather than other human body parts. To this end, we utilize a \textit{human parsing} model to obtain different human constituent parts and accurately output gesture regions, followed by characterizing the reflection differences in various regions of the human body. Meanwhile, a \textit{skeleton extraction} model is used to extract the skeleton points of the human body. Since the main reflection points are focused on the arm, we design a \textit{gesture reflection point generator}, which exploits the random interpolation method to expand the skeleton points of the arm (the number of interpolation points can be obtained according to the statistical results of real-world radar data), followed by concatenating skeleton points of other human body parts to obtain the human reflection points. Note that since the extracted skeleton points are 2D data, a \textit{depth prediction} model is used to supplement their depth information.

Based on the human reflection points, we can calculate their RCS and radial velocity. However, there will be serious multipath reflection and attenuation during the transmitting and receiving process of radar signals, especially when there are many reflectors. To address this problem, a \textit{signal simulation model} is designed, which utilizes RCS and depth information as inputs to simulate the propagation characteristics of radar signals to output the human intensity map, followed by concatenating the radial velocity to generate the initial radar data. Considering the changes in user postures, positions, and scenes, there are differences in number and distribution of points between the generated and real-world radar data. Therefore, an \textit{encoder-decoder model} is designed, which consists of a \textit{sampling} module and a \textit{fitting} module; the former utilizes graph convolution to obtain the number and distribution correlation of points between generated and real-world data, followed by using sampling layers to output generated points, and the latter employs matrix transformation to fit radial velocity and intensity, thus generating realistic radar data.
Besides, we train gesture recognition models using both a large amount of generated and a small amount of real-world radar data to further improve their generalization in practical scenes.
\section{Detailed design of \texttt{G\textsuperscript{3}R}}\label{sec4}
Now we introduce the detailed design of \texttt{G\textsuperscript{3}R}, which consists of \textit{human parsing}, \textit{skeleton extraction}, \textit{depth prediction}, \textit{gesture reflection point generator}, \textit{signal simulation model}, and \textit{encoder-decoder model}.

\subsection{Human Parsing}\label{sec4.1}
In general, when a user performs gestures, the arm moves more than the rest of the human body, causing more reflections. Therefore, to obtain fine-grained gesture data, it is crucial to precisely parse human constituent parts. The goal of human parsing is to partition a human in 2D videos into different constituent parts. However, existing methods \cite{zhang2020blended, zhang2022human} either rely on predefined human body hierarchies or accurate human postures, making it difficult to ensure generalization in scenes involving multiple humans or unexpected occlusions of human parts. Considering the hierarchical structure of the human body, each body part in an image can possess its unique position distribution characteristic. Therefore, we employ a state-of-the-art \textit{human parsing} model (CDGNet) \cite{liu2022cdgnet}, which generates instance class distributions by accumulating raw human parsing labels in horizontal and vertical directions, thereby providing valuable supervisory information. By leveraging these horizontal and vertical class distribution labels, CDGNet is guided to mining the intrinsic position distribution of each class. Finally, CDGNet combines two guided features into a spatial guidance map, which is subsequently superimposed on the baseline network through multiplication and concatenation to accurately parse human body parts.
\begin{figure}[t]
  \centering
  \centerline{\includegraphics[width=23mm]{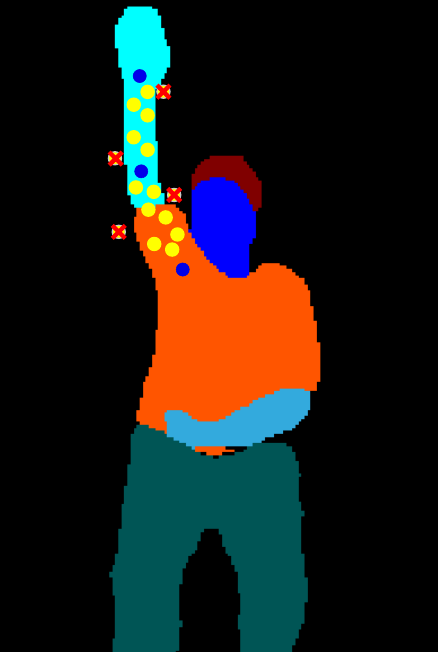}}
  \vspace{-1mm}
  \caption{Illustration of reflection point mapping. Blue, yellow, and crossed yellow dots represent the three initial skeleton points, randomly interpolated points, and overflow points, respectively.}\label{fig9}
   \vspace{-6mm}
\end{figure}

\subsection{Skeleton Extraction}\label{sec4.2}
Based on the input 2D videos, we adopt a \textit{skeleton extraction} model (RSN) \cite{cai2020learning} to extract the skeleton points of different human body parts. This process can avoid using highly complex human mesh models \cite{sun2022putting, guan2022out}; the main reason is that they are difficult to extract fine-grained hand features. As shown in Fig. \ref{fig8}, we visualize the extraction result of a state-of-the-art human mesh model (BEV) \cite{sun2022putting}. When the hand posture is a fist, the output of the model is an open hand, making it difficult to accurately map gesture features. Therefore, we use RSN to extract the skeleton points of different human constituent parts. RSN efficiently aggregates features with the same spatial size to obtain refined local representations while using an attention mechanism to weigh the output local and global features, enabling accurate localization and extraction of skeleton points.

\subsection{Depth Prediction}\label{sec4.3}
Given an input 2D video, we utilize a state-of-the-art \textit{depth prediction} model (ZoeDepth) \cite{bhat2023zoedepth} to obtain the depth information of humans. ZoeDepth first pre-trains an encoder-decoder architecture that is trained using relative depth information gathered from various datasets, and then adds domain-specific heads for metric bins module to the encoder-decoder architecture, followed by fine-tuning them on metric depth datasets. Moreover, in the inference process, ZoeDepth uses a classifier on encoder features to automatically route an image to the appropriate head for further improving the performance of depth estimation. In \texttt{G\textsuperscript{3}R}, considering the propagation characteristics of radar signals, the reflection points are mainly focused on the arm with a large movement amplitude. Therefore, we only need to accurately obtain the depth information of the arm. To this end, we average the depth values near the three skeleton points on the arm to further reduce the prediction error.

\subsection{Gesture Reflection Point Generator}\label{sec4.4}
Based on the obtained depth information, we perform a coordinate transformation and concatenate it to the 2D skeleton points to form 3D skeleton points, characterizing the corresponding reflection points of the human body. However, there are 19 human skeleton points extracted with only 3 skeleton points on the arm, making it difficult to characterize the real number and distribution of radar reflection points for each gesture. Therefore, we design a \textit{gesture reflection point generator}, which expands the reflection points of the arm based on skeleton points on the arm.
Specifically, we randomly interpolate the skeleton points of the arm to expand the reflection points on it to realistically simulate the movement characteristics of the arm. 
Note that as the default configuration for the radar sensor sets the number of points collected per frame to 64, we uniformly expand the original 3 skeleton points to 64 reflection points. During the random interpolation process, we observe that there is
a possibility of encountering reflection point overflow, which causes an incorrect correspondence between the reflection points and the arm.
The main reason is that the depth prediction will have some deviations with the movement of user gestures, resulting in significant errors for the interpolation points, as shown in Fig. \ref{fig9}. To solve this problem, we map human constituent parts partitioned by the \textit{human parsing} model with the reflection points of the human body, followed by checking the interpolation points to remove these reflection points that do not belong to the arm. Note that for the removed points, we will also re-execute the interpolation to supplement.

\subsection{Signal Simulation Model}\label{sec4.5}
\begin{figure}[t]
  \centering
  \centerline{\includegraphics[width=25mm]{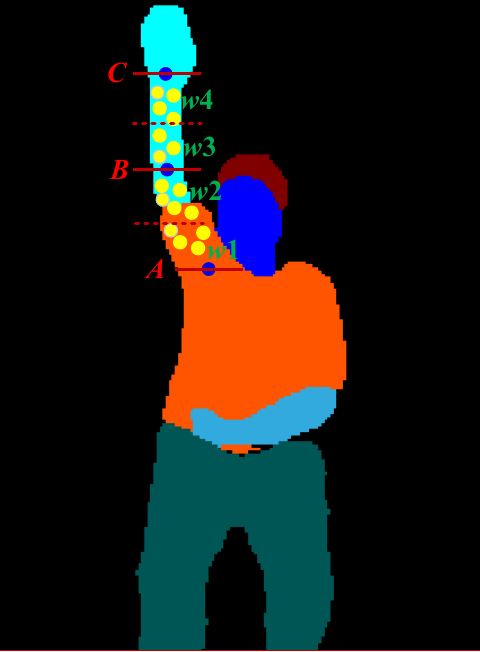}}
  \vspace{-2mm}
  \caption{Illustration of radial velocity of gesture reflection points.}\label{fig10}
   \vspace{-6mm}
\end{figure}

Based on the obtained human reflection points, we can calculate their RCS and radial velocity with respect to a radar sensor. For RCS, we convert the obtained reflection points into surface triangles, and the normal of all triangles is also calculated during the conversion process. Based on the obtained surface area and normal, we can calculate the RCS corresponding to each reflection point.
Meanwhile, we observe from the real-world radar data that the radial velocity of each gesture basically concentrates on a few values. Therefore, we perform a windowing strategy on the reflection points to calculate their radial velocities. As shown in Fig. \ref{fig10}, we can obtain their radial velocities based on the arm's three initial skeleton points ($A$, $B$, $C$). Note that the radial velocity of these 3 points can be obtained by looking back at the movement history (previous frames). Considering that during the movement of the arm, the speed of points $A$, $B$, and $C$ increases in order. Based on the statistical results of real-world data, we divide the points on the arm into four windows ($w$) on average, where the velocity of points within $w1$ aligns with point $A$, while the velocity of points within $w2$ and $w3$ corresponds to point $B$. Similarly, the velocity of points within $w4$ corresponds to point $C$. 
\begin{figure}[t]
  \centering
  \centerline{\includegraphics[width=63mm]{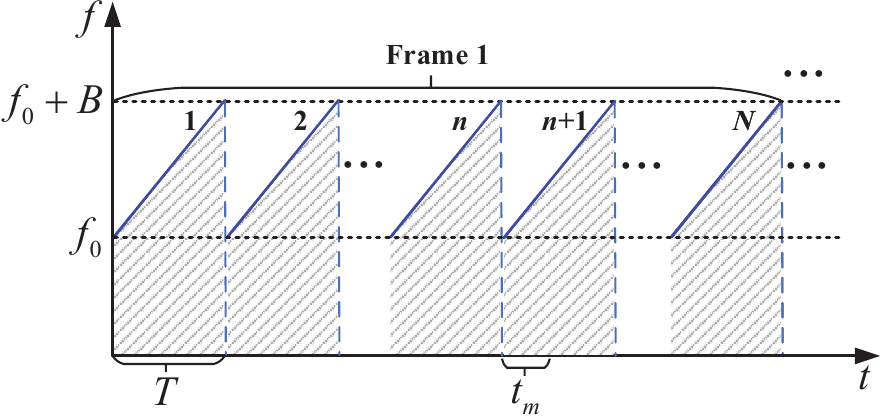}}
  \vspace{-3mm}
  \caption{Illustration of chirps in one frame. The shaded area represents the number of periods experienced since the starting of a frame.}\label{fig11}
   \vspace{-6mm}
\end{figure}
\begin{figure}[t]
  \centering
  \centerline{\includegraphics[width=56mm]{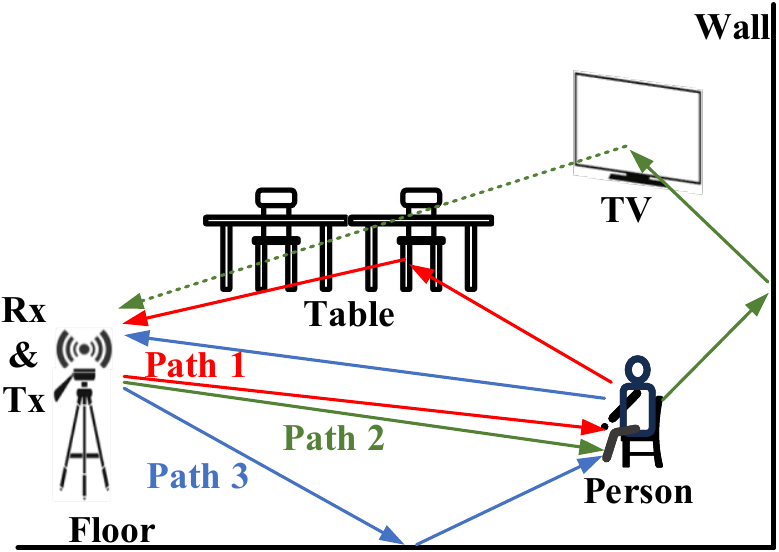}}
  \vspace{-2mm}
  \caption{Illustration of multipath reflection and attenuation of radar signals. Paths 1 and 3 represent the reflection process of radar signals among multiple reflectors. Path 2 represents an invalid radar signal due to signal attenuation.}\label{fig12}
  \vspace{-5mm}
\end{figure}

To simulate the multipath reflection and attenuation of radar signals, inspired by the ray tracing \cite{whitted2005improved} in computer graphics, we design a \textit{signal simulation model}. A typical application of ray tracing is to track the movement of visible light. Existing works \cite{yue2022cornerradar, scheiner2020seeing} have utilized the principle of multiple reflections of light to achieve object tracking and positioning. In addition, light propagation also involves common physical phenomena, i.e., reflection and attenuation, which are similar to the multipath reflection and attenuation of radar signals during the transmitting and receiving process. Therefore, the \textit{signal simulation model} takes depth information and RCS as inputs to simulate the propagation characteristics of radar signals. Specifically, for each transmitting ($Tx$) and receiving ($Rx$) antenna pair, the signal formed by a radar is sinusoidal-like \cite{ling1989shooting}. As shown in Fig. \ref{fig11}, to model it, we should consider the time $t$ across multiple chirps. Suppose $n$ chirps have been transmitted (0$\le$$n$$\le$$N$-1, where $N$ is the total number of chirps in a frame), and the current ($n+1$)$th$ chirp has been transmitted for $t_{n}$ time. Therefore, we can calculate the time $t$ and the instant frequency $f(t)$ at which the chirp is transmitted, i.e.,
\begin{equation}
    t = nT + {t_n}, and f(t) = {f_0} + \frac{{B{t_n}}}{T},
\end{equation}
where $T$ represents the period of each chirp, $B$ represents the sweep bandwidth. Based on this, we can mathematically express the radar emission signal ${T_{Tx}}(t) = A{e^{j\varphi}}$, where $A$ and $\varphi$ represent amplitude and phase, respectively.

As shown in Fig. \ref{fig11}, we can calculate the shaded area under all chirps in a frame, i.e., $\int_0^{nT + {t_n}} {f(x)dx}$, which represents the number of periods experienced. The signal phase $\varphi$ varies by 2 $\pi$ per period, which is calculated as follows:
\begin{equation}
 \begin{array}{l}
\varphi  = 2\pi (\int_0^{nT + {t_n}} {f(x) \cdot dx} ) + {\varphi _0} = 2\pi ({f_0} + \frac{{B(t_n^2 + n{T^2})}}{{2T}}) + {\varphi _0},
 \end{array} 
\end{equation}
where $\varphi_0$ represents initial phase. Then, $T_{Tx}(t)$ can be expressed as:
\begin{equation}
    {T_{Tx}}(t) = A{e^{j2\pi ({f_0} + \frac{{B(t_n^2 + n{T^2})}}{{2T}}) + {\varphi _0}}}.
\end{equation}

The transmitted signal $T_{Tx}(t)$ will bounce back to the receiving antenna ${R_{Rx}}(t)$ at the reflection point. The received signal can be seen as a delayed version of the transmitted signal and the latency is $\tau$. Therefore, ${R_{Rx}}(t)$ is expressed as:
\begin{equation}
   R_{Rx}(t) = {A^{'}}{e^{j(2\pi ({f_0}(t - \tau ) + \frac{{B({{({t_n} - \tau )}^2} + n{T^2})}}{{2T}}) + {\varphi _0})}},
\end{equation}
where ${A^{'}}$ represents the attenuated amplitude, which can be calculated according to the radar communication principle \cite{rao2017introduction}:
\begin{equation}
    {A^{'}}= \frac{{{G_{TX}}{G_{RX}}\lambda \sqrt {PR} }}{{{{(4\pi )}^{1.5}}{D^2}}},
\end{equation}
where ${G_{Tx/Rx}}$ represents antenna gains of $Tx/ Rx $. $\lambda$, $P$, $R$, and $D$ represent wavelength, transmission power, RCS, and the distance between reflection points and the $Rx$, respectively.

As shown in Fig. \ref{fig12}, to better simulate the propagation characteristics of radar signals in real-world scenes, we add some existing real reflectors, such as tables and TVs, when designing the \textit{signal simulation model}. 
Specifically, the voxels we selected are cubes with 0.05 m on each side, and the 3D space we constructed is divided into a grid with dimensions of 128 $\times$ 128 $\times$ 64, representing its length, width, and height, respectively.
During the simulation process, we will incorporate some common reflectors into the grid within the 3D space, followed by calculating the average signal intensity of all triangles associated with each vertex to derive the vertex's signal intensity information. Finally, by averaging the signal intensity of all vertices within each voxel grid, we can obtain the signal intensity information corresponding to the reflector in the grid. Note that the position of reflectors can be freely adjusted  to facilitate the simulation of diverse scenes; meanwhile, there will be a bias so that the model can better adapt to various scenes.
Apart from the reflection, we should also consider the energy loss of radar signals during propagation. Existing works \cite{yue2022cornerradar, deng2023midas} have also demonstrated that energy is essentially lost when a signal undergoes reflection more than three times. Therefore, when designing the \textit{signal simulation model}, we set the maximum threshold for the signal attenuation coefficient ($\alpha$) to 0.3. Moreover, we experimentally verify the impact of different $\alpha$ on the quality of generated radar data (see Section \ref{sec5.3.5}), where the interval is set to 0.05. Finally, in \texttt{G\textsuperscript{3}R}, $\alpha$ is set to 0.15, and the \textit{signal simulation model} outputs the human intensity map.

\subsection{Encoder-decoder Model}\label{sec4.6}
Based on the human intensity map and radial velocity, we can generate the initial radar data. However, such data seriously differs from real-world radar data in number and distribution. Therefore, an \textit{encoder-decoder model} is designed to generate realistic radar data, which contains a \textit{sampling module} and a \textit{fitting module}. Since point clouds are essentially an unordered point set and lack topological information, convolutional neural networks are unsuitable for processing this data type.
To this end, PointNet \cite{qi2017pointnet} is designed to process point cloud data without converting it to other data formats (e.g., voxel), but it ignores the geometric relationships between points, making it difficult to capture local features of gestures. DGCNN \cite{wang2019dynamic} incorporates a new neural network module (EdgeConv) based on PointNet, which explicitly constructs a locally connected graph and learns the embedding of edges, supporting the capture of local features while maintaining permutation invariance. This network architecture can produce richer contextual information and better learn the semantic information of the point set by dynamically updating the graph structure between layers, followed by accurately finding the mapping relationship between the input point cloud and the ground truth.
\begin{figure*}[t]
  \centering
  \centerline{\includegraphics[width=160mm]{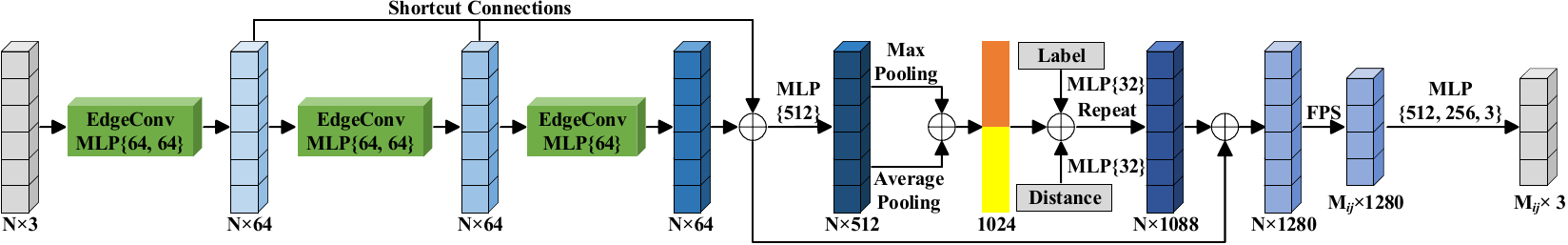}}
  \vspace{-2mm}
  \caption{The architecture of our \textit{sampling} module.}\label{fig13}
  \vspace{-6mm}
\end{figure*} 

Inspired by the DGCNN, as shown in Fig. \ref{fig13}, we design a \textit{sampling module}. The module takes the $x$, $y$, and $z$ coordinates of $N$ points as inputs and calculates an edge feature set of size $k$ for each point at an EdgeConv layer, where $N$=1920, $k$=20. After constructing the local connection graph, we use a multi-layer perceptron (MLP), whose parameters in parentheses are the number of output channels of the 2D convolutional layer. Note that a batch normalization (BN) and a leaky rectified linear unit (Leaky ReLU) are used after convolution. 
Shortcut connections are employed to incorporate the local features from all EdgeConv outputs to obtain (64 $\times$ 3)-dimensional point cloud features, followed by transforming them into 512-dimensional features by MLP. Meanwhile, max pooling and average pooling are performed on the obtained local features to obtain the global feature vector of point clouds (1 $\times$ (512+ 512) = 1024 dimensions). Moreover, the module encodes the gesture labels and distances into 32-dimensional feature vectors, respectively, followed by concatenating them into the global vector to obtain a 1088-dimensional feature vector. Finally, we repeat the obtained feature vectors for $N$ times to get the aggregated point-wise features, which are concatenated with local features (1920 $\times$ 1280), followed by employing farthest point sampling (FPS) layers to obtain the generated $M_{ij}$ points corresponding to different positions $i$ and different gestures $j$.

To guarantee near-optimal convergence, we construct two loss functions to measure the correlation between generated and real-world point clouds in number and distribution of points. To this end, we first use the Chamfer Distance (ChD) to measure the number difference between generated and real-world point sets \cite{fan2017point}. ChD is expressed as:
\begin{equation}
\begin{split}
  {L_{ChD}}({P_{G}},{P_{{\mathop{ R}} }}) = \frac{1}{{{N_{G}}}}\sum\limits_{x \in {P_{G}}} {\mathop {\min }\limits_{y \in {P_{{\mathop{ R}} }}} } ||x - y||_2^2 \\ + \frac{1}{{{N_{{\mathop{ R}} }}}}\sum\limits_{y \in {P_{{\mathop{ R}} }}} {\mathop {\min }\limits_{x \in {P_{G}}} } ||x - y||_2^2,  
  \end{split}
\end{equation}
where $P_{G}$ and ${P_{{\mathop{ R}} }}$ represent generated and real-world point sets, respectively. ${{N_{G}}}$ and  ${{N_{{\mathop{ R}}}}}$ represent the number of points in $P_{G}$ and ${P_{{\mathop{ R}} }}$, respectively. 
However, relying solely on the ChD loss does not effectively guarantee the accurate distribution of points; the main reason is that ChD only measures the distance between the nearest points, which may fail to capture the distribution differences between points. Therefore, to accurately obtain the point distribution, we introduce an Earth Movers' Distance (EMD) in the loss function, which is expressed as:
\begin{equation}
    {L_{EMD}}({P_{G}},{P_{{\mathop{ R}} }}) = \frac{1}{{{N_{G}}}}\mathop {\min }\limits_{\phi :{P_{G}} \to {P_{{\mathop{ R}} }}} \sum\limits_{x \in {P_{G}}} {||x - \phi (x)|{|_2}},
\end{equation}
where ${\phi :{P_{G}} \to {P_{{\mathop{ R}} }}}$ represents a bijection function to guarantee that every element of $P_{G}$ is paired with precisely one element of ${P_{{\mathop{ R}} }}$. Therefore, the final loss function is:
\begin{equation}
    L = \lambda {L_{ChD}} + (1-\lambda){L_{EMD}},
\end{equation}
where $\lambda$=0.5, which is a hyper-parameter to balance the weight of different components.

\begin{figure}[t]
  \centering
  \centerline{\includegraphics[width=70mm]{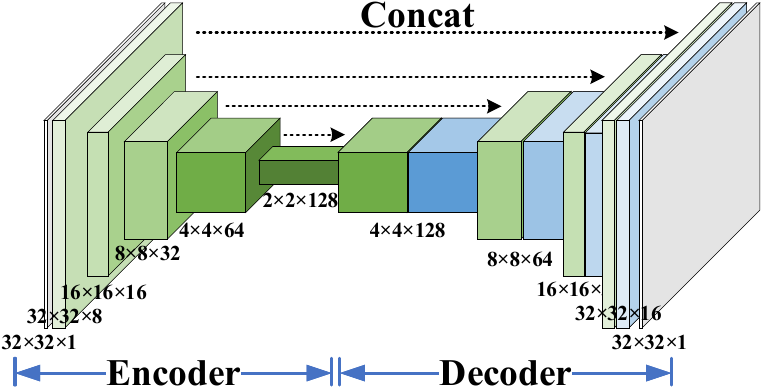}}
  \vspace{-2mm}
  \caption{The architecture of our \textit{fitting} module.}\label{fig14}
  \vspace{-5mm}
\end{figure}

In addition, considering the existing gesture recognition models will splice the radial velocity and intensity together to form a matrix similar to a grayscale map so that the model can better learn different gesture features. Therefore, we design a \textit{fitting module}, which adopts a \textit{U-Net} model with an encoder-decoder architecture to fit radial velocity and intensity between generated and real-world radar data for generating realistic radar data. As shown in Fig. \ref{fig14}, the \textit{U-Net} has a total of 23 convolutional layers and about 1.91M parameters. The encoder uses double convolution blocks (3$\times$3 kernel, stride=1, padding=1) for feature extraction. The decoder uses deconvolution layer (2$\times$2 kernel, strides=2, padding=0) and skip-connection for feature fusion. A Leaky ReLU and a BN are used for each convolutional layer. We employ the corresponding pairs of generated and real-world matrices to train the \textit{U-Net} model for 1000 epochs, using the \textit{Adam optimizer} \cite{kingma2014adam} with a learning rate of 0.001.
\begin{figure}[t]
  \centering
  \centerline{\includegraphics[width=25mm]{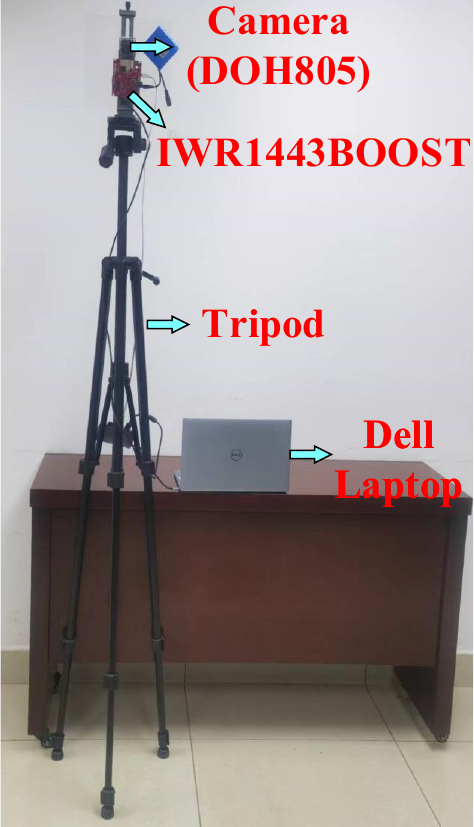}}
  \vspace{-2mm}
  \caption{Hardware platform for data collection.}\label{fig15}
  \vspace{-5mm}
\end{figure}
\section{Evaluation}\label{sec5}
\subsection{Implementation and Experimental Setup}\label{sec5.1}
\textbf{Implementation}. The hardware platform for data collection has been implemented, as depicted in Fig. \ref{fig15}. The platform includes a tripod to simulate the sensor installation position and a Dell notebook (XPS13, i7-1250U CPU, 16GB memory) for data storage and processing. The notebook is connected to both a commercial \textit{Frequency Modulated Continuous Wave} (FMCW) radar sensor (TI IWR1443BOOST) operating at 77 GHz and a global shutter camera (DOH805) for data collection via UART USB cables. By default, we employ the following FMCW parameters: idle time 7$\mu$s, ramp end time 114.29 $\mu$s, range resolution 4 cm, Doppler resolution 0.34 m/s, maximum unambiguous range 8.19 m, frame duration 100 ms, and maximum radial velocity ±2.67 m/s. In addition, we mount a camera next to the radar sensor to capture footage. The training of \texttt{G\textsuperscript{3}R} is performed on a server running on Ubuntu 20.08 OS with two RTX3080 GPUs. We implement different modules in \texttt{G\textsuperscript{3}R} using Python to facilitate integration with various radar-based sensing applications.

\textbf{Dataset Preparation}. 
\textit{Dataset1}. We place the hardware platform on a tripod with a height of 1.7 m to collect real-world radar data. We recruit 32 volunteers (23 males, 9 females, i.e., users) aging from 21 to 30 with heights ranging from 158 cm to 186 cm and weights ranging from 45 kg to 98 kg. Users sit and perform different gestures at a linear distance of 0.5-4.5 m in front of the collection device, resulting in a total of 720 samples per user. Meanwhile, we utilize a camera to record user gestures and label the radar dataset. Note that we link the frame of the camera with the nearest radar frame for temporal alignment and exploit coordinate transformation to achieve spatial alignment. Each user performs 5 gestures (PL, PS, CR, UP, and KO) in front of the collecting device (see Section \ref{sec2.1}). Each gesture takes about 1-2 s to complete. Thus, in total we collect roughly 9 hours of real-world radar data. For the collected dataset, we process them into point cloud data via 3D fast Fourier transform (FFT) for training gesture recognition models. Moreover, we aggregate 21 hours of 2D video data from five public datasets \cite{kuehne2011hmdb, abu2016youtube, perera2018uav, soomro2012ucf101, escalante2016chalearn}, which are structured with gesture labels, and unstructured available 2D video sources (YouTube, Bilibili) using queries related to our gesture set. These 2D videos serve as inputs of \texttt{G\textsuperscript{3}R}, enabling the generation of realistic radar data for model training. Note that when preparing the 2D video dataset, we manually eliminate some low-quality 2D videos (e.g., blurred samples).

\begin{figure}[t]
\centering
\subfloat[Meeting room (Scene 1)]{\vspace{-1mm}
\begin{minipage}[b]{0.3\textwidth}
\centering
\centerline{\includegraphics[width=45mm]{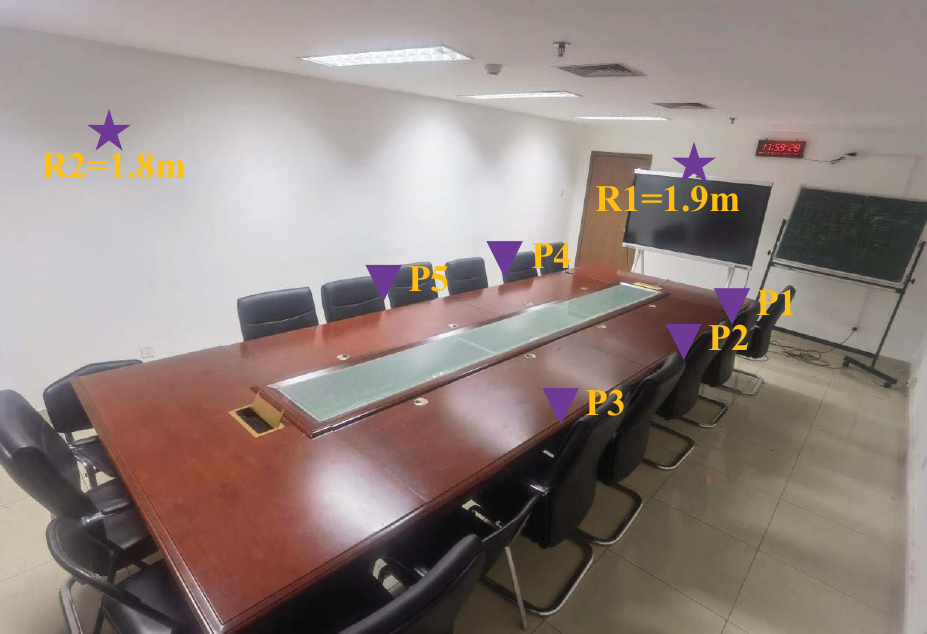}}
\end{minipage}
}
\vfill
\subfloat[Work room (Scene 2)]{\vspace{-1mm}
\begin{minipage}[b]{0.3\textwidth}
\centering
\centerline{\includegraphics[width=45mm]{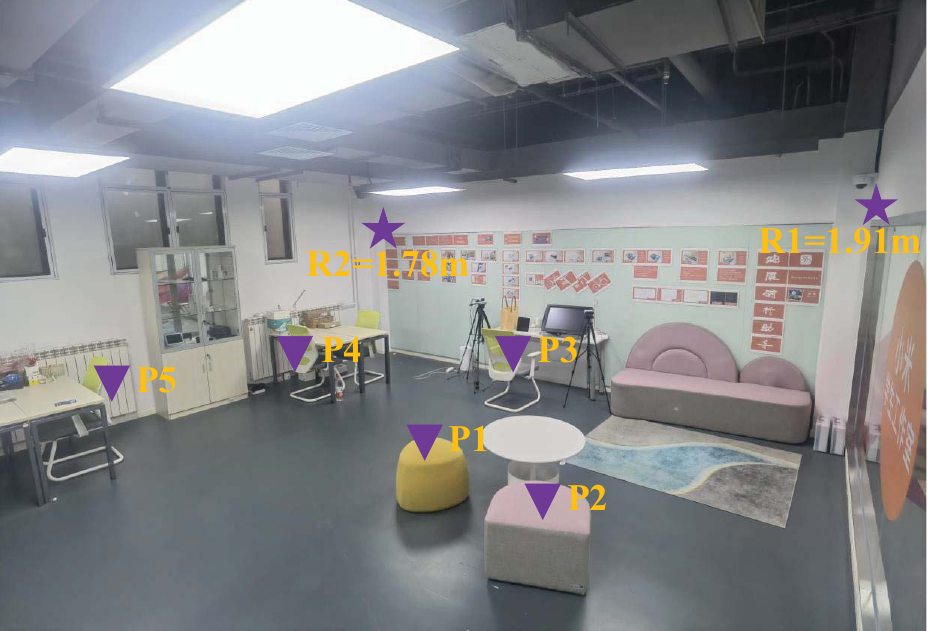}}
\end{minipage}
}
\vfill
\subfloat[Hotel room (Scene 3)]{\vspace{-1mm}
\begin{minipage}[b]{0.3\textwidth}
\centering
\centerline{\includegraphics[width=45mm]{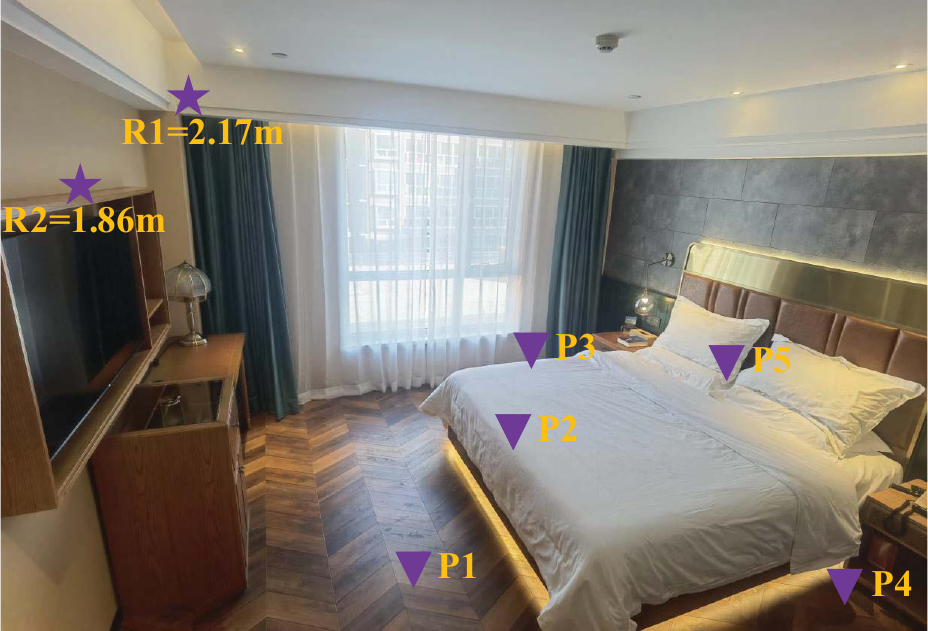}}
\end{minipage}
}
\vspace{-3mm}
\caption{Three different scenes. $\star$ represents different radar positions (R1 and R2), and $\blacktriangledown$ represents various user positions (P1-P5).}\label{fig16}
\vspace{-5mm}
\end{figure}

\textit{Dataset2}. To deeply evaluate the impact of various factors on recognition performance, we recruit 5 new users (3 males, 2 females) again. 
As shown in Fig. \ref{fig16}, we select three new scenes (meeting room, work room, and hotel room) for validation, where: (i) users perform 8 times per gesture in different postures (standing and sitting); (ii) users perform gestures in different positions (P1-P5) and the radar is also placed in different positions (R1 and R2). Thus, in total we collect 5 gestures $\times$ 5 users $\times$ 3 scenes $\times$ 8 times $\times$ 2 postures $\times$ 7 positions = 8400 samples.

\textbf{Baselines}. We verify the effectiveness of \texttt{G\textsuperscript{3}R} by comparing the following three state-of-the-art methods:
\begin{itemize}
    \item \textit{Vid2Doppler} \cite{ahuja2021vid2doppler}, a software pipeline that allows 2D videos to be transformed into radar data, which employs a neural network to directly extract human meshes, followed by generating signal reflection of each vertex with respect to a radar. Note that \textit{Vid2Doppler} ignores the multipath reflection and attenuation of radar signals while generating data that only represents the occupancy information of humans, resulting in a notable disparity from real-world radar data.
    \item \textit{SynMotion} \cite{zhang2022synthesized}, a method for generating radar data using human skeleton points. It employs a \textit{variant tracker} to obtain accurate skeletal point coordinates of humans, followed by generating radar signals reflected from these points. Note that \textit{SynMotion} does not consider the multipath reflection and attenuation of radar signals during the transmitting and receiving process, significantly compromising the quality of generated radar data.
    \item \textit{Midas} \cite{deng2023midas}, a method for generating radar data using 2D videos. It exploits several key modules, \textit{human region indexing}, \textit{human mesh fitting}, and \textit{multi-human reflection} model to simulate the multipath reflection and attenuation of radar signals, followed by using a \textit{Transformer} model to generate convertible and realistic radar data for various human sensing tasks. Note that while \textit{Midas} considers the multipath reflection and attenuation of radar signals, the point cloud data generated is coarse-grained and solely represents the occupancy information of humans, making it difficult to apply for fine-grained gesture recognition.
\end{itemize}
\begin{figure}[t]
  \centering
  \centerline{\includegraphics[width=58mm]{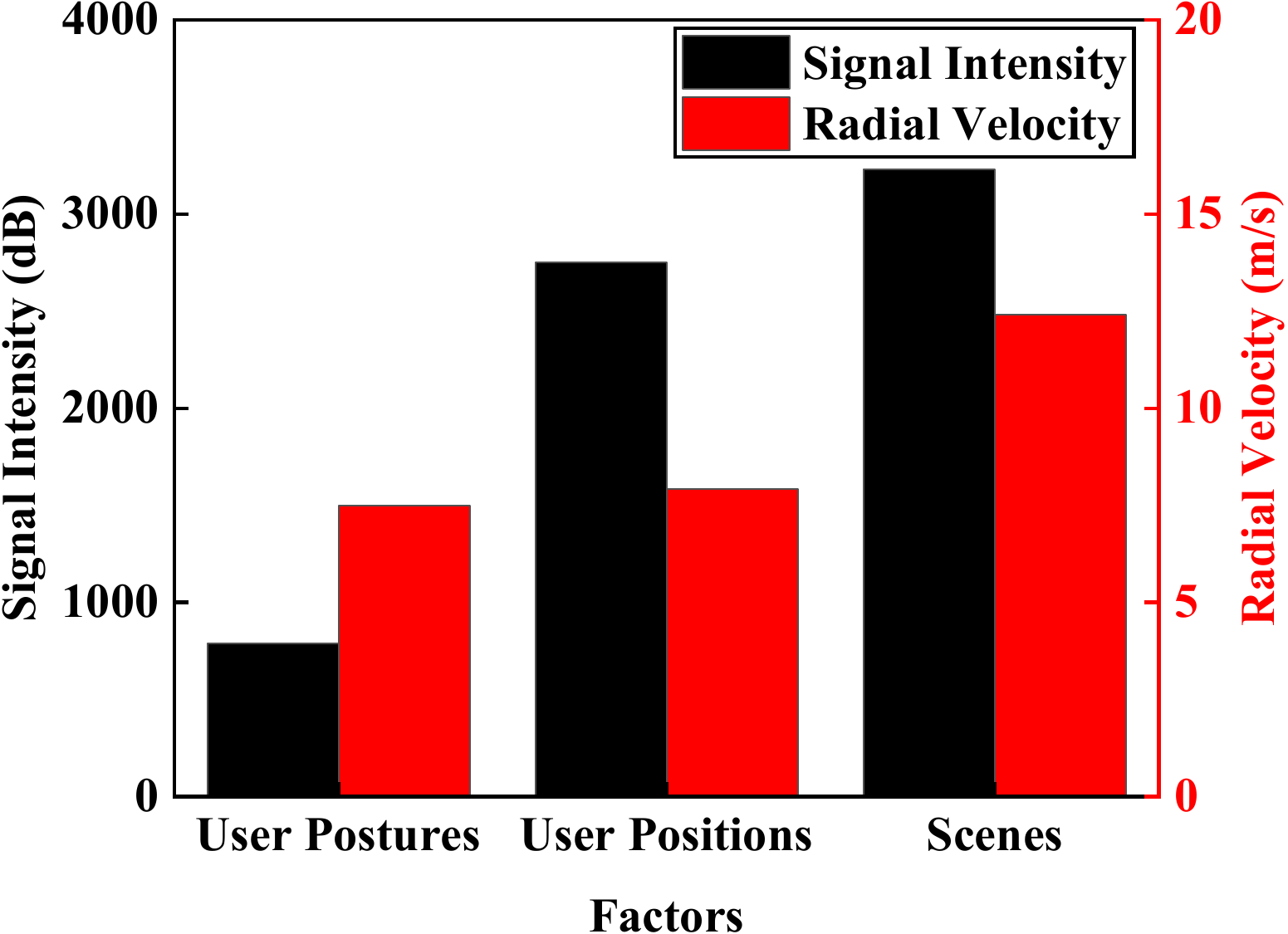}}
  \vspace{-3mm}
  \caption{Average cumulative errors of signal intensity and radial velocity for \texttt{G\textsuperscript{3}R} with various factors.}\label{fig17}
  \vspace{-4mm}
\end{figure}
\begin{figure}[t]
  \centering
  \centerline{\includegraphics[width=54mm]{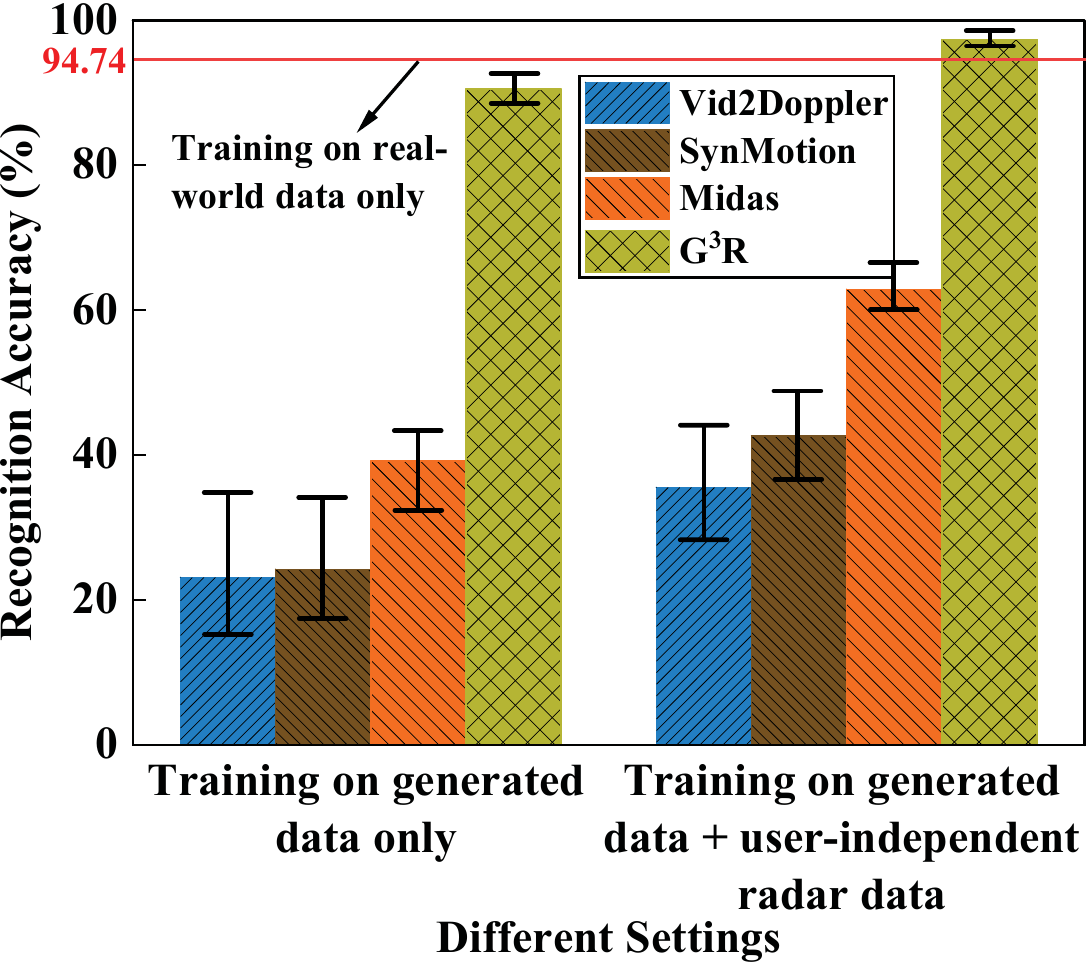}}
  \vspace{-2mm}
  \caption{Gesture recognition accuracy of different methods across different train/test settings.}\label{fig18}
  \vspace{-4mm}
\end{figure}
\begin{figure}[t]
  \centering
  \centerline{\includegraphics[width=46.5mm]{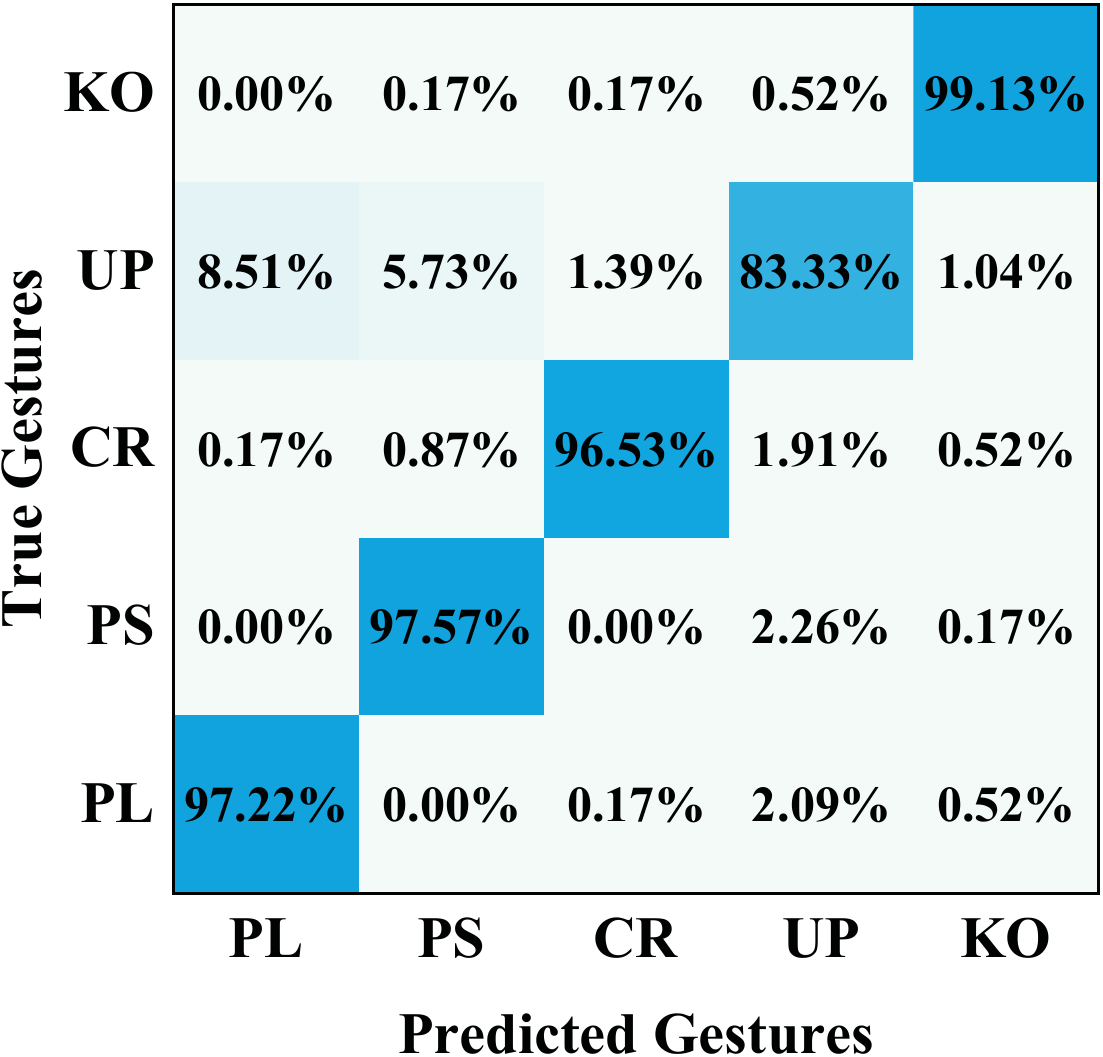}}
  \vspace{-2mm}
  \caption{Gesture recognition accuracy using real-world radar data.}\label{fig19}
  \vspace{-5mm}
\end{figure}

\textbf{Experimental Settings}. To fairly verify the accuracy of \texttt{G\textsuperscript{3}R} and three baselines, we evaluate their performance on dataset1 using three different experimental settings: (i) training on all generated radar data only and testing on real-world radar data from users, (ii) training on four folds of real-world radar data and testing on one fold of real-world radar data (all combinations, results averaged), and (iii) training on all generated data and four folds of real-world data, testing on one fold of real-world data (all combinations, results averaged) (see Section \ref{sec5.2}). Note that all generated radar data refers to 21 hours of 2D videos, and the real-world data is approximately 9 hours of gesture samples collected.
Meanwhile, we also use the first setting to conduct ablation experiments to demonstrate the effectiveness of \texttt{G\textsuperscript{3}R} (see Section \ref{sec5.3}). Moreover, to deeply evaluate the effectiveness of \texttt{G\textsuperscript{3}R} under various factors, we utilize the same training data from the first and third settings for model training, and use dataset2 for testing, further demonstrating the advantage of \texttt{G\textsuperscript{3}R} under various user postures, positions, and scenes (see Section \ref{sec5.4}). We use \textit{mTransSee} \cite{liu2022mtranssee} as the gesture recognition model.

\textbf{Evaluation Metrics}. To quantify the performance of \texttt{G\textsuperscript{3}R} and three baselines, the following evaluation metrics are defined:
\begin{itemize}
\item Average Cumulative Error. We run 2D videos through \texttt{G\textsuperscript{3}R} and compare the generated radar data to the real-world radar data by quantifying the difference values of signal intensity/radial velocity under various user postures, positions, and scenes, respectively, followed by averaging them. 
\item Recognition Accuracy. The recognition accuracy is defined as the probability that the model \cite{liu2022mtranssee} correctly recognizes a gesture.
\item Confusion Matrix. Each row corresponds to the true gestures in the confusion matrix, while each column corresponds to the predicted gestures. The entry of the $p$-th row and the $q$-th column represents the recognition accuracy.
\end{itemize}
\begin{figure}[t]
\centering
\subfloat[Vid2Doppler]{\vspace{-1mm}
\begin{minipage}[b]{0.3\textwidth}
\centering
\centerline{\includegraphics[width=49mm]{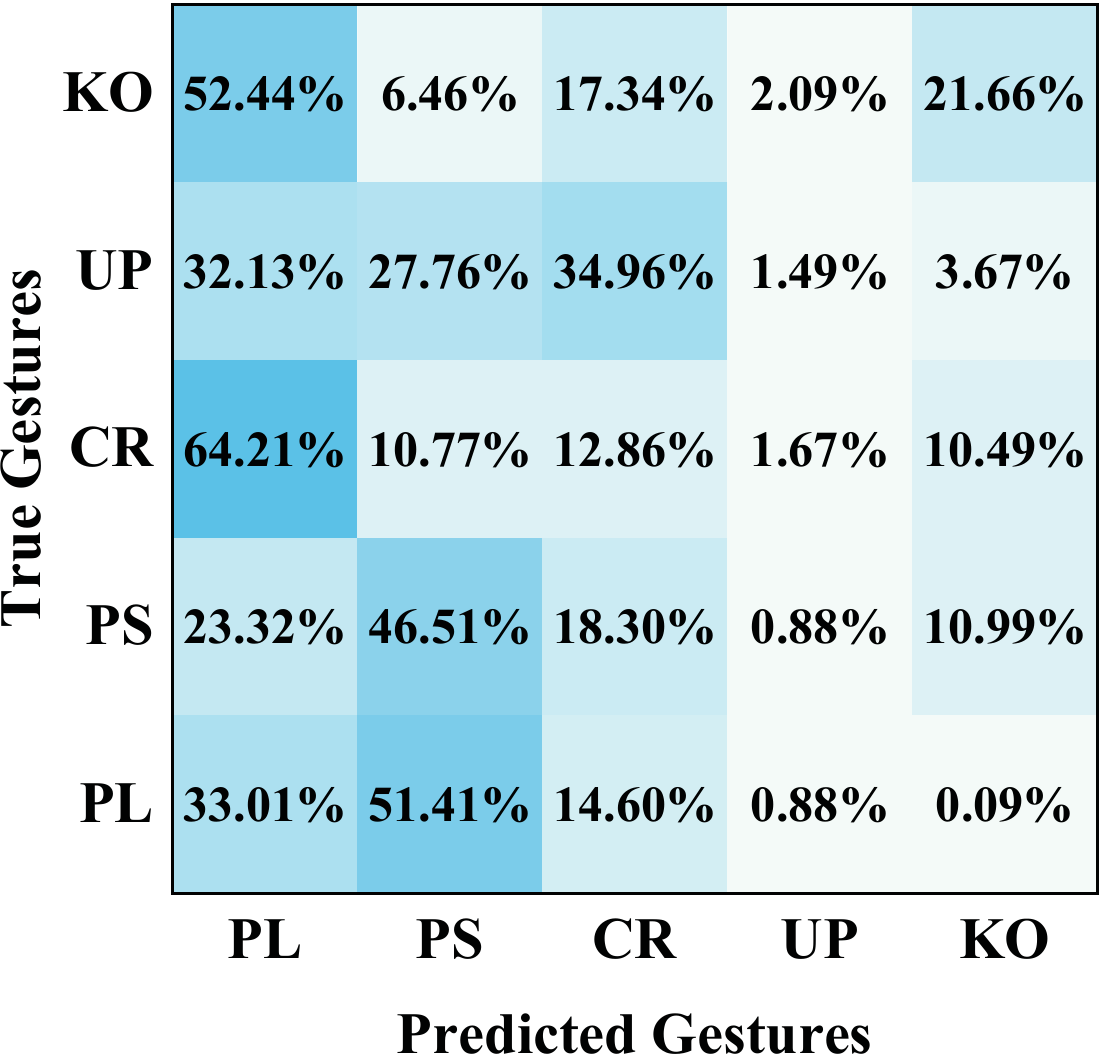}}
\end{minipage}
}\hspace{15mm}
\subfloat[SynMotion]{\vspace{-1mm}
\begin{minipage}[b]{0.3\textwidth}
\centering
\centerline{\includegraphics[width=49mm]{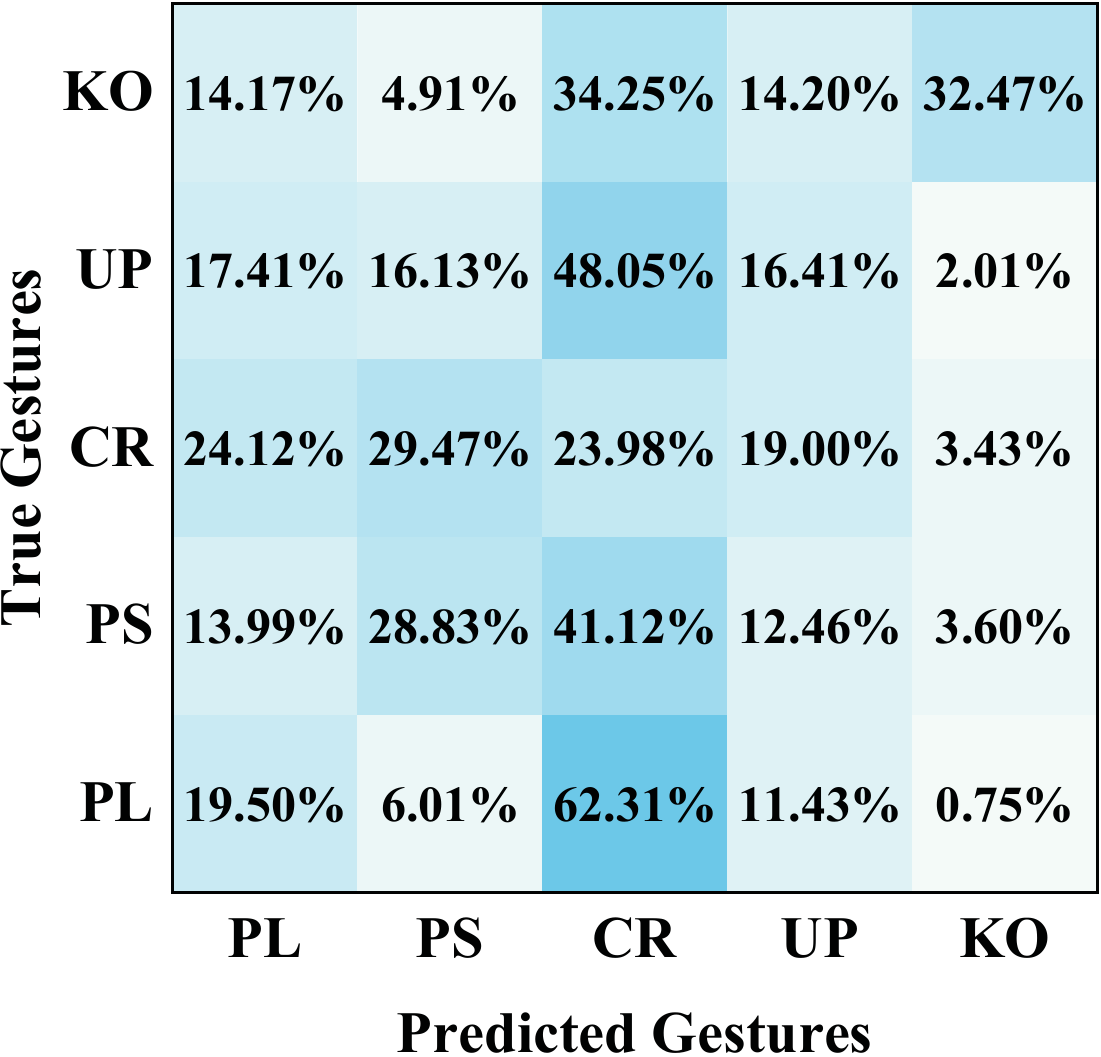}}
\end{minipage}
}
\vfill
\subfloat[Midas]{\vspace{-1mm}
\begin{minipage}[b]{0.3\textwidth}
\centering
\centerline{\includegraphics[width=49mm]{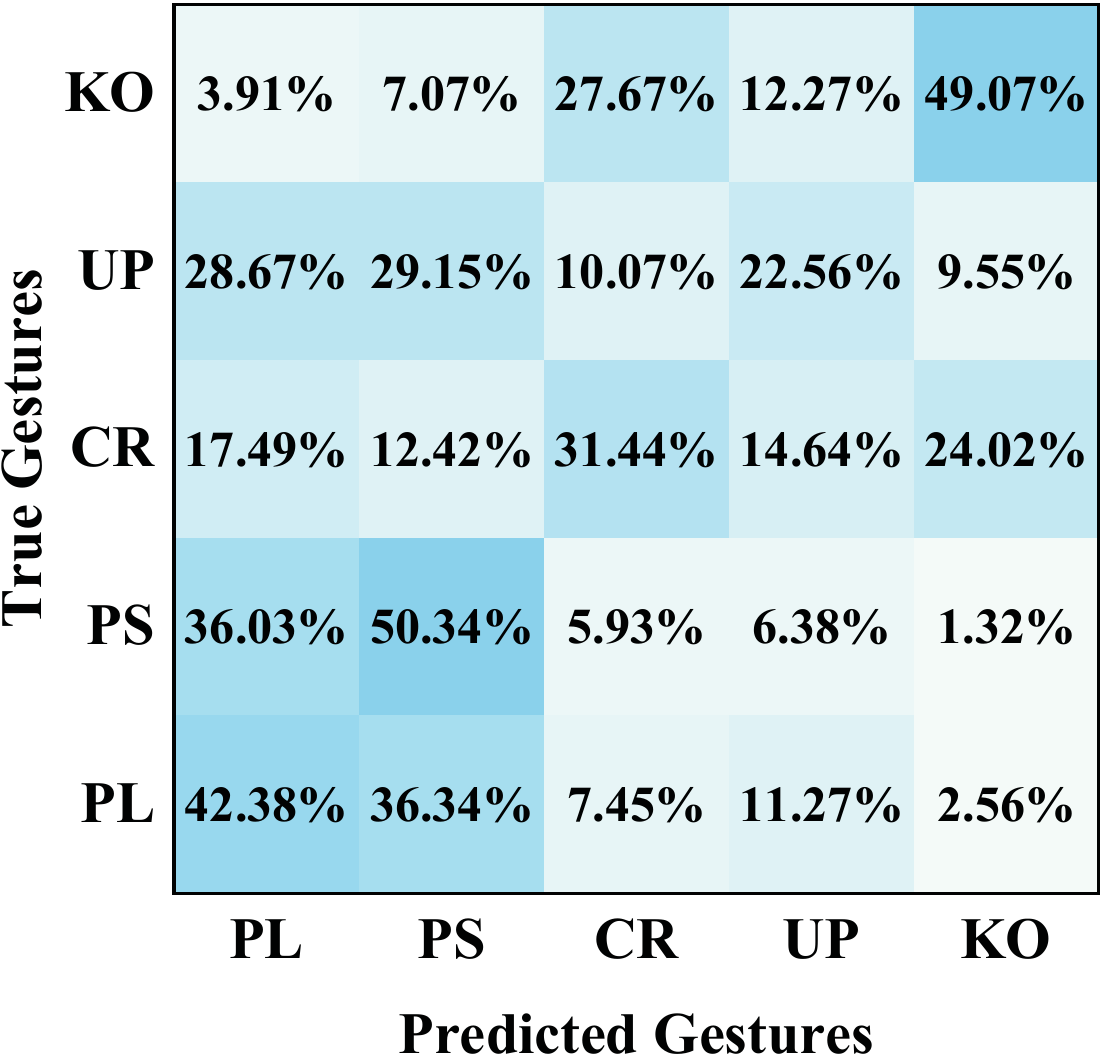}}
\end{minipage}
}
\hspace{15mm}
\subfloat[\texttt{G\textsuperscript{3}R}]{\vspace{-1mm}
\begin{minipage}[b]{0.3\textwidth}
\centering
\centerline{\includegraphics[width=49mm]{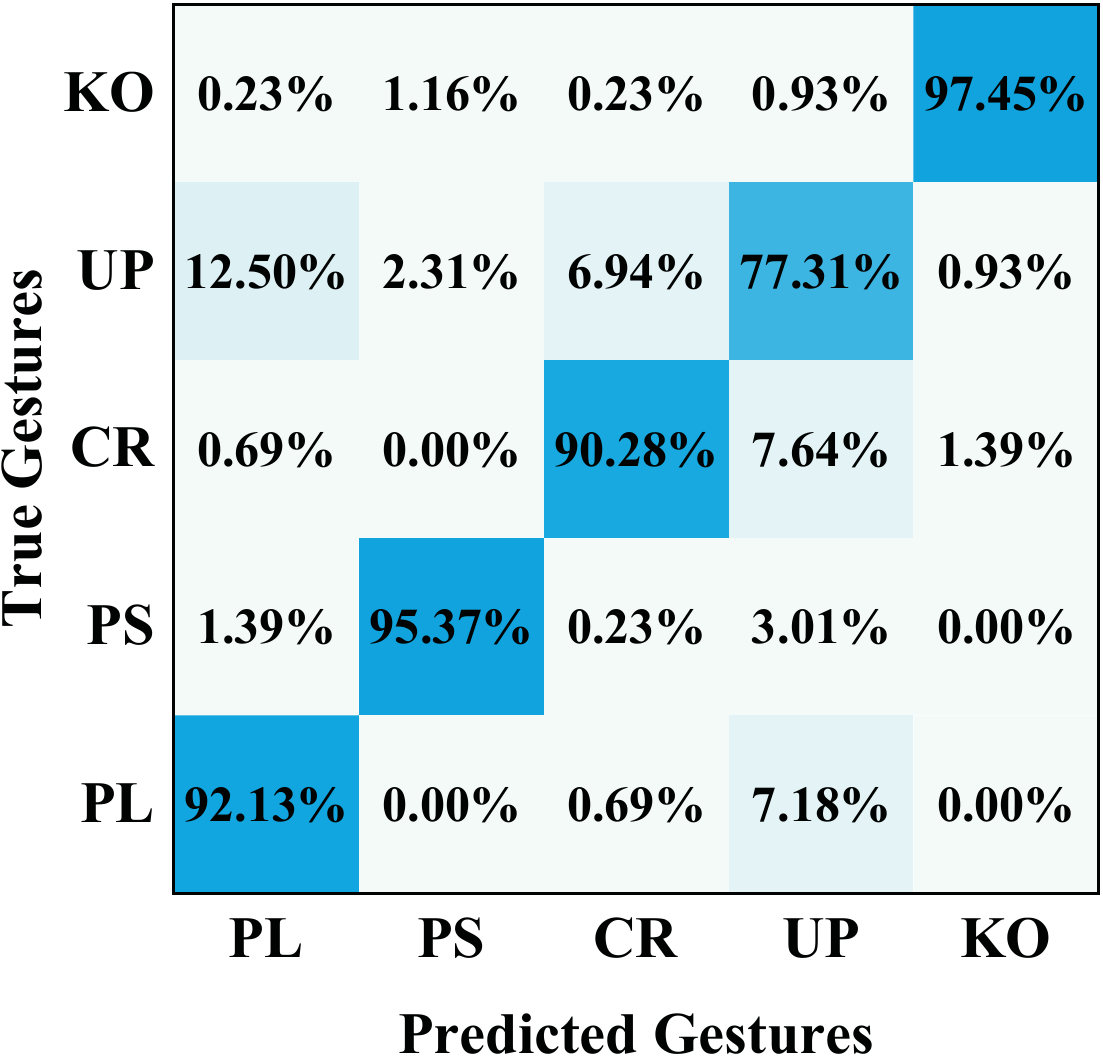}}
\end{minipage}
}
\vspace{-4mm}
\caption{Gesture recognition accuracy using all generated radar data.}\label{fig20}
\vspace{-7mm}
\end{figure}

\subsection{Evaluation on Gesture Recognition}\label{sec5.2}
\subsubsection{Quality of Generated vs. Real-world Radar Data}\label{sec5.2.1}
To demonstrate the effectiveness of \texttt{G\textsuperscript{3}R}, we use
our collected 2D videos and corresponding real-world radar data for verification.
Specifically, we measure the average cumulative errors between generated and real-world radar data regarding signal intensity/radial velocity under various user postures, positions, and scenes. Fig. \ref{fig17} shows that the errors are 789 dB/7.5 m/s, 2752 dB/7.92 m/s, and 3232 dB/12.41 m/s under three different factors, respectively. \texttt{G\textsuperscript{3}R} reduces the whole average cumulative errors of signal intensity/radial velocity by 10.65x/5.02x, 8.76x/4.63x, and 7.31x/3.86x compared with \textit{Vid2Doppler}, \textit{SynMotion}, and \textit{Midas} (see Section \ref{sec2.3}), respectively, demonstrating that the radar data generated by \texttt{G\textsuperscript{3}R} is closer to real-world data.

\subsubsection{Gesture Recognition Accuracy}\label{sec5.2.2} 
As shown in Fig. \ref{fig18}, \texttt{G\textsuperscript{3}R} achieves 90.51\% accuracy, 67.41 pp, 66.28 pp, and 51.36 pp higher than that of \textit{Vid2Doppler}, \textit{SynMotion}, and \textit{Midas}, respectively, when using the first experimental setting, demonstrating the advantage of \texttt{G\textsuperscript{3}R}. The second setting achieves 94.76\%. While a 4.25 pp difference exists in recognition accuracy, the model trained on all generated radar data can be comparable to a model trained solely on real-world radar data. Moreover, \texttt{G\textsuperscript{3}R} achieves the best average recognition accuracy, 97.32\% under the third setting; meanwhile, it improves the accuracy by 61.80 pp, 54.67 pp, and 34.45 pp compared with \textit{Vid2Doppler}, \textit{SynMotion}, and \textit{Midas}, respectively.
\begin{figure}[t]
\centering
\subfloat[Vid2Doppler]{\vspace{-1mm}
\begin{minipage}[b]{0.3\textwidth}
\centering
\centerline{\includegraphics[width=49mm]{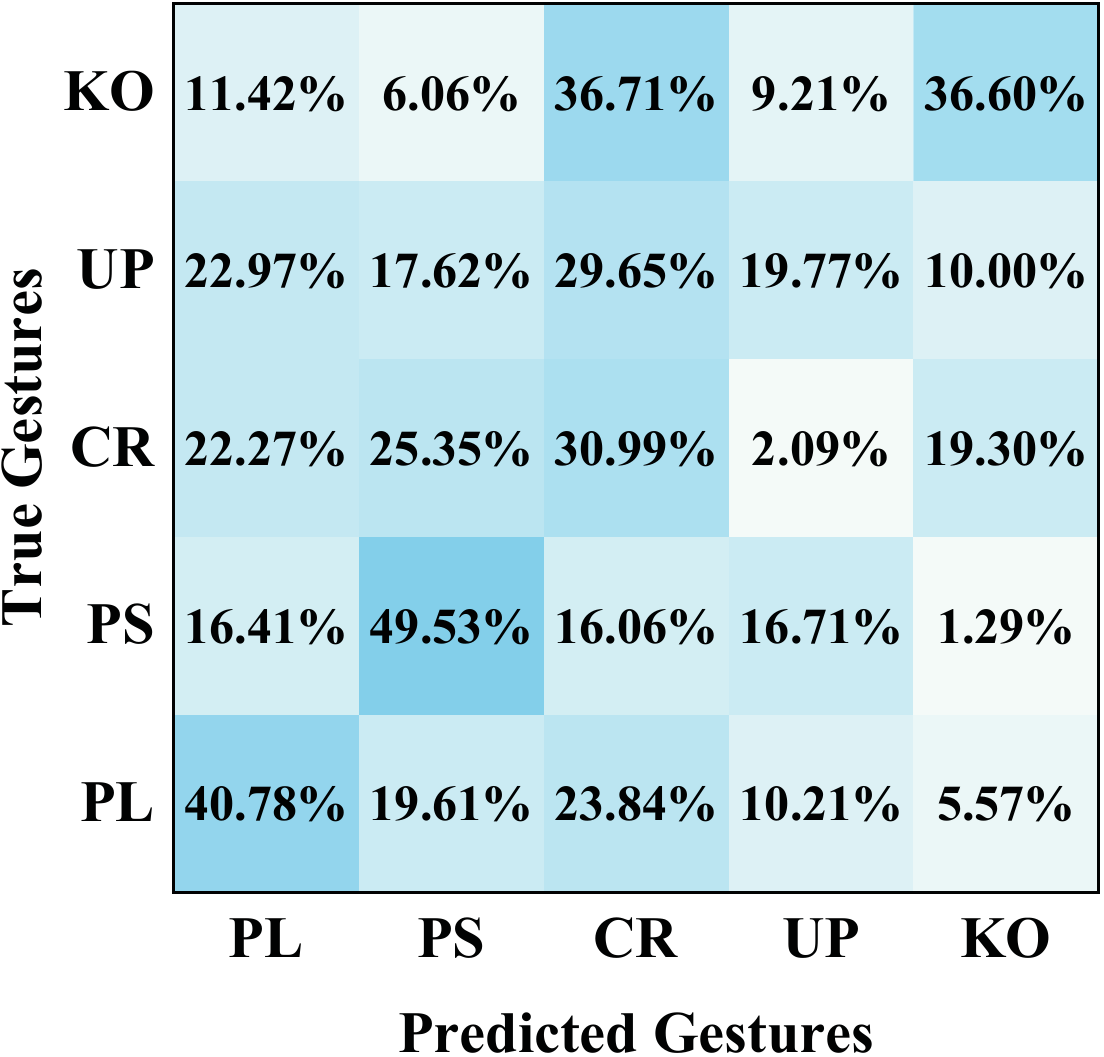}}
\end{minipage}
}\hspace{15mm}
\subfloat[SynMotion]{\vspace{-1mm}
\begin{minipage}[b]{0.3\textwidth}
\centering
\centerline{\includegraphics[width=49mm]{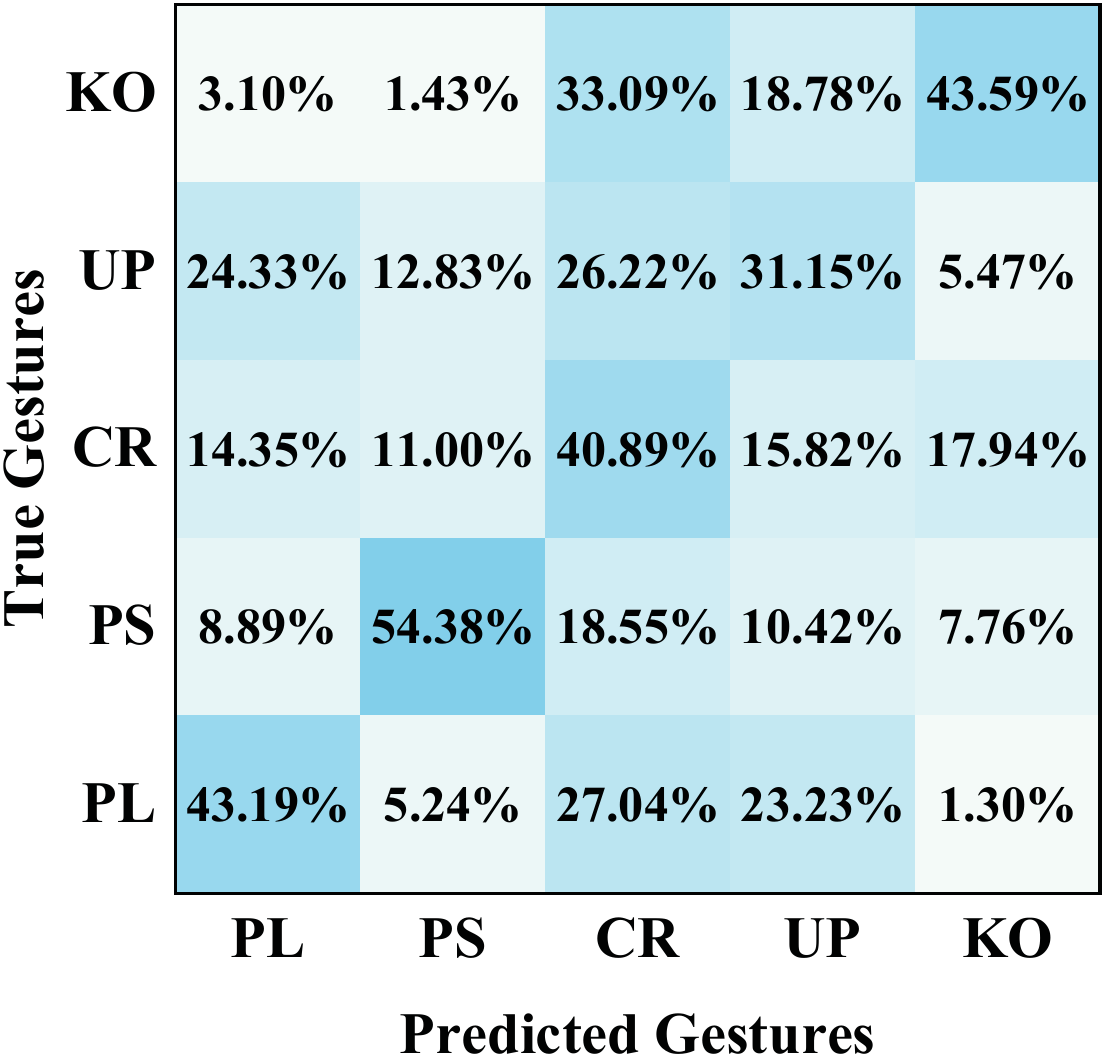}}
\end{minipage}
}
\vfill
\subfloat[Midas]{\vspace{-1mm}
\begin{minipage}[b]{0.3\textwidth}
\centering
\centerline{\includegraphics[width=49mm]{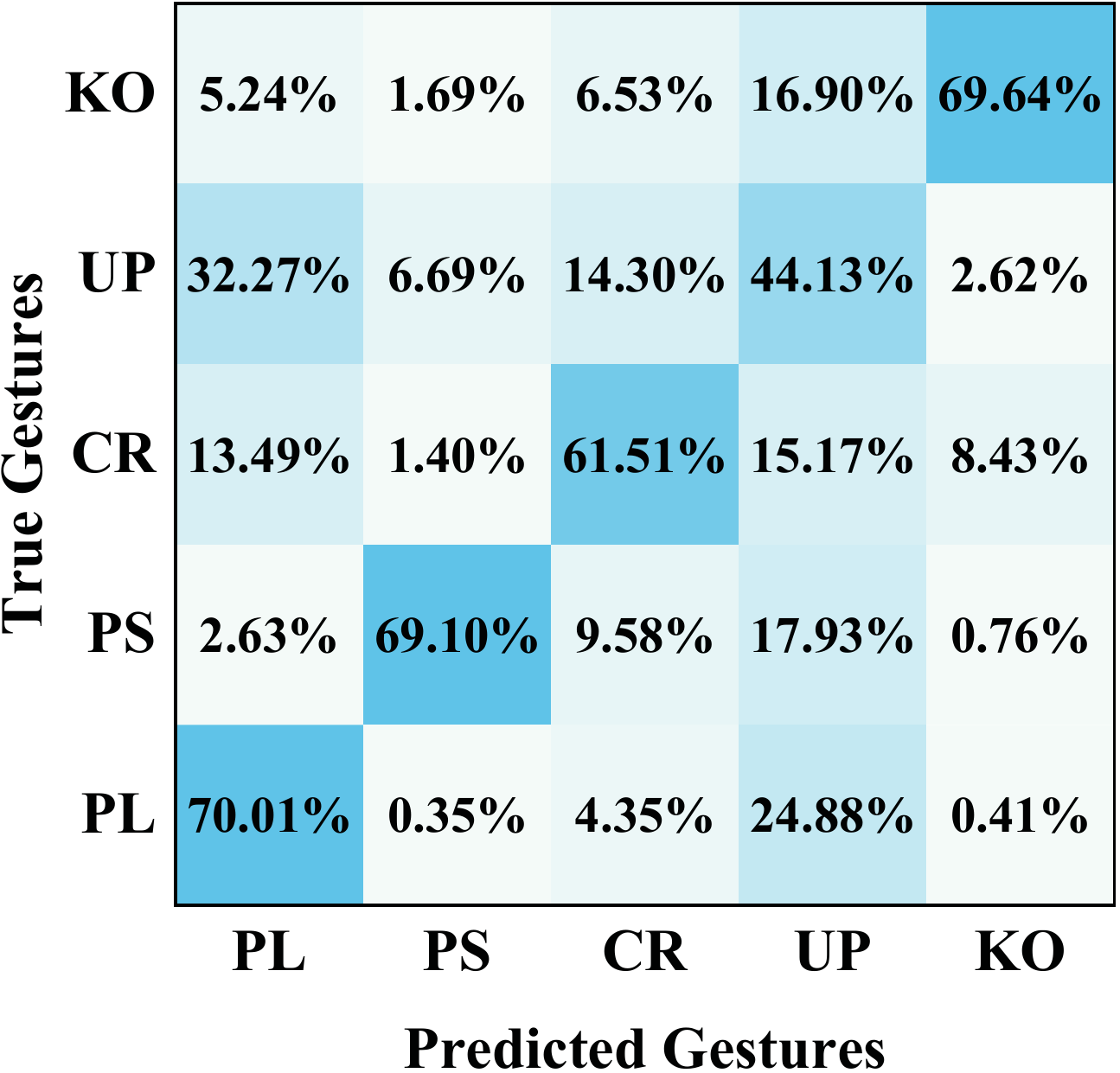}}
\end{minipage}
}
\hspace{15mm}
\subfloat[\texttt{G\textsuperscript{3}R}]{\vspace{-1mm}
\begin{minipage}[b]{0.3\textwidth}
\centering
\centerline{\includegraphics[width=49mm]{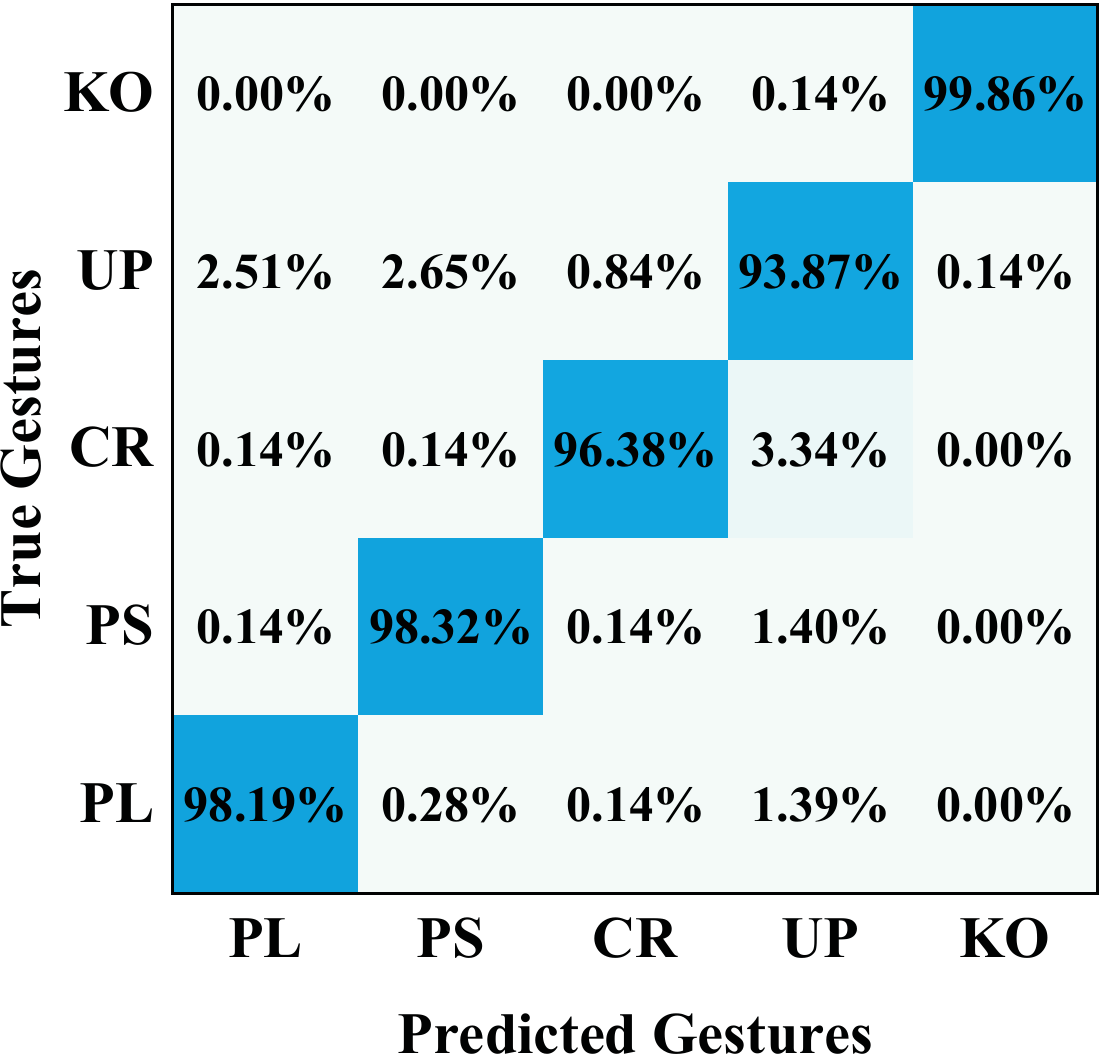}}
\end{minipage}
}
\vspace{-4mm}
\caption{Gesture recognition accuracy using all generated data + user-independent
radar data.}\label{fig21}
\vspace{-7mm}
\end{figure}

\begin{figure}[t]
  \centering
  \centerline{\includegraphics[width=55mm]{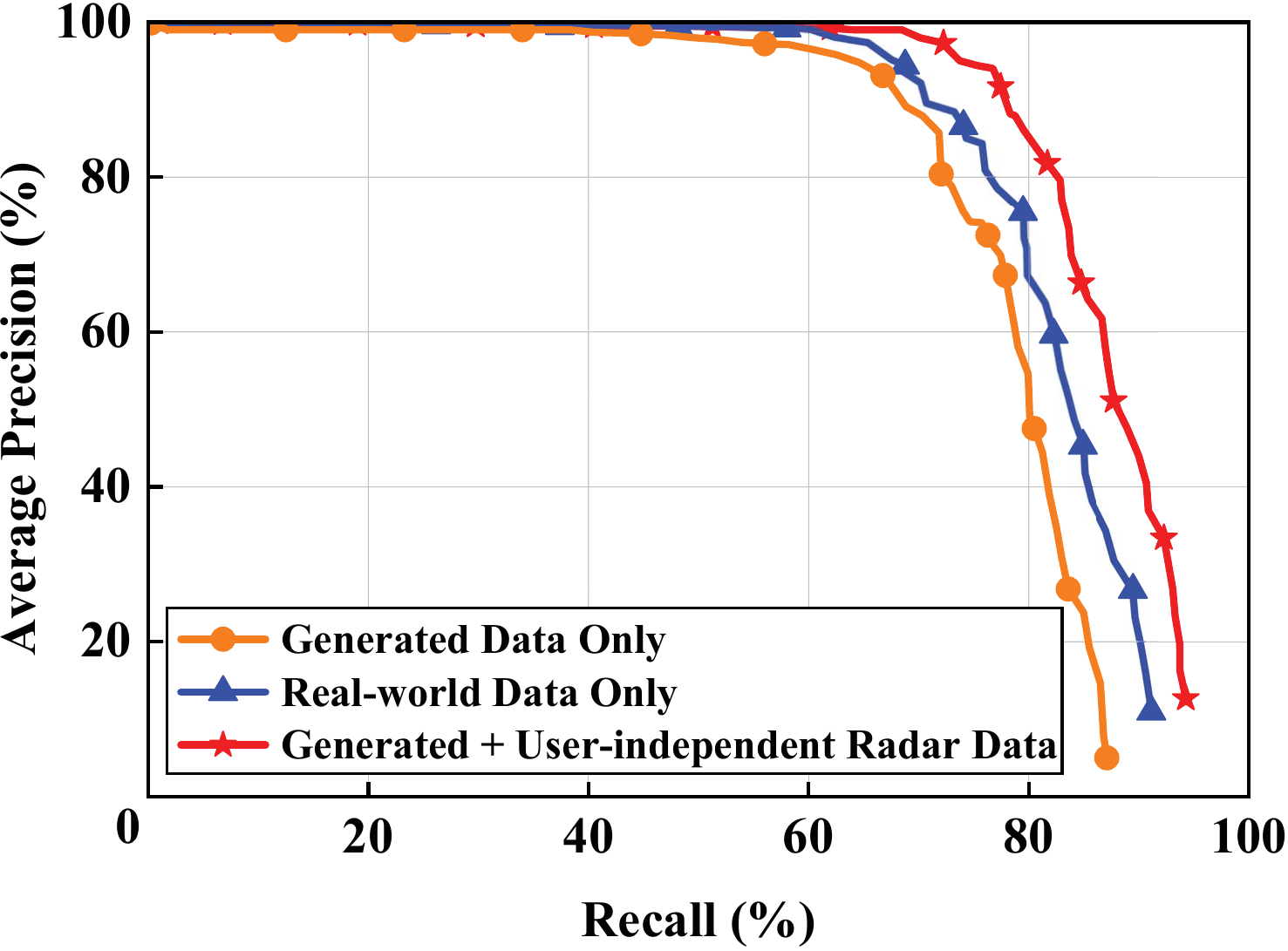}}
  \vspace{-3mm}
  \caption{Precision-recall curves of three settings.}\label{fig22}
  \vspace{-5mm}
\end{figure}

We then refine the recognition accuracy of each gesture under three different experimental settings. Fig. \ref{fig20} shows that: (i) \texttt{G\textsuperscript{3}R} achieves an accuracy of over 90\% for four gestures, and only one gesture is below 90\%; (ii) the recognition accuracy is below 50\% for each gesture in \textit{Vid2Doppler}, \textit{SynMotion}, and \textit{Midas}; the main reason is that they are difficult to simulate diversified and fine-grained gesture reflection properties. Moreover, we refine the recognition accuracy of each gesture for the second setting, as shown in Fig. \ref{fig19}. We observe that the recognition accuracy of the model trained on generated data only closely approximates the performance of a model trained on real-world radar data. Similarly, we also refine the recognition accuracy of each gesture for the third setting. Fig. \ref{fig21} shows that: (i) \texttt{G\textsuperscript{3}R} achieves an accuracy of over 93\% for all gestures, and even approaches 100\% accuracy for some gestures; (ii) \texttt{G\textsuperscript{3}R} improves the accuracy by 57.41 pp/48.79 pp/65.39 pp/74.10 pp/63.26 pp, 55.00 pp/43.94 pp/55.49 pp/62.72 pp/56.27 pp, and 28.18 pp/29.22 pp/34.87 pp/49.74 pp/30.22 pp for PL/PS/CR/UP/KO gestures compared with \textit{Vid2Doppler}, \textit{SynMotion}, and \textit{Midas}, respectively, demonstrating the advantage of \texttt{G\textsuperscript{3}R} in generating fine-grained radar data.
Furthermore, Fig. \ref{fig22} shows the detailed precision-recall curves under  three experimental settings, demonstrating that
\texttt{G\textsuperscript{3}R} is a promising system to achieve higher accuracy for gesture recognition.

\begin{figure}[t]
\centering
\subfloat[Training on generated data only]{\vspace{-1mm}
\begin{minipage}[b]{0.4\textwidth}
\centering
\centerline{\includegraphics[width=51mm]{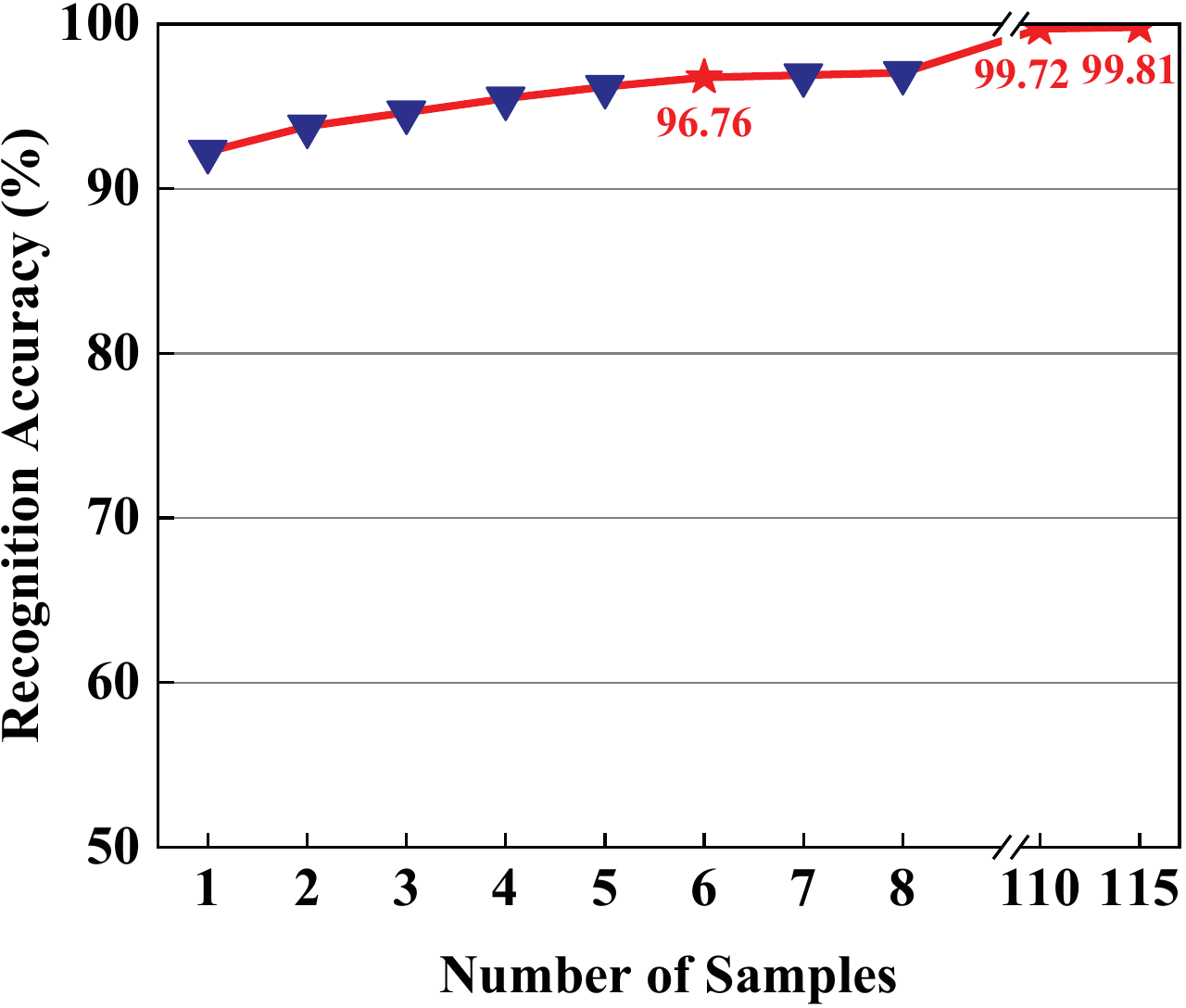}}
\end{minipage}
}
\vfill
\subfloat[Training on generated data + user-independent radar data]{\vspace{-1mm}
\begin{minipage}[b]{0.48\textwidth}
\centering
\centerline{\includegraphics[width=51mm]{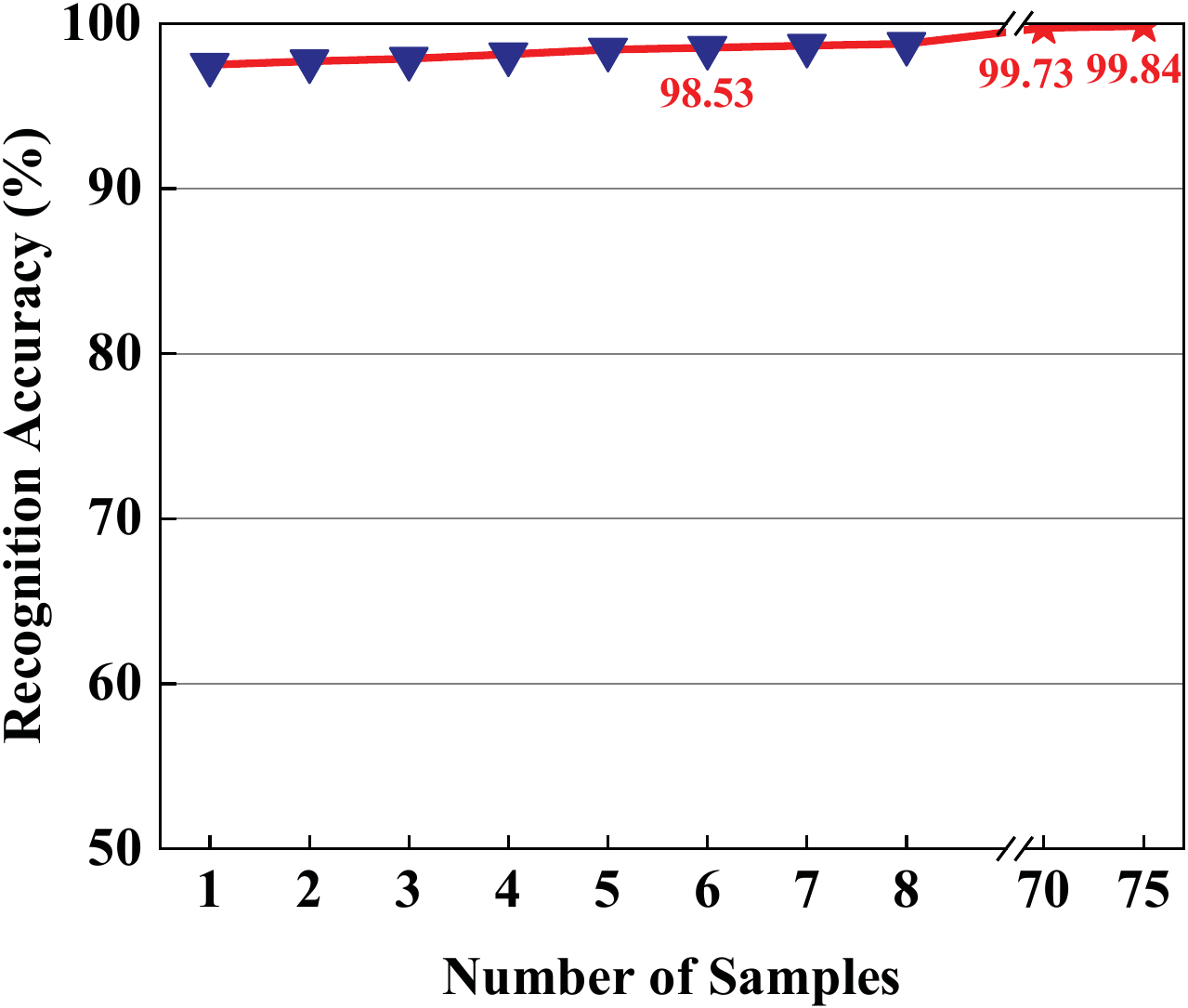}}
\end{minipage}
}
\vspace{-2mm}
\caption{Gesture recognition accuracy changes with the increase of real-world samples from new users.}\label{fig23}
\vspace{-5mm}
\end{figure}

Note that in the above three settings, the users in the test set are entirely excluded from the model's training process. However, it is common for gesture recognition systems to collect some real-world radar data from new users. Therefore, we verify the first and third settings respectively, and continuously add samples from new users to train the model. From Fig. \ref{fig23}(a) we observe that once the number of new user samples reaches 6, the recognition accuracy (96.76\%, solely training with generated data) exceeds the claimed result of the existing work, \textit{mTransSee} (96.7\%, training with large-scale real-world data, at least 8 new user samples) \cite{liu2022mtranssee}. Meanwhile, when users utilize the gesture recognition system, the corresponding radar data can be stored to iteratively update the model, thus achieving more accuracy gesture recognition.
Once the number of samples reaches 115, the recognition accuracy is infinitely close to 100\%, demonstrating the effectiveness of \texttt{G\textsuperscript{3}R}. Note that stored radar data can be automatically annotated according to the recognition results. 
Moreover, Fig. \ref{fig23}(b) shows that when adding 6 samples, the recognition accuracy can reach 98.53\% under the third setting; meanwhile, as new user samples continue to increase, the recognition accuracy can quickly approach 100\%, further demonstrating that \texttt{G\textsuperscript{3}R} provides data support for improving the performance of gesture recognition.

\subsection{Ablation Study}\label{sec5.3}
\subsubsection{Impact of the \textit{Human Parsing}}\label{sec5.3.1}
\begin{figure}[t]
  \centering
  \centerline{\includegraphics[width=85mm]{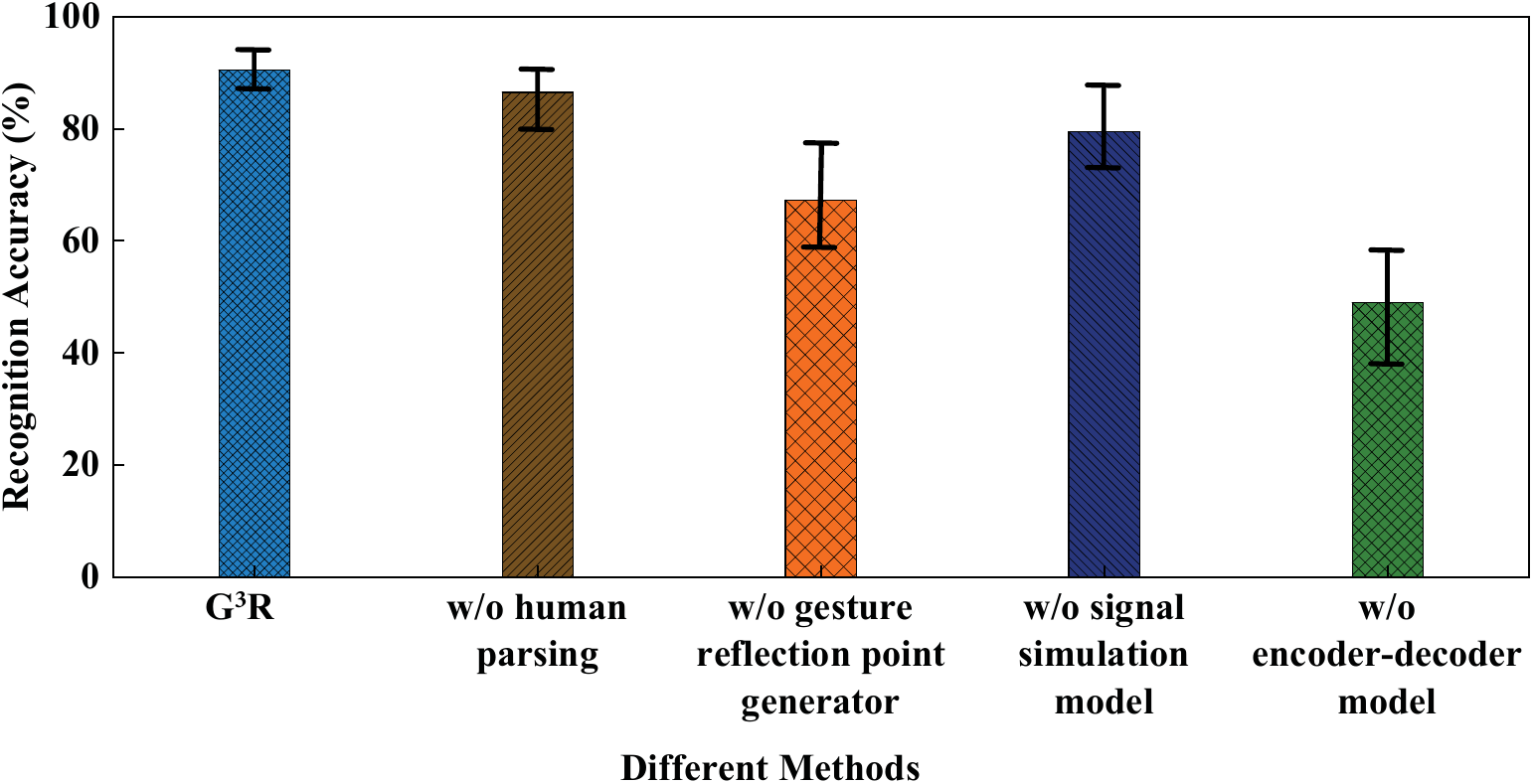}}
  \vspace{-1mm}
  \caption{Impact of individual modules on the recognition accuracy.}\label{fig24}
   \vspace{-3mm}
\end{figure}

To evaluate the contribution of the \textit{human parsing} module, we do not parse and map human constituent parts. Fig. \ref{fig24} shows that \texttt{G\textsuperscript{3}R} improves the recognition accuracy by 3.97 pp compared to \texttt{G\textsuperscript{3}R} w/o \textit{human parsing} module. The results show that \textit{human parsing} module can effectively filter out overflow points and guarantee accurate mapping of reflection points to enhance the quality of generated radar data.
\begin{figure}[t]
  \centering
  \centerline{\includegraphics[width=56mm]{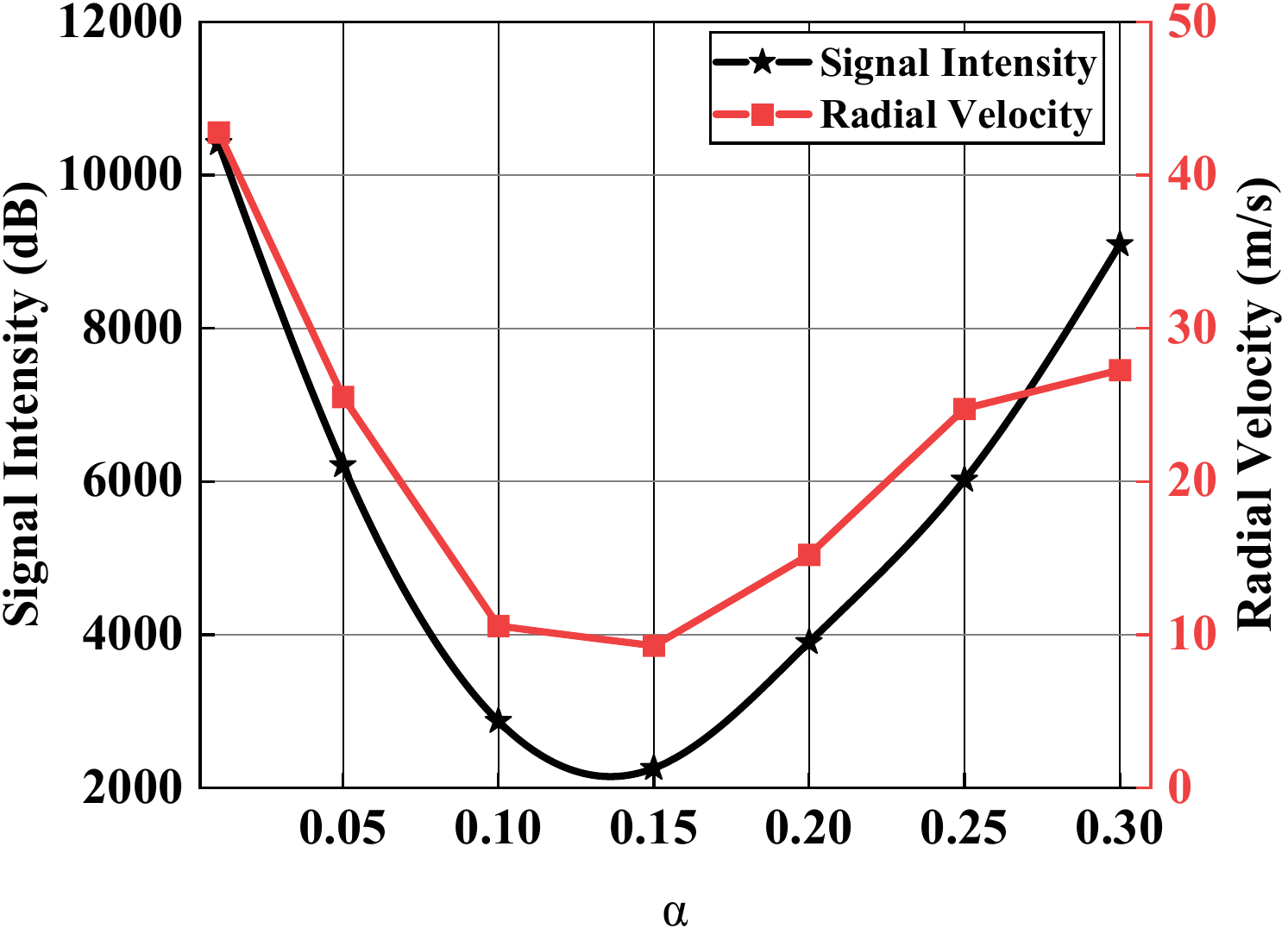}}
  \vspace{-1mm}
  \caption{Average cumulative errors of signal intensity and radial velocity with various attenuation coefficients.}\label{fig25}
   \vspace{-6mm}
\end{figure}

\subsubsection{Impact of the \textit{Gesture Reflection Point Generator}}\label{sec5.3.2}
We then evaluate the benefits of the \textit{gesture reflection point generator} module. Fig. \ref{fig24} shows that \texttt{G\textsuperscript{3}R} improves the recognition accuracy by 23.22 pp compared to \texttt{G\textsuperscript{3}R} w/o \textit{gesture reflection point generator} module. The main reason is that there are few reflection points on the arm, making it difficult to characterize fine-grained gesture features. If the reflective points on the arm are not expanded, it will make the generated radar data tend to the reflection properties of the whole human body, which makes it difficult to simulate the fine-grained gesture data.

\subsubsection{Impact of the \textit{Signal Simulation Model}}\label{sec5.3.3}
To evaluate the benefits of the \textit{signal simulation model}, we directly use the calculated RCS and radial velocity. From Fig. \ref{fig24} we observe that \texttt{G\textsuperscript{3}R} improves the recognition accuracy by 11.04 pp compared to \texttt{G\textsuperscript{3}R} w/o \textit{signal simulation model}. The results show that \textit{signal simulation model} can effectively simulate the multipath reflection and attenuation of radar signals during transmission to improve the quality of generated radar data.

\subsubsection{Impact of the \textit{Encoder-decoder Model}}\label{sec5.3.4}
We also conduct extensive experiments to verify the necessity of the \textit{encoder-decoder model}. We directly concatenate human intensity map and radial velocity to obtain the generated radar data without performing any fitting operation.
Fig. \ref{fig24} shows that \texttt{G\textsuperscript{3}R} improves the recognition accuracy by 41.49 pp compared to \texttt{G\textsuperscript{3}R} w/o \textit{encoder-decoder model}. The results show that \textit{encoder-decoder model} can effectively simulate real-world data characteristics and improve the fidelity of generated radar data.

\subsubsection{Impact of the Attenuation Coefficient}\label{sec5.3.5}
Following the propagation characteristics of radar signals, we vary the attenuation coefficient within the range of 0 to 0.3 to evaluate its impact on the generated radar data. From Fig. \ref{fig25} we observe that average cumulative errors of signal intensity and radial velocity almost reach the minimum when $\alpha$ = 0.15. Meanwhile, we find that: (i) when $\alpha$ is too small, the signal that should be lost reaches the receiver, resulting in excessive noise; (ii) when $\alpha$ is too large, the signal that should be retained does not reach the receiver, making the generated radar data too sparse. 
\begin{figure}[t]
  \centering
  \centerline{\includegraphics[width=49mm]{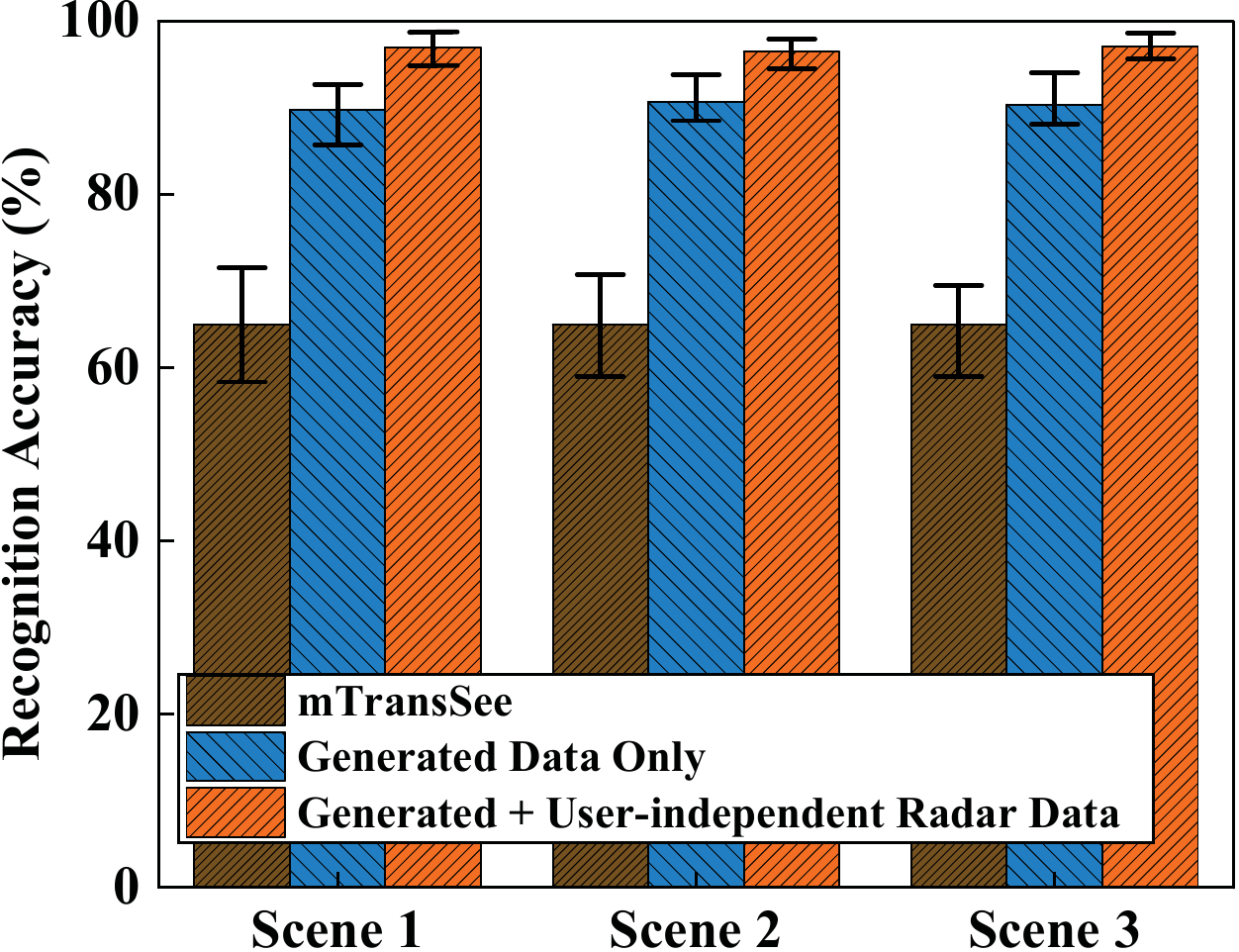}}
  \vspace{-3mm}
  \caption{Gesture recognition accuracy of three settings for different user postures under three different scenes.}\label{fig26}
   \vspace{-5mm}
\end{figure}

\begin{figure}[t]
\centering
\subfloat[Scene 1]{\vspace{-1mm}
\begin{minipage}[b]{0.3\textwidth}
\centering
\centerline{\includegraphics[width=49mm]{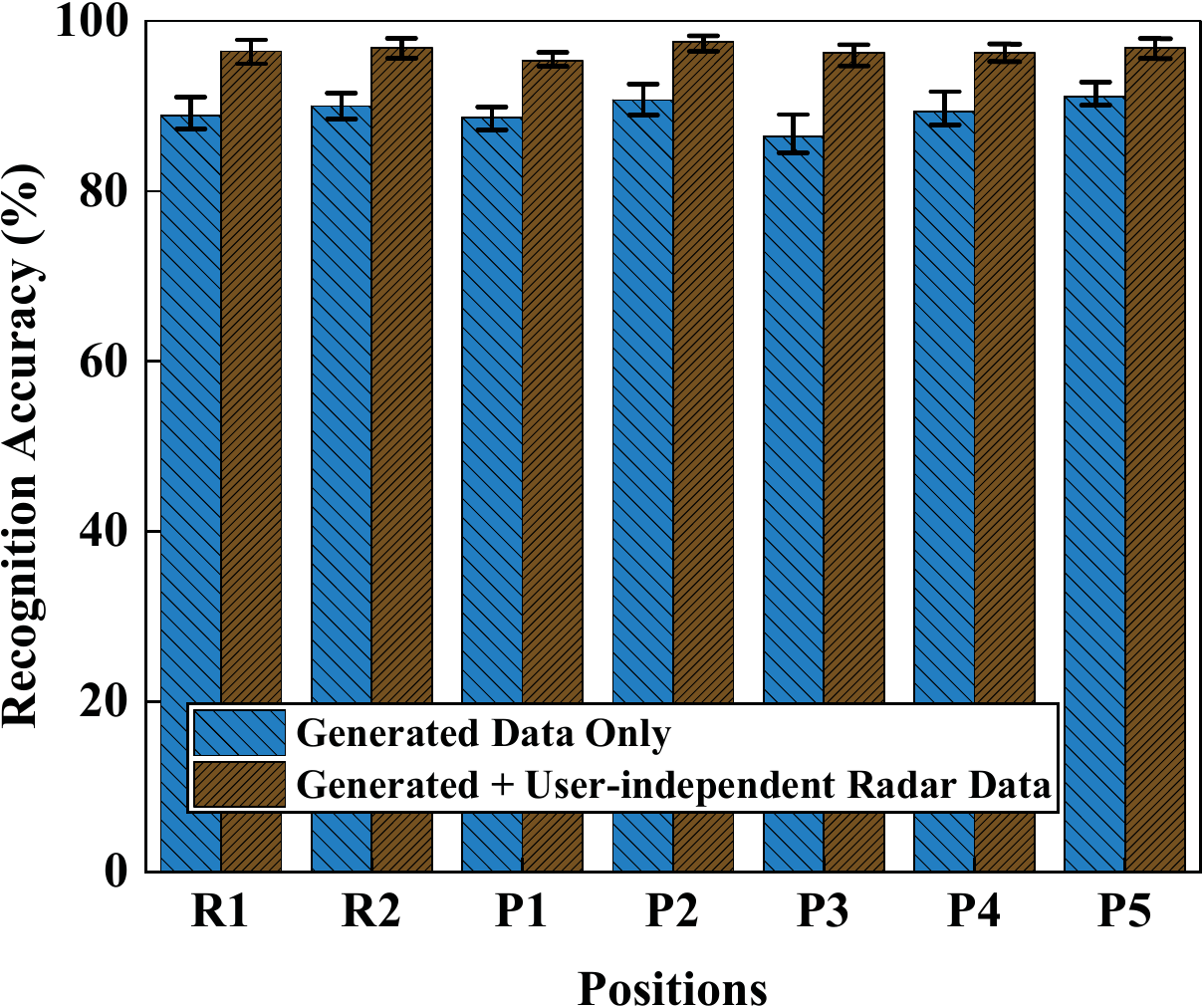}}
\end{minipage}
}
\vfill
\subfloat[Scene 2]{\vspace{-1mm}
\begin{minipage}[b]{0.3\textwidth}
\centering
\centerline{\includegraphics[width=49mm]{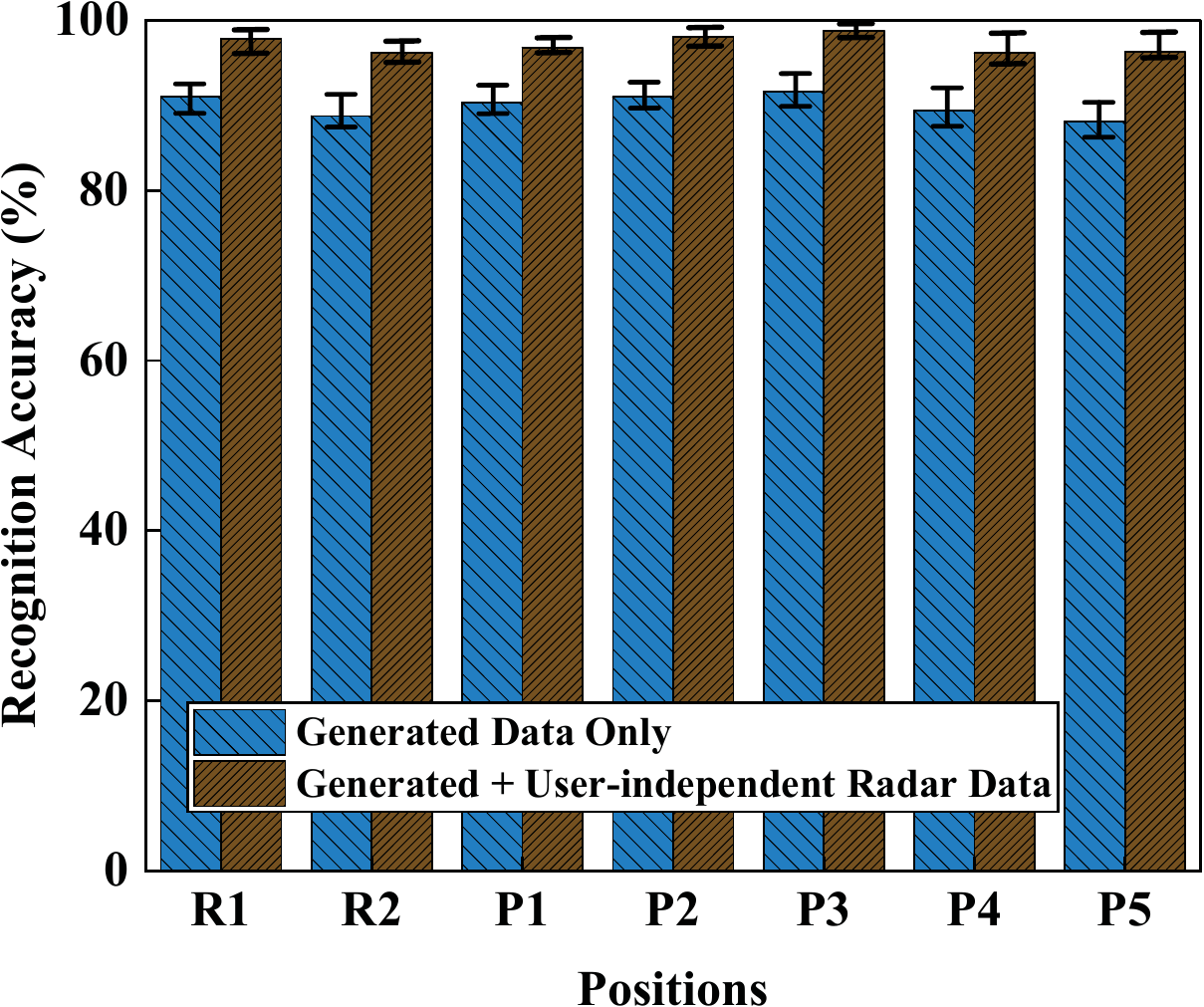}}
\end{minipage}
}
\vfill
\subfloat[Scene 3]{\vspace{-1mm}
\begin{minipage}[b]{0.3\textwidth}
\centering
\centerline{\includegraphics[width=49mm]{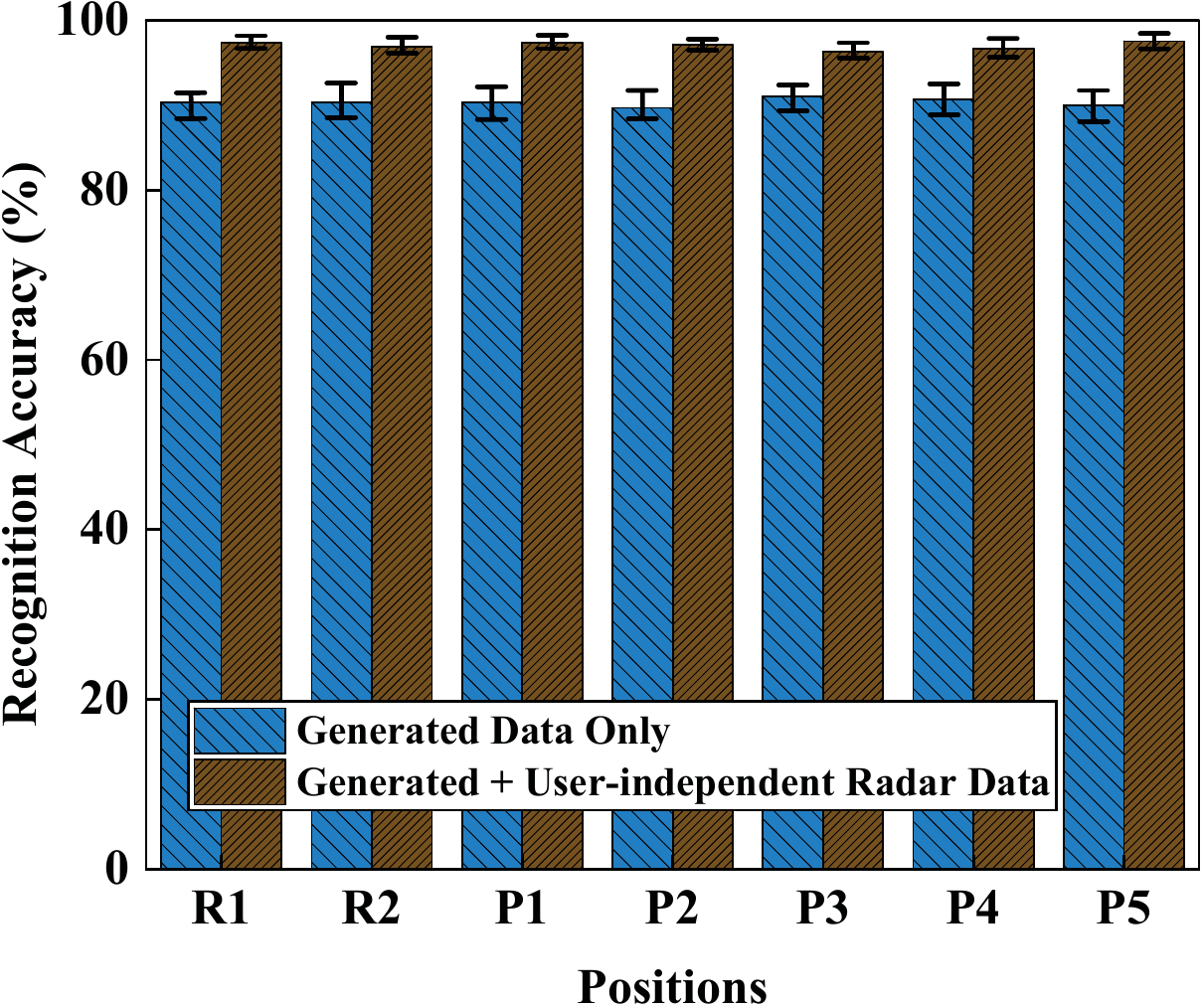}}
\end{minipage}
}
\vspace{-3mm}
\caption{Gesture recognition accuracy of different settings for various positions under three different scenes.}\label{fig27}
\vspace{-5mm}
\end{figure}

\subsection{In-depth Evaluation on Impact of Various Factors}\label{sec5.4} 
\subsubsection{Impact of User Postures}\label{sec5.4.1}
To evaluate the performance of \texttt{G\textsuperscript{3}R} for different postures, we conduct extensive experiments under the first and third settings; meanwhile, we also test the performance of \textit{mTransSee} \cite{liu2022mtranssee} that leverages a transfer learning strategy to learn the knowledge of large-scale users who perform gestures while standing for improving generalization. 
From Fig. \ref{fig26} we observe that: (i) \textit{mTransSee} only achieves an average recognition accuracy of 64.93\% for standing and sitting postures across three different scenes; the main reason is that different user postures would cause differences in radar data due to the reflection properties of signals; (ii) \texttt{G\textsuperscript{3}R} achieves an average recognition accuracy of 90.22\% for standing and sitting postures across three different scenes under the first setting, still maintaining a similar accuracy (90.51\%, see Section \ref{sec5.2.2}); meanwhile, it improves the average recognition accuracy by 25.29 pp compared with \textit{mTransSee}; the main reason is that the generated radar data covers more various user postures, enabling the model to learn more knowledge; (iii) \texttt{G\textsuperscript{3}R} achieves the best average recognition accuracy, 96.78\% under the third setting; meanwhile, it improves the accuracy by 31.85 pp compared with \textit{mTransSee}, demonstrating the advantage of \texttt{G\textsuperscript{3}R} under different postures.

\subsubsection{Impact of User and Radar Positions}\label{sec5.4.2}
We then evaluate the impact of user and radar positions on recognition performance under different scenes. We select 5 different user positions and 2 different radar positions for testing to verify the effectiveness of \texttt{G\textsuperscript{3}R} under the first and third settings. Fig. \ref{fig27} shows that: (i) \texttt{G\textsuperscript{3}R} achieves an average recognition accuracy of 89.31\%/90.02\%/90.33\% and 96.50\%/97.14\%/97.02\% across all positions under scene 1/scene 2/scene 3 under the first and third settings, respectively, demonstrating that \texttt{G\textsuperscript{3}R} can fully adapt to changes in user and radar positions; (ii) the recognition accuracy will experience slight fluctuations in different user and radar positions; the main reason is that changes in radar's heights and user's distances will have a certain impact on the quality of generated radar data.
\begin{figure}[t]
  \centering
  \centerline{\includegraphics[width=49mm]{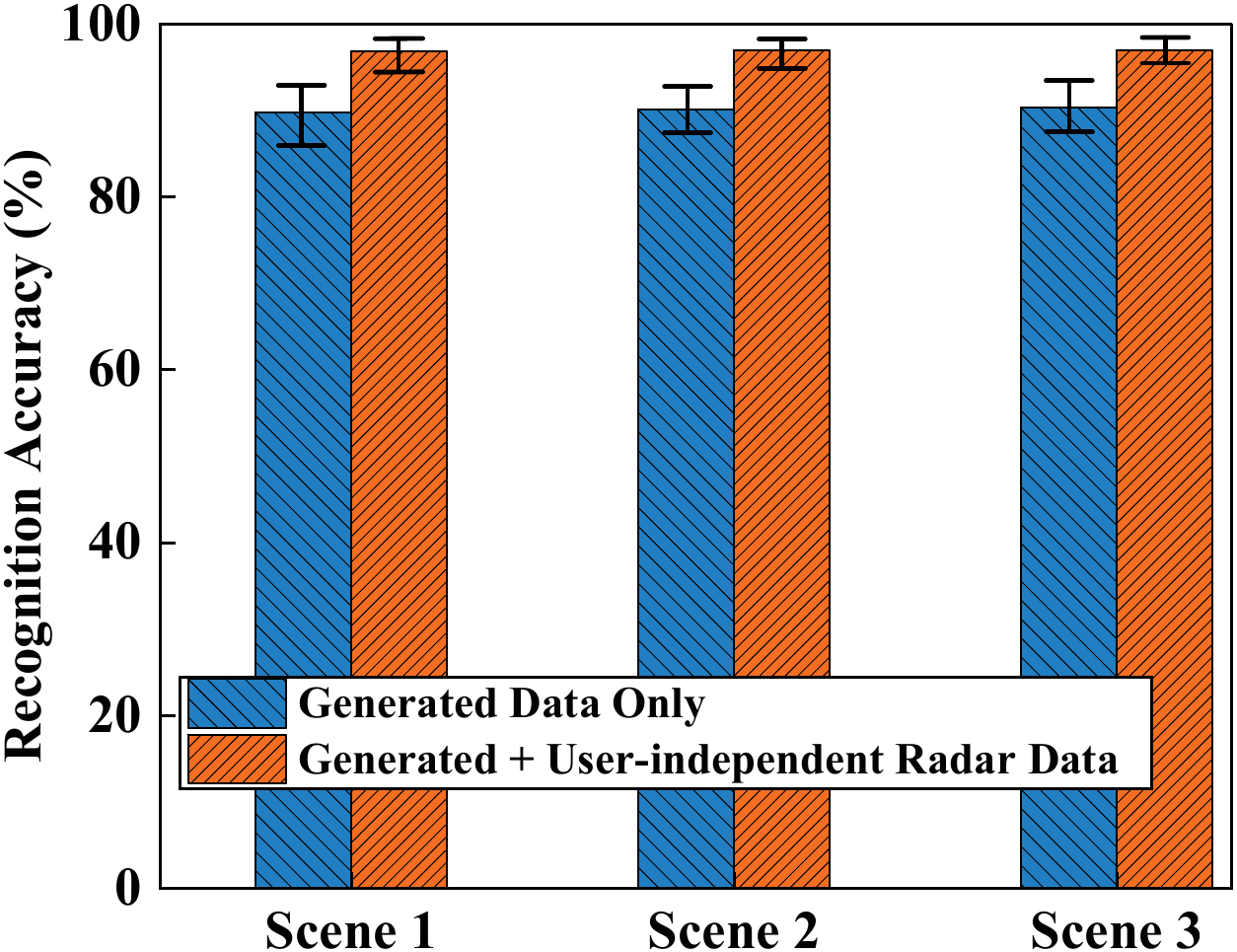}}
  \vspace{-2mm}
  \caption{Gesture recognition accuracy of two settings under three different scenes.}\label{fig28}
   \vspace{-2mm}
\end{figure}
\begin{figure}[t]
\centering
\subfloat[Scene 1]{\vspace{-1mm}
\begin{minipage}[b]{0.3\textwidth}
\centering
\centerline{\includegraphics[width=49mm]{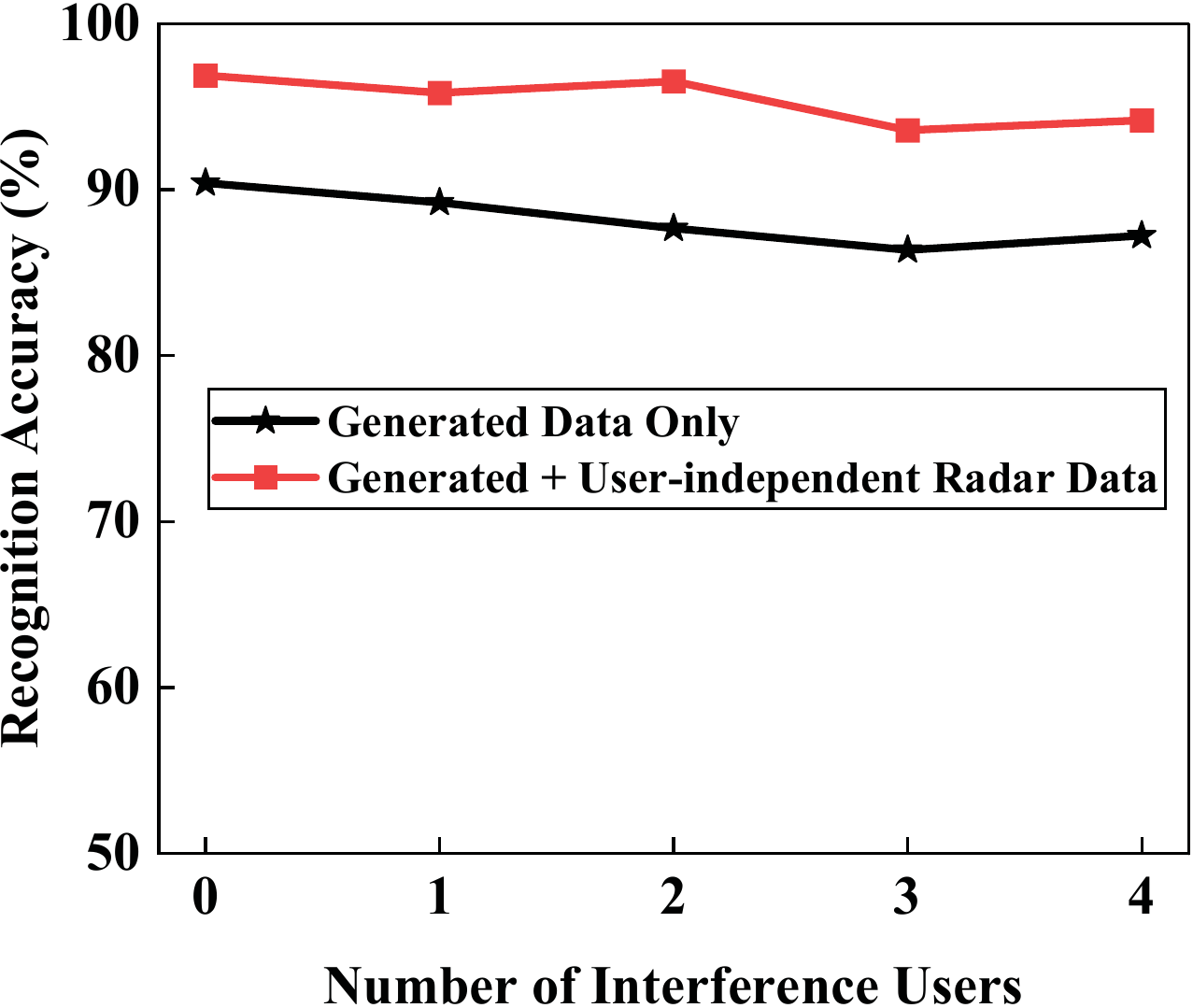}}
\end{minipage}
}
\vfill
\subfloat[Scene 2]{\vspace{-1mm}
\begin{minipage}[b]{0.3\textwidth}
\centering
\centerline{\includegraphics[width=49mm]{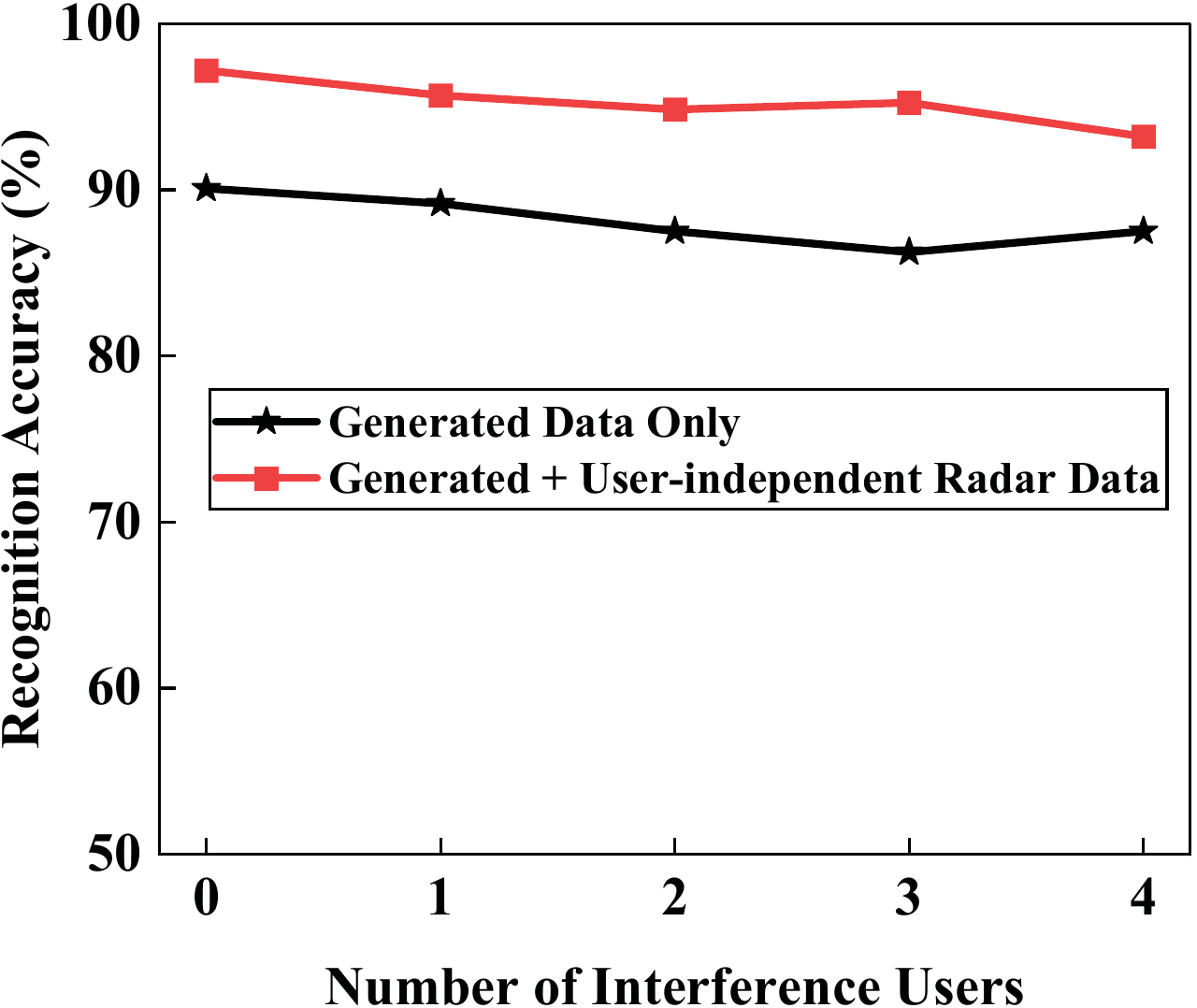}}
\end{minipage}
}
\vfill
\subfloat[Scene 3]{\vspace{-1mm}
\begin{minipage}[b]{0.3\textwidth}
\centering
\centerline{\includegraphics[width=49mm]{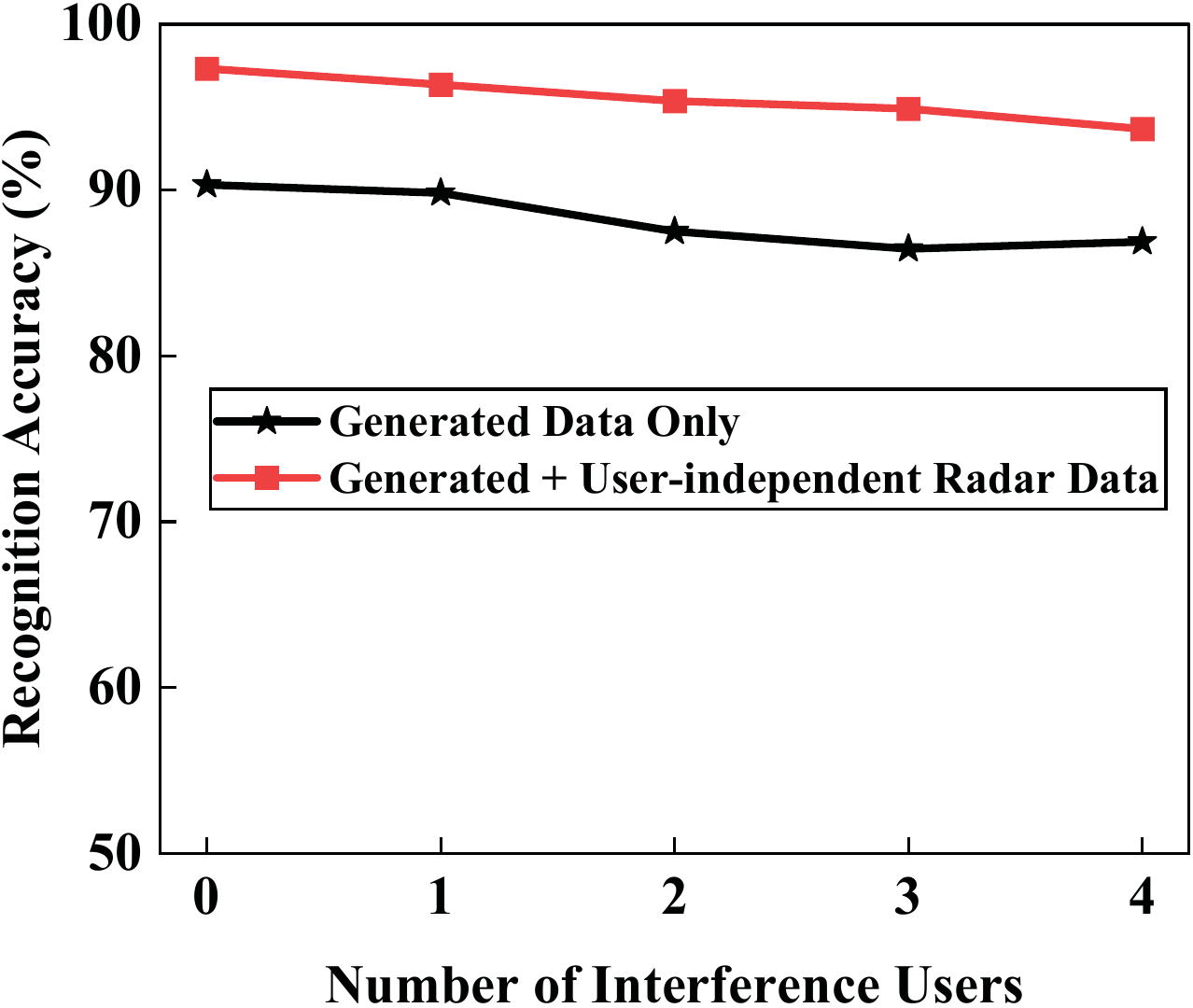}}
\end{minipage}
}
\vspace{-3mm}
\caption{Gesture recognition accuracy of different settings for multi-user coexistence under three different scenes.}\label{fig29}
\vspace{-5mm}
\end{figure}

\subsubsection{Impact of Different Scenes}\label{sec5.4.3}
We measure the overall average recognition accuracy under three different scenes. Fig. \ref{fig28} shows that: (i) \texttt{G\textsuperscript{3}R} achieves an average accuracy of 90.06\% and 96.99\% across various factors under the first and third settings, respectively,  still achieving a similar accuracy (90.51\% and 97.32\%, see Section \ref{sec5.2.2}); (ii) the model's accuracy remains relatively stable across three scenes, with a maximum difference of only 0.58 pp and 0.11 pp under the first and third settings, respectively, demonstrating that \texttt{G\textsuperscript{3}R} is a prospective system to achieve generalized gesture recognition.

\subsubsection{Impact of Multi-user Coexistence}\label{sec5.4.4}
We further evaluate the recognition performance of \texttt{G\textsuperscript{3}R} under multi-user coexistence scenes. Specifically, we allow the interference users (1-4 persons) to freely move within the radar sensing range while a target user performs corresponding gestures. Meanwhile, we ask the target user to perform each gesture in turn in the case of 1-4 interference users (5 gestures $\times$ 8 times $\times$ 2 postures $\times$ 3 positions $\times$ 4 interference cases $\times$ 3 scenes = 2880 samples in total), followed by testing the recognition accuracy. Similarly, we also adopt the first and third settings for evaluation. From Fig. \ref{fig29} we observe that: (i) the recognition accuracy only drops by an average of 2.55 pp and 2.18 pp when there are 1-4 interference users across three scenes under the first and third settings, respectively; (ii) as the number of interference users increases, the recognition accuracy slightly jitters; the main reason is that interference users may perform some activities close to gestures during free movement; (iii) even if 5 users coexist, the recognition accuracy still surpasses 86\% and 93\% across three scenes under the first and third settings, respectively, demonstrating the effectiveness of \texttt{G\textsuperscript{3}R}.

\subsubsection{Impact of Different Gestures}\label{sec5.4.5}
\begin{figure}[t]
  \centering
  \centerline{\includegraphics[width=49mm]{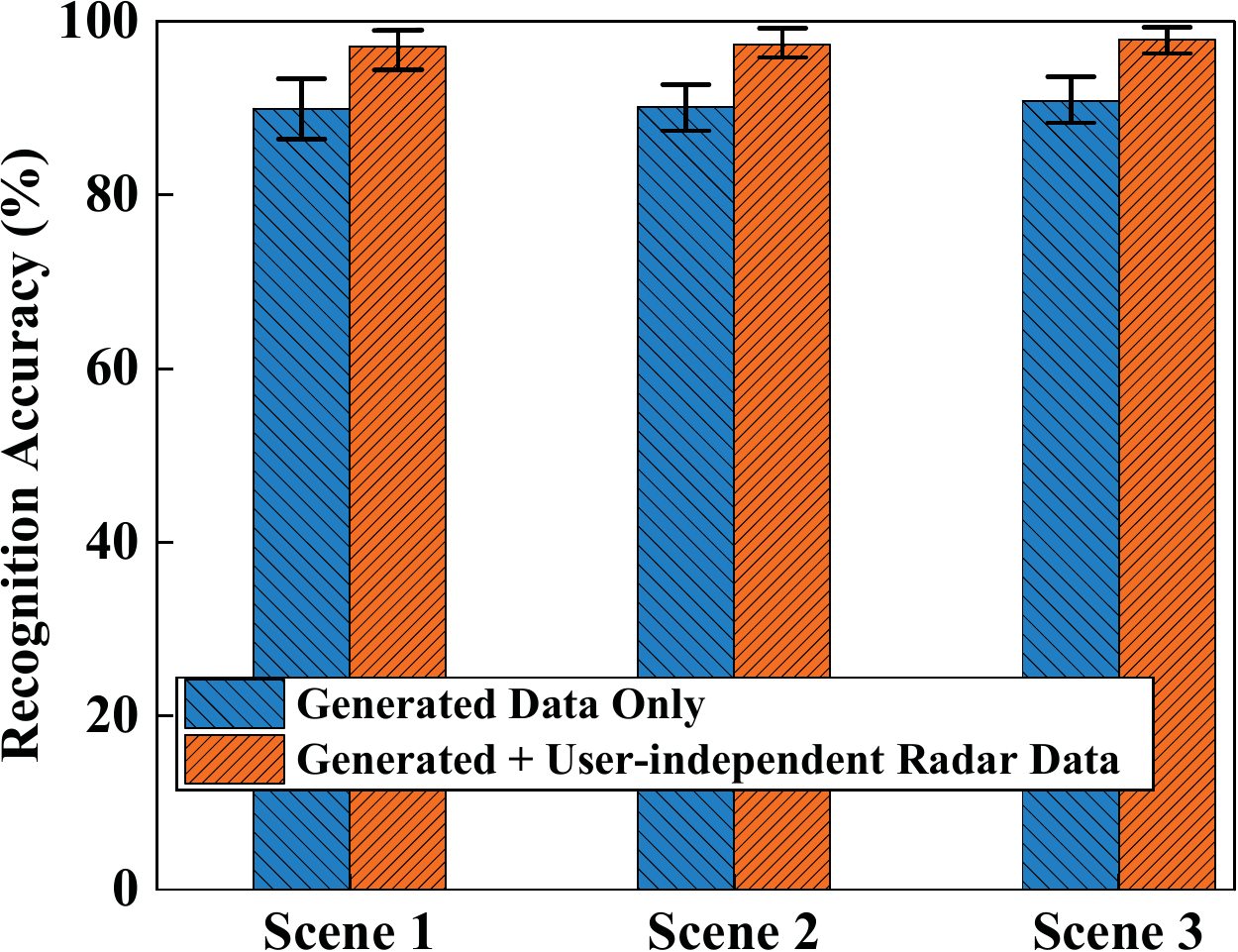}}
  \vspace{-2mm}
  \caption{Gesture recognition accuracy of two settings for two different gestures under three different scenes.}\label{fig30}
   \vspace{-4mm}
\end{figure}

We further evaluate the recognition performance of \texttt{G\textsuperscript{3}R} under different user gestures. Specifically, we additionally collect two different gesture data, i.e., users waving from left (right) to right (left), to verify the generalization of \texttt{G\textsuperscript{3}R}. Thus, in total we collect 5 users $\times$ 2 gestures $\times$ 3 scenes $\times$ 8 times $\times$ 2 postures $\times$ 3 positions = 1440 samples. Fig. \ref{fig30} shows that: (i) \texttt{G\textsuperscript{3}R} achieves an average accuracy of 90.32\% and 97.45\% across various factors under the first and third settings, respectively, still achieving an accuracy comparable to other gestures (90.51\% and 97.32\%, see Section \ref{sec5.2.2}); (ii) the model's accuracy remains relatively stable across three scenes, with a maximum difference of only 0.87 pp and 0.76 pp under the first and third settings, respectively, demonstrating that \texttt{G\textsuperscript{3}R} is a prospective system to achieve generalized recognition for arbitrary gestures.

\section{Related Work}\label{sec6}
\textbf{Contact-based Gesture Recognition}.
Some early works on gesture recognition mainly utilize wearable sensors (such as MoCap \cite{zhu2023stmt, barnachon2014ongoing}, IMUs \cite{villani2023general, sharma2023sparseimu}, and RFID tags \cite{bu2018rf, li2015idsense}). STMT \cite{zhu2023stmt} employs a novel hierarchical transformer architecture that encodes intrinsic and extrinsic representations, incorporating both intra- and inter-frame attention to enhance spatial-temporal mesh modeling to better distinguish between nuanced gestures.
GPOGR \cite{villani2023general} recognizes user gestures by extracting motion information from IMU data. RF-Dial \cite{bu2018rf} attaches two antennas and two RFID tags to users, followed by using translation and rotation to track their trajectories to perform gesture recognition. However, these methods rely on physical contact, necessitating users to wear or attach sensors or RFID tags, significantly impacting user experience. In contrast, we focus on accurate gesture recognition using contactless radar.

\textbf{Contactless Gesture Recognition}.
In recent years, contactless gesture recognition is gaining traction due to the fact that it does not require attaching sensors/tags to users. These sensors can be sound \cite{li2022room, xu2023finger}, RGB camera \cite{mukherjee2019fingertip}, depth camera \cite{alam2021implementation}, WiFi \cite{zhang2021widar3, feng2022wi}, and radar \cite{liu2022mtranssee}. SpeakerGesture \cite{li2022room} adopts smart speakers to realize room-scale gesture recognition without any hardware deployment. AO-Finger \cite{li2022room} proposes a fast fine-grained gesture recognition system based on acoustic-optic sensor fusion. However, sound-based gesture recognition methods are easily affected by environmental noise while involving user privacy. FDT \cite{mukherjee2019fingertip} enables fast recognition by detecting and segmenting user gesture features in 2D videos. ICRS \cite{alam2021implementation} tracks gesture movements through 3D information to achieve precise human-computer interaction. However, vision-based gesture recognition methods will face problems of user privacy leakage, environmental lighting, and non-line-of-sight conditions. Compared to vision and sound technologies, WiFi technologies have a larger operating area and can work in environments with poor illumination. Widar3.0 \cite{zhang2021widar3} extracts domain independent features of gestures to explore a zero-effort corss-domain recognition system. However, it is difficult for WiFi to capture fine-grained gesture features due to the larger wavelength; meanwhile, existing WiFi-based gesture recognition methods \cite{xiao2021onefi, feng2022wi} cannot adapt well to changes in user positions and scenes. Recently, mTransSee \cite{liu2022mtranssee} deploys a radar to collect data, followed by utilizing transfer learning to adapt to environmental changes to support accurate gesture recognition. A commonality in previous work is the need for a large amount of real-world data to train gesture recognition models with the changes in user postures, user positions, and scenes. However, collecting and labeling data is a time-consuming and labor-intensive task.

\textbf{Generating Radar Data}.
Some works have generated multiple types of sensing data, such as IMU data \cite{santhalingam2023synthetic, kwon2020imutube, kwon2021approaching}, sound data \cite{liang2019audio}, depth camera data \cite{planche2017depthsynth}, and radar data \cite{seyfioglu2018diversified, zhang2022synthesized, deng2023midas}, for human sensing models using other data, such as 2D videos, MoCap data, and animated 3D models. The idea of generating radar data using other data sources is not new. In particular, DRM \cite{seyfioglu2018diversified} and FRS \cite{li2019kinect} use MoCap and depth camera data to generate Doppler signals for activity recognition, respectively. However, either the generated radar data is coarse or they are missing some common gestures. Moreover, some works \cite{rahman2023self, rahman2022physics} utilize GANs to augment existing radar datasets, but these methods may generate confusing radar data when users perform similar gestures. Recently, Vid2Doppler \cite{ahuja2021vid2doppler} employs a neural network to directly extract human meshes from 2D videos, followed by generating signal reflection of each vertex to output Doppler radar data. 
SynMotion \cite{zhang2022synthesized} adopts a signal synthesizer to emulate the radar sensing procedure during the transmitting and receiving process of radar chirps, followed by using a signature generator to generate Doppler signals.
Similarly, Midas \cite{deng2023midas} further simulates multipath reflection and attenuation of radar signals to generate realistic radar data. Although Midas can generate some points to characterize the occupancy information of users, these points are randomly selected and coarse-grained, which cannot accurately characterize the spatial features of fine-grained gestures. In contrast, \texttt{G\textsuperscript{3}R} is designed for generalized gesture recognition, which addresses the challenge of simulating diversified and fine-grained reflection properties of user gestures.
\section{Discussion}\label{sec7}
\textbf{Generality}. Our \texttt{G\textsuperscript{3}R} is a unique software pipeline that converts wealthy 2D videos into realistic radar data to address the gap of data limitation for training a generalized gesture recognition across various user postures, positions, and scenes. The results show that training the model with the generated radar data can achieve comparable performance in gesture recognition as training the model with real-world radar data. Meanwhile, \texttt{G\textsuperscript{3}R} guarantees the generalization of the gesture recognition model, enabling it to effectively adapt to changes in user postures, user positions, and scenes. If the video data only focuses on the gesture part in a near distance rather than the whole body, \texttt{G\textsuperscript{3}R} can still generate large-scale radar data by simply replacing the human parsing model with a hand detection model, further demonstrating the advantages of modular design.
Moreover, \texttt{G\textsuperscript{3}R} is compatible with different hardwares and operating systems, and the gesture recognition model has a low complexity, occupying only 20 M of memory, which makes it effortless to deploy on end devices with weak computing power. Besides, with advances in \textit{human parsing}, \textit{skeleton extraction}, and \textit{depth prediction} techniques, it is possible to generate higher-quality radar data.

\textbf{Limitations}. There exist several key technical limitations that need to be addressed. \textit{First}, \texttt{G\textsuperscript{3}R} has not yet attempted to generate radar data beyond distances of 4.5 m; the main reason is that current test distances universally satisfy the needs of normal households. For longer distance scenes, the performance of \texttt{G\textsuperscript{3}R} could experience a decrease.
\textit{Second}, \texttt{G\textsuperscript{3}R} cannot yet recognize who a user gesture corresponds to in different positions, but it is possible to achieve the correspondences by exploiting user behavioral features. \textit{Third}, \texttt{G\textsuperscript{3}R} may fail when a user or the radar is in motion, but it is possible to recognize fine-grained gestures by leveraging radial velocity and signal intensity to partition the reflection points of different human body parts. \textit{Fourth}, some postures (e.g., users perform gestures while lying down) have fewer 2D video sources, it is possible to address it by employing the latest AI-generated content (AIGC) model \cite{du2023exploring}, thereby enhancing the capabilities of \texttt{G\textsuperscript{3}R}. \textit{Fifth}, for unseen gestures (videos), \texttt{G\textsuperscript{3}R} may fail to generate realistic radar data, but it is possible to generate rich and fine-grained radar data using few-shot domain adaptation.
\textit{Sixth}, although the radar data generated by \texttt{G\textsuperscript{3}R} protects user privacy, recording gesture data in itself may bring privacy concerns. This issue is a long-standing research topic in the sensing computing community, and radar sensors will also face unique scrutiny due to their increasing popularity.
\section{Conclusion}\label{sec8}
In this work, we design a software pipeline that exploits wealthy 2D videos to generate rich and fine-grained radar data to support research for gesture recognition. In \texttt{G\textsuperscript{3}R}, a \textit{gesture reflection point generator} expands reflection points of the arm; a \textit{signal simulation model} simulates the multipath reflection and attenuation of radar signals to output the human intensity map; an \textit{encoder-decoder model} combines a \textit{sampling module} and a \textit{fitting module} to generate realistic radar data. We implement and evaluate \texttt{G\textsuperscript{3}R} for gesture recognition using 2D videos from five public datasets, YouTube, and Bilibili, as well as self-collected real-world radar data from 32 volunteers. 
The results demonstrate that \texttt{G\textsuperscript{3}R} achieves 90.51\% accuracy when training with all generated radar data; if we add a small amount of real-world radar data, \texttt{G\textsuperscript{3}R} can achieve 97.32\% accuracy.
Moreover, when facing a new user, the model only requires 6 samples per gesture to achieve 96.76\% and 98.53\% accuracy when training with all generated radar data and all generated radar data with a small amount of real-world radar data, respectively.
Besides, based on the above two training settings, \texttt{G\textsuperscript{3}R} achieves 90.06\% and 96.99\% accuracy across various user postures, positions using self-collected real-world radar data from 5 volunteers, respectively.


\section*{Acknowledgement}
The work is supported by the Innovation Research Group Project of NSFC (61921003), in part by the National Natural Science Foundation of China under No. 62222202, No. 62232004, Beijing Natural Science Foundation (L223002), and in part by the 111 Project under No. B18008.

\bibliographystyle{IEEEtran}
\bibliography{references}

\begin{thebibliography}{10}
\providecommand{\url}[1]{#1}
\csname url@samestyle\endcsname
\providecommand{\newblock}{\relax}
\providecommand{\bibinfo}[2]{#2}
\providecommand{\BIBentrySTDinterwordspacing}{\spaceskip=0pt\relax}
\providecommand{\BIBentryALTinterwordstretchfactor}{4}
\providecommand{\BIBentryALTinterwordspacing}{\spaceskip=\fontdimen2\font plus
\BIBentryALTinterwordstretchfactor\fontdimen3\font minus \fontdimen4\font\relax}
\providecommand{\BIBforeignlanguage}[2]{{%
\expandafter\ifx\csname l@#1\endcsname\relax
\typeout{** WARNING: IEEEtran.bst: No hyphenation pattern has been}%
\typeout{** loaded for the language `#1'. Using the pattern for}%
\typeout{** the default language instead.}%
\else
\language=\csname l@#1\endcsname
\fi
#2}}
\providecommand{\BIBdecl}{\relax}
\BIBdecl

\bibitem{deng2022geryon}
K.~Deng, D.~Zhao, Q.~Han, S.~Wang, Z.~Zhang, A.~Zhou, and H.~Ma, ``Geryon: Edge assisted real-time and robust object detection on drones via mmwave radar and camera fusion,'' \emph{Proc. of ACM IMWUT}, vol.~6, no.~3, pp. 1--27, 2022.

\bibitem{deng2022global}
K.~Deng, D.~Zhao, Q.~Han, Z.~Zhang, S.~Wang, and H.~Ma, ``Global-local feature enhancement network for robust object detection using mmwave radar and camera,'' in \emph{Proc. of IEEE ICASSP}, 2022, pp. 4708--4712.

\bibitem{liu2020real}
H.~Liu, Y.~Wang, A.~Zhou, H.~He, W.~Wang, K.~Wang, P.~Pan, Y.~Lu, L.~Liu, and H.~Ma, ``Real-time arm gesture recognition in smart home scenarios via millimeter wave sensing,'' \emph{Proc. of ACM IMWUT}, vol.~4, no.~4, pp. 1--28, 2020.

\bibitem{liu2022mtranssee}
H.~Liu, K.~Cui, K.~Hu, Y.~Wang, A.~Zhou, L.~Liu, and H.~Ma, ``Mtranssee: Enabling environment-independent mmwave sensing based gesture recognition via transfer learning,'' \emph{Proc. of ACM IMWUT}, vol.~6, no.~1, pp. 1--28, 2022.

\bibitem{ahuja2021vid2doppler}
K.~Ahuja, Y.~Jiang, M.~Goel, and C.~Harrison, ``Vid2doppler: Synthesizing doppler radar data from videos for training privacy-preserving activity recognition,'' in \emph{Proc. of ACM CHI}, 2021, pp. 1--10.

\bibitem{deng2023midas}
K.~Deng, D.~Zhao, Q.~Han, Z.~Zhang, S.~Wang, A.~Zhou, and H.~Ma, ``Midas: Generating mmwave radar data from videos for training pervasive and privacy-preserving human sensing tasks,'' \emph{Proc. of ACM IMWUT}, vol.~7, no.~1, pp. 1--26, 2023.

\bibitem{zhang2022synthesized}
X.~Zhang, Z.~Li, and J.~Zhang, ``Synthesized millimeter-waves for human motion sensing,'' in \emph{Proc. of ACM SenSys}, 2022, pp. 377--390.

\bibitem{liu2021m}
H.~Liu, A.~Zhou, Z.~Dong, Y.~Sun, J.~Zhang, L.~Liu, H.~Ma, J.~Liu, and N.~Yang, ``M-gesture: Person-independent real-time in-air gesture recognition using commodity millimeter wave radar,'' \emph{IEEE Internet of Things Journal}, vol.~9, no.~5, pp. 3397--3415, 2021.

\bibitem{palipana2021pantomime}
S.~Palipana, D.~Salami, L.~A. Leiva, and S.~Sigg, ``Pantomime: Mid-air gesture recognition with sparse millimeter-wave radar point clouds,'' \emph{Proc. of ACM IMWUT}, vol.~5, no.~1, pp. 1--27, 2021.

\bibitem{shen2022ml}
X.~Shen, H.~Zheng, X.~Feng, and J.~Hu, ``Ml-hgr-net: A meta-learning network for fmcw radar based hand gesture recognition,'' \emph{IEEE Sensors Journal}, vol.~22, no.~11, pp. 10\,808--10\,817, 2022.

\bibitem{ling2020ultragesture}
K.~Ling, H.~Dai, Y.~Liu, A.~X. Liu, W.~Wang, and Q.~Gu, ``Ultragesture: Fine-grained gesture sensing and recognition,'' \emph{IEEE Transactions on Mobile Computing}, vol.~21, no.~7, pp. 2620--2636, 2020.

\bibitem{waghmare2023z}
A.~Waghmare, Y.~Ben~Taleb, I.~Chatterjee, A.~Narendra, and S.~Patel, ``Z-ring: Single-point bio-impedance sensing for gesture, touch, object and user recognition,'' in \emph{Proc. of ACM CHI}, 2023, pp. 1--18.

\bibitem{xia2022time}
Z.~Xia and F.~Xu, ``Time-space dimension reduction of millimeter-wave radar point-clouds for smart-home hand-gesture recognition,'' \emph{IEEE Sensors Journal}, vol.~22, no.~5, pp. 4425--4437, 2022.

\bibitem{chen2021sensecollect}
W.~Chen, S.~Lin, E.~Thompson, and J.~Stankovic, ``Sensecollect: We need efficient ways to collect on-body sensor-based human activity data!'' \emph{Proc. of ACM IMWUT}, vol.~5, no.~3, pp. 1--27, 2021.

\bibitem{kuehne2011hmdb}
H.~Kuehne, H.~Jhuang, E.~Garrote, T.~Poggio, and T.~Serre, ``Hmdb: a large video database for human motion recognition,'' in \emph{Proc. of IEEE ICCV}, 2011, pp. 2556--2563.

\bibitem{abu2016youtube}
S.~Abu-El-Haija, N.~Kothari, J.~Lee, P.~Natsev, G.~Toderici, B.~Varadarajan, and S.~Vijayanarasimhan, ``Youtube-8m: A large-scale video classification benchmark,'' \emph{arXiv preprint arXiv:1609.08675}, 2016.

\bibitem{perera2018uav}
A.~G. Perera, Y.~Wei~Law, and J.~Chahl, ``Uav-gesture: A dataset for uav control and gesture recognition,'' in \emph{Proc. of ECCV Workshops}, 2018, pp. 0--0.

\bibitem{soomro2012ucf101}
K.~Soomro, A.~R. Zamir, and M.~Shah, ``Ucf101: A dataset of 101 human actions classes from videos in the wild,'' \emph{arXiv preprint arXiv:1212.0402}, 2012.

\bibitem{escalante2016chalearn}
H.~J. Escalante, V.~Ponce-L{\'o}pez, J.~Wan, M.~A. Riegler, B.~Chen, A.~Clap{\'e}s, S.~Escalera, I.~Guyon, X.~Bar{\'o}, P.~Halvorsen \emph{et~al.}, ``Chalearn joint contest on multimedia challenges beyond visual analysis: An overview,'' in \emph{Proc. of IEEE ICPR}, 2016, pp. 67--73.

\bibitem{kwon2020imutube}
H.~Kwon, C.~Tong, H.~Haresamudram, Y.~Gao, G.~D. Abowd, N.~D. Lane, and T.~Ploetz, ``Imutube: Automatic extraction of virtual on-body accelerometry from video for human activity recognition,'' \emph{Proc. of ACM IMWUT}, vol.~4, no.~3, pp. 1--29, 2020.

\bibitem{kwon2021approaching}
H.~Kwon, B.~Wang, G.~D. Abowd, and T.~Pl{\"o}tz, ``Approaching the real-world: Supporting activity recognition training with virtual imu data,'' \emph{Proc. of ACM IMWUT}, vol.~5, no.~3, pp. 1--32, 2021.

\bibitem{liang2019audio}
D.~Liang and E.~Thomaz, ``Audio-based activities of daily living (adl) recognition with large-scale acoustic embeddings from online videos,'' \emph{Proc. of ACM IMWUT}, vol.~3, no.~1, pp. 1--18, 2019.

\bibitem{santhalingam2023synthetic}
P.~S. Santhalingam, P.~Pathak, H.~Rangwala, and J.~Kosecka, ``Synthetic smartwatch imu data generation from in-the-wild asl videos,'' \emph{Proc. of ACM IMWUT}, vol.~7, no.~2, pp. 1--34, 2023.

\bibitem{lin2017performance}
Y.~Lin and J.~Le~Kernec, ``Performance analysis of classification algorithms for activity recognition using micro-doppler feature,'' in \emph{Proc. of IEEE CIS}, 2017, pp. 480--483.

\bibitem{seyfioglu2018diversified}
M.~S. Seyfioglu, B.~Erol, S.~Z. Gurbuz, and M.~G. Amin, ``Diversified radar micro-doppler simulations as training data for deep residual neural networks,'' in \emph{Proc. of IEEE radarConf}, 2018, pp. 0612--0617.

\bibitem{erol2015kinect}
B.~Erol and S.~Z. Gurbuz, ``A kinect-based human micro-doppler simulator,'' \emph{IEEE AESM}, vol.~30, no.~5, pp. 6--17, 2015.

\bibitem{li2019kinect}
J.~Li, A.~Shrestha, J.~Le~Kernec, and F.~Fioranelli, ``From kinect skeleton data to hand gesture recognition with radar,'' \emph{The Journal of Engineering}, vol. 2019, no.~20, pp. 6914--6919, 2019.

\bibitem{erol2019gan}
B.~Erol, S.~Z. Gurbuz, and M.~G. Amin, ``Gan-based synthetic radar micro-doppler augmentations for improved human activity recognition,'' in \emph{Proc. of IEEE radarConf}, 2019, pp. 1--5.

\bibitem{rahman2021physics}
M.~M. Rahman, S.~Z. Gurbuz, and M.~G. Amin, ``Physics-aware design of multi-branch gan for human rf micro-doppler signature synthesis,'' in \emph{Proc. of IEEE radarConf}, 2021, pp. 1--6.

\bibitem{rohling1983radar}
H.~Rohling, ``Radar cfar thresholding in clutter and multiple target situations,'' \emph{IEEE TAES}, no.~4, pp. 608--621, 1983.

\bibitem{ester1996density}
M.~Ester, H.-P. Kriegel, J.~Sander, X.~Xu \emph{et~al.}, ``A density-based algorithm for discovering clusters in large spatial databases with noise,'' in \emph{Proc. of ACM SIGKDD}, vol.~96, no.~34, 1996, pp. 226--231.

\bibitem{statista}
\BIBentryALTinterwordspacing
Statista. (2020, February) Hours of video uploaded to youtube every minute as of february. [Online]. Available: \url{https://www.statista.com/statistics/259477/hours-of-video-uploaded-to-youtube-every-minute/}
\BIBentrySTDinterwordspacing

\bibitem{zhang2020blended}
X.~Zhang, Y.~Chen, B.~Zhu, J.~Wang, and M.~Tang, ``Blended grammar network for human parsing,'' in \emph{Proc. of ECCV}, 2020, pp. 189--205.

\bibitem{zhang2022human}
X.~Zhang, Y.~Chen, M.~Tang, J.~Wang, X.~Zhu, and Z.~Lei, ``Human parsing with part-aware relation modeling,'' \emph{IEEE TMM}, 2022.

\bibitem{liu2022cdgnet}
K.~Liu, O.~Choi, J.~Wang, and W.~Hwang, ``Cdgnet: Class distribution guided network for human parsing,'' in \emph{Proc. of IEEE CVPR}, 2022, pp. 4473--4482.

\bibitem{cai2020learning}
Y.~Cai, Z.~Wang, Z.~Luo, B.~Yin, A.~Du, H.~Wang, X.~Zhang, X.~Zhou, E.~Zhou, and J.~Sun, ``Learning delicate local representations for multi-person pose estimation,'' in \emph{Proc. of ECCV}, 2020, pp. 455--472.

\bibitem{sun2022putting}
Y.~Sun, W.~Liu, Q.~Bao, Y.~Fu, T.~Mei, and M.~J. Black, ``Putting people in their place: Monocular regression of 3d people in depth,'' in \emph{Proc. of IEEE CVPR}, 2022, pp. 13\,243--13\,252.

\bibitem{guan2022out}
S.~Guan, J.~Xu, M.~Z. He, Y.~Wang, B.~Ni, and X.~Yang, ``Out-of-domain human mesh reconstruction via dynamic bilevel online adaptation,'' \emph{IEEE TPAMI}, vol.~45, no.~4, pp. 5070--5086, 2022.

\bibitem{bhat2023zoedepth}
S.~F. Bhat, R.~Birkl, D.~Wofk, P.~Wonka, and M.~M{\"u}ller, ``Zoedepth: Zero-shot transfer by combining relative and metric depth,'' \emph{arXiv preprint arXiv:2302.12288}, 2023.

\bibitem{whitted2005improved}
T.~Whitted, ``An improved illumination model for shaded display,'' in \emph{ACM Siggraph 2005 Courses}, 2005, pp. 4--es.

\bibitem{yue2022cornerradar}
S.~Yue, H.~He, P.~Cao, K.~Zha, M.~Koizumi, and D.~Katabi, ``Cornerradar: Rf-based indoor localization around corners,'' \emph{Proc. of ACM IMWUT}, vol.~6, no.~1, pp. 1--24, 2022.

\bibitem{scheiner2020seeing}
N.~Scheiner, F.~Kraus, F.~Wei, B.~Phan, F.~Mannan, N.~Appenrodt, W.~Ritter, J.~Dickmann, K.~Dietmayer, B.~Sick \emph{et~al.}, ``Seeing around street corners: Non-line-of-sight detection and tracking in-the-wild using doppler radar,'' in \emph{Proc. of IEEE CVPR}, 2020, pp. 2068--2077.

\bibitem{ling1989shooting}
H.~Ling, R.-C. Chou, and S.-W. Lee, ``Shooting and bouncing rays: Calculating the rcs of an arbitrarily shaped cavity,'' \emph{IEEE TAP}, vol.~37, no.~2, pp. 194--205, 1989.

\bibitem{rao2017introduction}
S.~Rao, ``Introduction to mmwave sensing: Fmcw radars,'' \emph{Texas Instruments (TI) mmWave Training Series}, pp. 1--11, 2017.

\bibitem{qi2017pointnet}
C.~R. Qi, H.~Su, K.~Mo, and L.~J. Guibas, ``Pointnet: Deep learning on point sets for 3d classification and segmentation,'' in \emph{Proc. of IEEE CVPR}, 2017, pp. 652--660.

\bibitem{wang2019dynamic}
Y.~Wang, Y.~Sun, Z.~Liu, S.~E. Sarma, M.~M. Bronstein, and J.~M. Solomon, ``Dynamic graph cnn for learning on point clouds,'' \emph{ACM TOG}, vol.~38, no.~5, pp. 1--12, 2019.

\bibitem{fan2017point}
H.~Fan, H.~Su, and L.~J. Guibas, ``A point set generation network for 3d object reconstruction from a single image,'' in \emph{Proc. of IEEE CVPR}, 2017, pp. 605--613.

\bibitem{kingma2014adam}
D.~P. Kingma and J.~Ba, ``Adam: A method for stochastic optimization,'' in \emph{Proc. of ICLR}, 2015.

\bibitem{zhu2023stmt}
X.~Zhu, P.-Y. Huang, J.~Liang, C.~M. de~Melo, and A.~G. Hauptmann, ``Stmt: A spatial-temporal mesh transformer for mocap-based action recognition,'' in \emph{Proc. of IEEE CVPR}, 2023, pp. 1526--1536.

\bibitem{barnachon2014ongoing}
M.~Barnachon, S.~Bouakaz, B.~Boufama, and E.~Guillou, ``Ongoing human action recognition with motion capture,'' \emph{Pattern Recognition}, vol.~47, no.~1, pp. 238--247, 2014.

\bibitem{villani2023general}
V.~Villani, C.~Secchi, M.~Lippi, and L.~Sabattini, ``A general pipeline for online gesture recognition in human--robot interaction,'' \emph{IEEE Transactions on Human-Machine Systems}, vol.~53, no.~2, pp. 315--324, 2023.

\bibitem{sharma2023sparseimu}
A.~Sharma, C.~Salchow-H{\"o}mmen, V.~S. Mollyn, A.~S. Nittala, M.~A. Hedderich, M.~Koelle, T.~Seel, and J.~Steimle, ``Sparseimu: Computational design of sparse imu layouts for sensing fine-grained finger microgestures,'' \emph{ACM TOCHI}, vol.~30, no.~3, pp. 1--40, 2023.

\bibitem{bu2018rf}
Y.~Bu, L.~Xie, Y.~Gong, C.~Wang, L.~Yang, J.~Liu, and S.~Lu, ``Rf-dial: An rfid-based 2d human-computer interaction via tag array,'' in \emph{Proc. of IEEE INFOCOM}, 2018, pp. 837--845.

\bibitem{li2015idsense}
H.~Li, C.~Ye, and A.~P. Sample, ``Idsense: A human object interaction detection system based on passive uhf rfid,'' in \emph{Proc. of ACM CHI}, 2015, pp. 2555--2564.

\bibitem{li2022room}
D.~Li, J.~Liu, S.~I. Lee, and J.~Xiong, ``Room-scale hand gesture recognition using smart speakers,'' in \emph{Proc. of ACM SenSys}, 2022, pp. 462--475.

\bibitem{xu2023finger}
C.~Xu, B.~Zhou, G.~Krishnan, and S.~Nayar, ``Ao-finger: Hands-free fine-grained finger gesture recognition via acoustic-optic sensor fusing,'' in \emph{Proc. of ACM CHI}, 2023, pp. 1--14.

\bibitem{mukherjee2019fingertip}
S.~Mukherjee, S.~A. Ahmed, D.~P. Dogra, S.~Kar, and P.~P. Roy, ``Fingertip detection and tracking for recognition of air-writing in videos,'' \emph{ESWA}, vol. 136, pp. 217--229, 2019.

\bibitem{alam2021implementation}
M.~S. Alam, K.-C. Kwon, and N.~Kim, ``Implementation of a character recognition system based on finger-joint tracking using a depth camera,'' \emph{IEEE Transactions on Human-Machine Systems}, vol.~51, no.~3, pp. 229--241, 2021.

\bibitem{zhang2021widar3}
Y.~Zhang, Y.~Zheng, K.~Qian, G.~Zhang, Y.~Liu, C.~Wu, and Z.~Yang, ``Widar3. 0: Zero-effort cross-domain gesture recognition with wi-fi,'' \emph{IEEE TPAMI}, vol.~44, no.~11, pp. 8671--8688, 2021.

\bibitem{feng2022wi}
C.~Feng, N.~Wang, Y.~Jiang, X.~Zheng, K.~Li, Z.~Wang, and X.~Chen, ``Wi-learner: Towards one-shot learning for cross-domain wi-fi based gesture recognition,'' \emph{Proc. of ACM IMWUT}, vol.~6, no.~3, pp. 1--27, 2022.

\bibitem{xiao2021onefi}
R.~Xiao, J.~Liu, J.~Han, and K.~Ren, ``Onefi: One-shot recognition for unseen gesture via cots wifi,'' in \emph{Proc. of ACM SenSys}, 2021, pp. 206--219.

\bibitem{planche2017depthsynth}
B.~Planche, Z.~Wu, K.~Ma, S.~Sun, S.~Kluckner, O.~Lehmann, T.~Chen, A.~Hutter, S.~Zakharov, H.~Kosch \emph{et~al.}, ``Depthsynth: Real-time realistic synthetic data generation from cad models for 2.5 d recognition,'' in \emph{Proc. of IEEE 3DV}, 2017, pp. 1--10.

\bibitem{rahman2023self}
M.~M. Rahman and S.~Z. Gurbuz, ``Self-supervised contrastive learning for radar-based human activity recognition,'' in \emph{Proc. of IEEE RadarConf}, 2023, pp. 1--6.

\bibitem{rahman2022physics}
M.~M. Rahman, S.~Z. Gurbuz, and M.~G. Amin, ``Physics-aware generative adversarial networks for radar-based human activity recognition,'' \emph{IEEE TAES}, 2022.

\bibitem{du2023exploring}
H.~Du, R.~Zhang, D.~Niyato, J.~Kang, Z.~Xiong, D.~I. Kim, X.~S. Shen, and H.~V. Poor, ``Exploring collaborative distributed diffusion-based ai-generated content (aigc) in wireless networks,'' \emph{IEEE Network}, no.~99, pp. 1--8, 2023.

\end{thebibliography}

\begin{IEEEbiography}[{\includegraphics[width=1in,height=1.25in,clip,keepaspectratio]{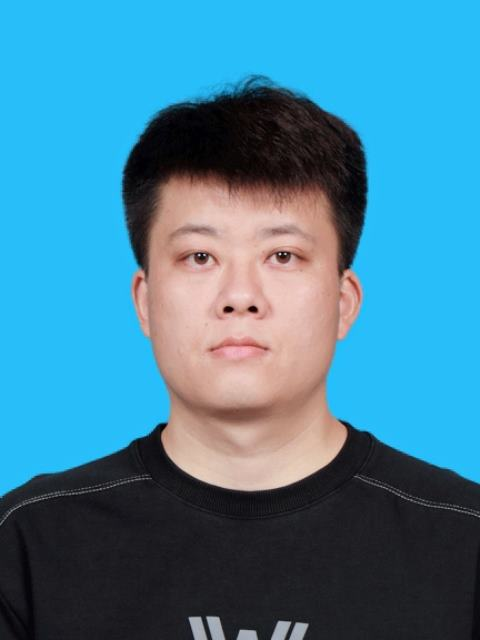}}]{Kaikai Deng} (Student Member, IEEE) is currently pursuing the Ph.D. degree with the School of Computer Science and the Beijing Key Laboratory of Intelligent Telecommunications Software and Multimedia, Beijing University of Posts and Telecommunications, China. He serves as a reviewer for many top conferences and journals, such as IMWUT/UbiComp (Outstanding reviewer), Knowledge-Based Systems, Signal Processing, and Frontiers of Computer Science. His research interests include Internet of Things, edge computing, and multimodal sensing computing.
\end{IEEEbiography}

\begin{IEEEbiography}[{\includegraphics[width=1in,height=1.25in,clip,keepaspectratio]{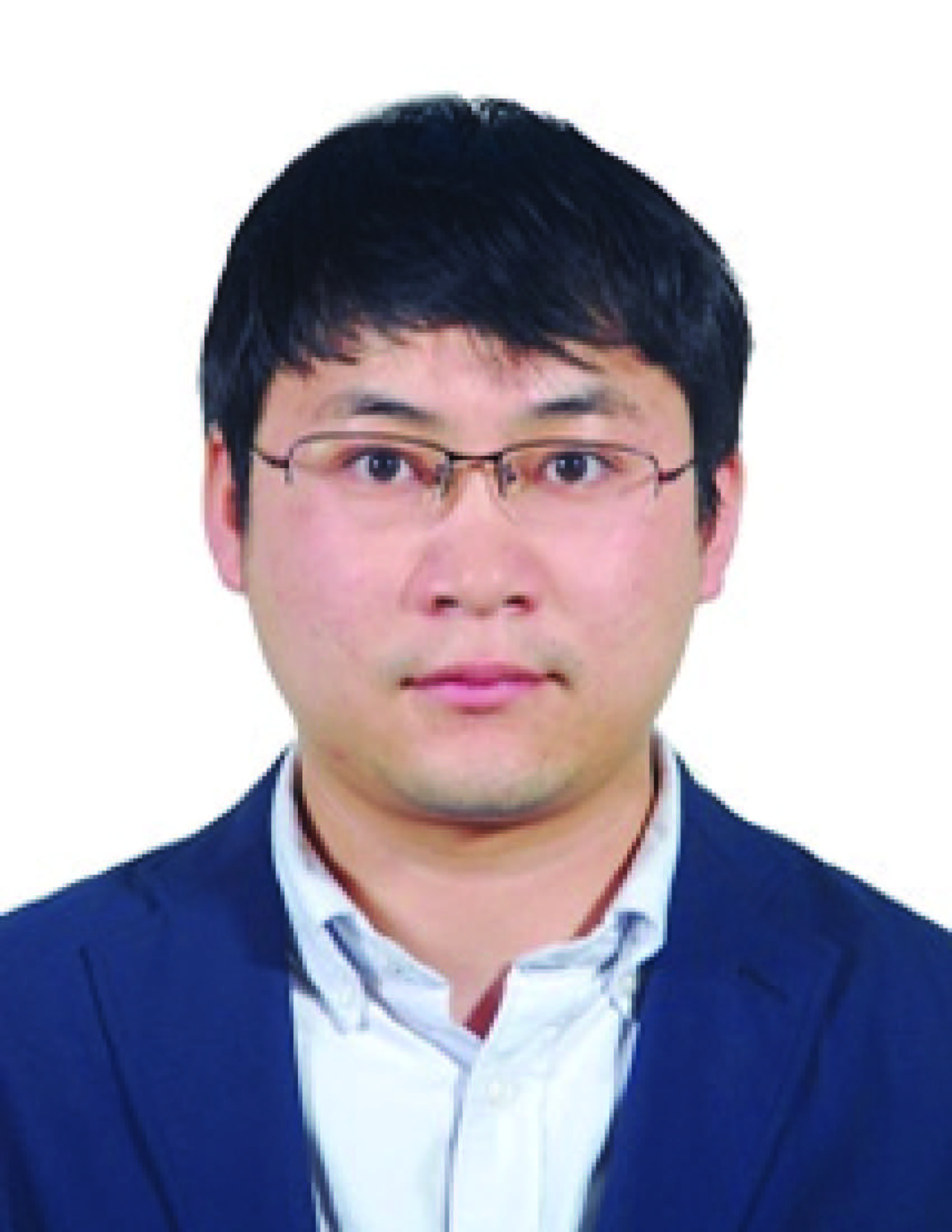}}]{Dong Zhao} (Member, IEEE) received the B.S. degree from the Department of Computer Science and Technology, Henan University, Kaifeng, China, in 2008, and the Ph.D. degree from the School of Computer Science, Beijing University of Posts and Telecommunications, Beijing, China, in 2014.
He is a Professor with Beijing Key Laboratory of Intelligent Telecommunications Software and Multimedia, Beijing University of Posts and Telecommunications. He was a visiting Ph.D. student with Illinois Institute of Technology, Chicago, IL, USA, from 2012 to 2013, and was a Visiting Scholar with Rutgers University, New Brunswick, NJ, USA, from 2019 to 2020. His research interests include Internet of Things, mobile crowd sensing, urban computing, and data science, and he has published over 60 papers and two books on these fields. Dr. Zhao was awarded the China Computer Federation (CCF) Outstanding Doctoral Dissertation Award in 2015, the ACM Beijing Doctoral Dissertation Award in 2015, and the Natural Science Award of the Ministry of Education, China, in 2017.
\end{IEEEbiography}
\vspace{-13mm}

\begin{IEEEbiography}[{\includegraphics[width=1in,height=1.25in,clip,keepaspectratio]{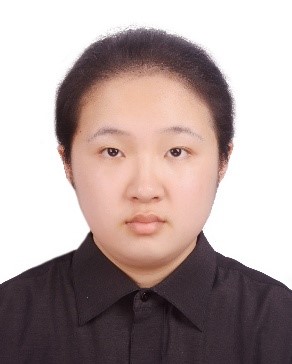}}]{Wenxin Zheng} is currently working toward the M.S. degree with the Beijing Key Lab of Intelligent Telecommunications Software and Multimedia, Beijing University of Posts and Telecommunications, China. Her research interests include Internet of Things, data generation, and infrastructure-assisted autonomous driving.
\end{IEEEbiography}
\vspace{-13mm}

\begin{IEEEbiography}[{\includegraphics[width=1in,height=1.25in,clip,keepaspectratio]{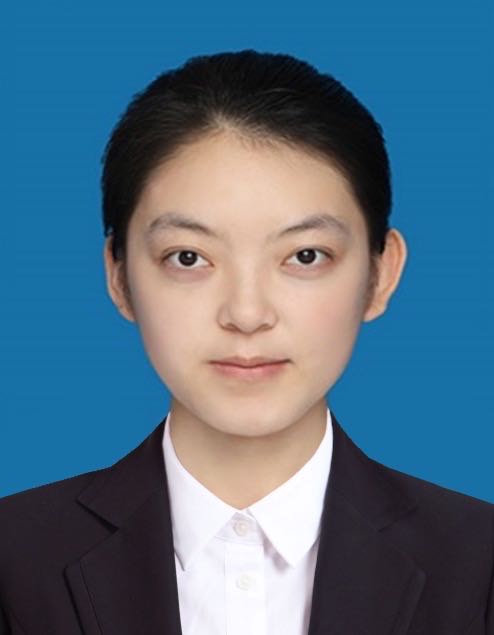}}]{Yue Ling} is currently working toward the PhD degree with the School of Computer Science and the Beijing Key Laboratory of Intelligent Telecommunications Software and Multimedia, Beijing University of Posts and Telecommunications, China. Her research interests include Internet of Things, AIGC, and multimodal data augmentation.
\end{IEEEbiography}
\vspace{-13mm}

\begin{IEEEbiography}[{\includegraphics[width=1in,height=1.25in,clip,keepaspectratio]{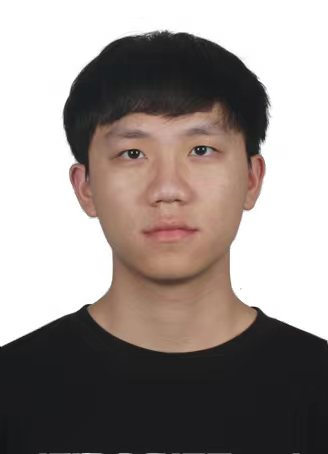}}]{Kangwen Yin} is currently working toward the Master degree with the School of Computer Science and the Beijing Key Laboratory of Intelligent Telecommunications Software and Multimedia, Beijing University of Posts and Telecommunications, China. His research interests include Internet of Things and multimodal data generation.
\end{IEEEbiography}
\vspace{-13mm}

\begin{IEEEbiography}[{\includegraphics[width=1in,height=1.25in,clip,keepaspectratio]{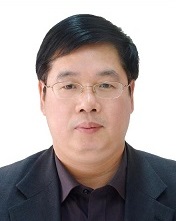}}] {Huadong Ma} (Fellow, IEEE) received the B.S. degree in mathematics from Henan Normal University, Xinxiang, China, in 1984, the M.S. degree in computer science from Shenyang Institute of Computing Technology, Chinese Academy of Science, China, in 1990, and the Ph.D. degree in
computer science from the Institute of Computing Technology, Beijing, Chinese Academy of Science in 1995. He is currently a Professor in School of Computer Science, Beijing University of Posts and Telecommunications (BUPT), China. From 1999 to 2000, he held a Visiting Position with the University of Michigan, Ann Arbor, MI, USA. His current research interests include Internet of Things and sensor networks, multimedia computing, and he has published more than 300 papers in prestigious journals (such as ACM/IEEE Transactions) or conferences (such as ACM SIGCOMM, ACM MobiCom/MM, IEEE INFOCOM) and five books. Dr. Ma received the first class prize of the Natural Science Award of the Ministry of Education, China, in 2017. He received the 2019 Prize Paper Award of IEEE TRANSACTIONS ON MULTIMEDIA, 2018 Best Paper Award from IEEE MULTIMEDIA, Best Paper Award in IEEE ICPADS2010, and Best Student Paper Award in IEEE ICME2016 for his coauthored papers. He received the National Funds for Distinguished Young Scientists in 2009. He was/is an Editorial Board Member of the IEEE Transactions on Multimedia, IEEE Internet of Things Journal, ACM Transactions on Internet of Things, and Multimedia Tools and Applications. He serves as Chair of ACM SIGMOBILE China.
\end{IEEEbiography}

\end{document}